\documentclass[11pt]{article}
\usepackage[utf8]{inputenc}
\usepackage[letterpaper,bindingoffset=0.5in,%
            left=1  in,right=1in,top=1in,bottom=1in,%
            footskip=.25in]{geometry}
\usepackage{standalone}
\usepackage{amsmath}
\usepackage{graphicx}
\usepackage{amssymb}
\usepackage{epstopdf}
\usepackage{pdflscape}
\usepackage{afterpage}
\usepackage{times}
\usepackage{indentfirst}
\usepackage{float}
\usepackage{xcolor}
\usepackage{bm}

\usepackage[labelfont=bf]{caption}
\usepackage{hhline}
\usepackage{boldline}
\usepackage{setspace}
\usepackage{url}
\usepackage{subcaption}
\usepackage{epstopdf}
\usepackage[toc,page]{appendix}

\usepackage{listings}
\usepackage{color}

\usepackage[parfill]{parskip} 
\usepackage{graphicx}
\usepackage{amssymb}
\usepackage{epstopdf}

\usepackage{framed}

\usepackage{etoolbox}
\usepackage{setspace}
\usepackage{url}
\usepackage{array}
\usepackage{longtable}
\usepackage{tabu}

\usepackage{pdflscape}
\usepackage{rotating}
\usepackage[section]{placeins}
\usepackage{enumitem}

\usepackage{pdfpages}

\usepackage{hyperref}
\hypersetup{breaklinks=true}
\newcounter{rowcntr}[table]
\renewcommand{\therowcntr}{\thetable.\arabic{rowcntr}}

\newcommand{\flb}{\nonumber\\}
\newcommand{\ensavg}[1]{\left\langle {#1} \right\rangle}
\newcommand{\rmsub}[2]{#1_\mathrm{#2}}

\newcommand{\oa}{\hat{a}}
\newcommand{\oad}{{\hat{a}^\dag}}
\newcommand{\od}{\hat{d}}
\newcommand{\odd}{{\hat{d}^\dag}}
\newcommand{\odoxi}{\od_\xi}
\newcommand{\oddoxi}{\odd_\xi}
\newcommand{\oc}{\hat{c}}
\newcommand{\ocd}{{\hat{c}^\dag}}

\newcommand{\oadet}{\rmsub{\oa}{det}}
\newcommand{\oaddet}{\rmsub{\oad}{det}}
\newcommand{\oaoutpt}{\rmsub{\oa}{out,pt}}
\newcommand{\oadoutpt}{\rmsub{\oad}{out,pt}}

\newcommand{\odout}{\rmsub{\od}{out}}
\newcommand{\oddout}{\rmsub{\odd}{out}}
\newcommand{\odpt}{\rmsub{\od}{pt}}
\newcommand{\oddpt}{\rmsub{\odd}{pt}}

\newcommand{\oxi}{\hat{\xi}}
\newcommand{\oxid}{{\oxi^\dag}}
\newcommand{\oxiint}{\rmsub{\oxi}{int}}
\newcommand{\oxiext}{\rmsub{\oxi}{ext}}
\newcommand{\oxipt}{\rmsub{\oxi}{pt}}
\newcommand{\oxidpt}{\rmsub{\oxid}{pt}}

\newcommand{\oeta}{\hat{\eta}}
\newcommand{\oetad}{\oeta^\dag}

\newcommand{\oFRPSN}{\rmsub{\hat{F}}{RPSN}}
\newcommand{\oFpt}{\rmsub{\hat{F}}{pt}}
\newcommand{\oFopt}{\rmsub{\hat{F}}{opt}}
\newcommand{\oFdopt}{\rmsub{\hat{F}^\dag}{opt}}

\newcommand{\oHam}{{\hat{\mathcal{H}}}}
\newcommand{\oHamenv}{\rmsub{\oHam}{env}}

\newcommand{\oIoutpt}{\rmsub{\hat{I}}{out,pt}}
\newcommand{\Apt}{\rmsub{A}{pt}}
\newcommand{\Apteff}{\rmsub{A}{pt,eff}}
\newcommand{\taupt}{\rmsub{\tau}{pt}}

\newcommand{\aOLO}{\rmsub{a}{LO}}
\newcommand{\aext}{\rmsub{a}{ext}}
\newcommand{\aextu}{\rmsub{a}{ext,u}}
\newcommand{\aextl}{\rmsub{a}{ext,\ell}}
\newcommand{\ano}{a_0}
\newcommand{\anl}{\rmsub{a}{0,\ell}}
\newcommand{\anu}{\rmsub{a}{0,u}}
\newcommand{\arelbb}{a^\mathrm{(bb)}_\mathrm{rel}}
\newcommand{\adrbb}{a^\mathrm{(bb)}_\mathrm{dr}}
\newcommand{\cno}{c_0}
\newcommand{\zno}{z_0}

\newcommand{\ib}{\rmsub{i}{b}}
\newcommand{\ir}{\rmsub{i}{r}}

\newcommand{\grp}{g^{(0)}}
\newcommand{\gpt}{\grp_\mathrm{pt}}
\newcommand{\gtot}{\grp_\mathrm{tot}}

\newcommand{\nm}{\rmsub{n}{ac}}
\newcommand{\nth}{\rmsub{n}{th}}
\newcommand{\nc}{\rmsub{n}{c}}
\newcommand{\nopt}{\rmsub{n}{OFSN}}
\newcommand{\nRPSN}{\rmsub{n}{RPSN}}
\newcommand{\nadd}{\rmsub{n}{add}}
\newcommand{\ncirc}{\bar{n}}

\newcommand{\omegam}{\rmsub{\omega}{ac}}
\newcommand{\omegameff}{\rmsub{\omega}{ac,eff}}
\newcommand{\omegac}{\rmsub{\omega}{opt}}
\newcommand{\omegab}{\rmsub{\omega}{b}}
\newcommand{\omegar}{\rmsub{\omega}{r}}
\newcommand{\omegaLO}{\rmsub{\omega}{LO}}
\newcommand{\domega}{\delta\omega}

\newcommand{\gammam}{\rmsub{\gamma}{ac}}
\newcommand{\gammameff}{\rmsub{\gamma}{ac,eff}}
\newcommand{\domegamopt}{\rmsub{\omega}{ac,opt}}
\newcommand{\dgammamopt}{\rmsub{\gamma}{ac,opt}}
\newcommand{\gammameasl}{\rmsub{\Gamma}{meas,\ell}}
\newcommand{\gammameasu}{\rmsub{\Gamma}{meas,u}}

\newcommand{\Deltau}{\rmsub{\Delta}{u}}
\newcommand{\Deltal}{\rmsub{\Delta}{\ell}}

\newcommand{\kappaext}{\rmsub{\kappa}{ext}}
\newcommand{\kappaint}{\rmsub{\kappa}{int}}
\newcommand{\kappapt}{\rmsub{\kappa}{pt}}

\newcommand{\chic}{\rmsub{\chi}{c}}
\newcommand{\chimeff}{\rmsub{\chi}{ac,eff}}

\newcommand{\Cii}{C_{ii}}
\newcommand{\Sii}{S_{ii}}
\newcommand{\Soddod}{S_{\odd\od}}
\newcommand{\Socdoc}{S_{\ocd\oc}}
\newcommand{\Sococd}{S_{\oc\ocd}}
\newcommand{\Siirr}{S_{ii}^\mathrm{(rr)}}
\newcommand{\Siirb}{S_{ii}^\mathrm{(rb)}}
\newcommand{\Siibb}{S_{ii}^\mathrm{(bb)}}
\newcommand{\SFFdth}{S_{\hat{F}\hat{F}^\dag}^\mathrm{th}}
\newcommand{\SFdFth}{S_{\hat{F}^\dag\hat{F}}^\mathrm{th}}
\newcommand{\SFFRPSN}{S_{\hat{F}\hat{F}}^\mathrm{RPSN}}
\newcommand{\SFFdopt}{S_{\hat{F}\hat{F}^\dag}^\mathrm{opt}}
\newcommand{\SFdFopt}{S_{\hat{F}^\dag\hat{F}}^\mathrm{opt}}

\newcommand{\kB}{\rmsub{k}{B}}

\newcommand{\etak}{\rmsub{\eta}{\kappa}}
\newcommand{\etal}{\rmsub{\eta}{\ell}}
\newcommand{\etan}{\rmsub{\eta}{n}}
\newcommand{\etar}{\rmsub{\eta}{r}}
\newcommand{\etat}{\rmsub{\eta}{t}}

\newcolumntype{N}{>{\refstepcounter{rowcntr}\therowcntr}l}

\AtBeginEnvironment{tabular}{\setcounter{rowcntr}{0}}

\usepackage[cm]{fullpage}
\title{Quantum optomechanical effects in a liquid}
\author
{ A.B. Shkarin,$^{1}$ A.D. Kashkanova,$^{1}$ C. D. Brown,$^{1}$ \\
 K. Ott,$^{2}$  S. Garcia,$^{2}$ J. Reichel,$^{3}$ and J. G. E. Harris$^{1,3}$\\
\\
\normalsize{$^{1}$Department of Physics, Yale University, New Haven, CT, 06511, USA}\\
\normalsize{$^{2}$Laboratoire Kastler Brossel, ENS/UPMC-Paris 6/CNRS, F-75005 Paris, France}\\
\normalsize{$^{3}$Department of Applied Physics, Yale University, New Haven, CT, 06511, USA}\\
}
\date{} 
\begin{document}



**Quantum optomechanics in a liquid**

A. B. Shkarin,[1,†] A. D. Kashkanova,[1,†] C. D. Brown,[1] S. Garcia,[2] K. Ott,[2] J. Reichel,[2] J. G. E. Harris[1,3,4,*]

[1] *Department of Physics, Yale University, New Haven, CT, 06520, USA*
[2] *Laboratoire Kastler Brossel, ENS-PSL Research University, CNRS,*
*UPMC-Sorbonne Universités, Collège de France, F-75005 Paris, France*
[3] *Department of Applied Physics, Yale University, New Haven, CT, 06520, USA*
[4] *Yale Quantum Institute, Yale University, New Haven, CT, 06520, USA*
[†] Equal contributions
[*] Corresponding author: jack.harris@yale.edu

Optomechanical systems provide a means for studying and controlling quantum effects in the motion of macroscopic objects. To date, quantum optomechanical effects have been studied in objects made from solids and gases. Here we describe measurements of quantum behavior in the vibrations of a liquid body. Specifically, we monitor the fluctuations of an individual acoustic standing wave in superfluid liquid helium, and find that it displays the characteristic signatures of zero-point motion and measurement back-action. This opens the possibility of exploiting the properties of liquids in general (and superfluid helium in particular) to access qualitatively new regimes of quantum optomechanics.

The fundamental interactions between light and matter are reversible. However, a macroscopic object possesses so many degrees of freedom that interactions with light typically produce complex excitations within the object. Such processes are not usually reversible by practical means, and so effectively destroy the light's quantum state and preclude access to the object's quantum dynamics. This limitation may be overcome by identifying objects in which a collective degree of freedom interacts strongly with the electromagnetic (EM) field but is well-isolated from other degrees of freedom. This can be achieved, for example, in superconducting circuits (1), atomic gases (2), solid-state magnets (3), or objects whose vibrations couple to an EM cavity (4). In the lattermost case (known as an optomechanical system) the vibrational mode is isolated by its high quality factor and the environment's low temperature, while its interaction with the EM field results from radiation pressure, electrostriction, or other reversible processes (4).

Optomechanical systems provide a means for studying quantum effects on macroscopic scales and for realizing new forms of dynamical control in both the quantum and classical domains. In addition, mechanical oscillators interact reversibly with many other physical systems and so can be used to build hybrid quantum devices with efficient EM readout. Quantum effects have been demonstrated in mechanical devices as massive as ~100 ng (5), at temperatures as high as ~300 K (6), using EM fields in the microwave (7,8,9) and near-infrared domains (10,11,12,13,5,6), and with mechanical oscillators formed from solids (5,6,7,8,9,10,11,13) and from ultracold gases (12). They have been used to realize hybrid quantum systems with superconducting qubits (7,14,15), atoms (16), and solid-state impurities (17), and show considerable promise for coherent optical-to-microwave conversion (18,19).

Here we describe quantum optomechanical behavior in a system where the mechanical oscillator is a liquid. Specifically, we measure the sidebands imprinted on a laser beam that has interacted with a cavity full of superfluid liquid He. These sidebands are produced by the fluctuations of a single acoustic mode within the liquid, and they show the characteristic asymmetry and cross-correlations that result from the acoustic mode's zero-point motion and the quantum backaction of the light used to monitor it (20,21,22). While these features have been observed in a number of optomechanical systems (5,9,10,21,23,24), their observation in superfluid He is significant for several reasons. First, liquids in general (and superfluid He in particular) offer mechanical degrees of freedom that differ qualitatively from solids. These include motion with unbounded displacement (e.g., in the form of rotational flow), rather than simply harmonic

oscillations about an equilibrium (*25*). Second, the presence of a free surface allows a liquid body's geometry and topology to be reconfigured *in situ* and to serve as a dynamical degree of freedom. Third, superfluid He can host a number of atom-like impurities (such as electrons, ions, and $\text{He}_2^*$ excimers) that are potentially suitable for hybrid quantum systems (*26*). Fourth, superfluid helium's remarkable physical properties help to address some of the outstanding technical challenges in optomechanics: its exceptional thermal conductivity allows for effective cooling by conventional refrigerators, its acoustic damping can be calculated *a priori* (*27,28*), and its ability to conformally fill or coat a cryogenic EM resonator (*27,28,29,30*) means that such devices require no *in situ* alignment. Lastly, this type of device offers the possibility of applying precision optical measurements to address outstanding questions regarding the fundamental properties of superfluid He (*31,32*).

Some of these features can be explored by optomechanical systems in the classical regime [using normal fluids (*33,34*) or superfluid He (*27,28,29,30*)]. However the quantum regime of fluid-based optomechanical systems remains largely unexplored by theory or experiment.

The device used in this study is similar to the one described in Refs. (*28,35*), and is shown in Fig. 1. It consists of a cavity formed between the end faces of two optical fibers. The fibers are laser-machined to have concave surfaces, which are coated with high-reflectivity distributed Bragg reflectors (DBRs) (*36*). The fibers are aligned in a pair of glass ferrules (*37*) mounted in a Cu cell which is connected to the mixing chamber (MC) of a dilution refrigerator. In contrast with Ref. (*28*), the use of two separate ferrules greatly improves the thermal link between the He in the cavity and the MC. The cavity length is $L = 69.1$ μm. The two DBRs' reflectivities are $R_1 = 0.99995$ and $R_2 = 0.99999$. Light is coupled to and collected from the cavity via the lower-reflectivity DBR. The cavity linewidth, external coupling rate, and finesse are measured to be $\kappa = 2\pi \times 21$ MHz, $\kappa_{\text{ext}} = 2\pi \times 10$ MHz, and $F = 9.5 \times 10^4$ respectively. The frequency of the optical mode used in these measurements is $\omega_{\text{opt}} = 2\pi \times 196.0$ THz.

When the cavity is filled with liquid He, the fiber ends also confine acoustic modes. The acoustic modes' density variations alter the index of refraction experienced by the optical modes. Equivalently, the optical modes' intensity variations exert a force that can excite the acoustic modes. As discussed in Refs. [*28,38*], this leads to optomechanical coupling of the conventional (*4*) form $H_{\text{OM}} = g^{(0)} a^\dagger a (c^\dagger + c)$ where $a$ and $c$ are the annihilation operators for cavity photons and phonons, respectively. Straightforward geometric considerations show that the single-

quantum optomechanical coupling rate $g^{(0)}$ is maximized for an acoustic mode with half the wavelength of the optical mode (28,38). As a result, the optical mode used in this experiment couples to an acoustic mode with resonant frequency $\omega_{ac} = 2\pi \times 319.2$ MHz.

The device was characterized using optomechanically induced transparency / amplification (OMIT/A). For OMIT/A measurements, three laser beams drive the cavity: a weak ("probe") beam, a stronger ("control") beam, and a far-detuned local oscillator ("LO") beam (Fig. 2A). Beating between the probe and control beams causes the intracavity optical intensity to oscillate, thereby driving the acoustic mode. The resulting oscillation of the acoustic mode then adds sidebands to these beams. One sideband from the control beam interferes with the probe beam; as a result, a heterodyne measurement of the reflected probe (made using the LO) provides information about the acoustic mode's driven motion (39,28) (the OMIT/A measurement and analysis are described in detail in the Supplementary Materials).

Figure 2B shows a typical OMIT/A spectrum in which the probe beam's detuning is varied while keeping the control beam fixed, thereby varying the frequency $\omega$ of the drive applied to the acoustic mode. The heterodyne signal is fit to a Lorentzian with a constant background: $a_{OMIT}[\omega] = a_{bg} + a_{lor}/(\gamma_{ac,eff}/2 + i(\omega - \omega_{ac,eff}))$, where $a_{bg}$ and $a_{lor}$ are both complex. This fit returns values for $a_{bg}$, $a_{lor}$, and the acoustic linewidth $\gamma_{ac,eff}$ and frequency $\omega_{ac,eff}$ [the subscript "eff" indicates that the optomechanical interaction $H_{OM}$ can shift the oscillator's effective resonance frequency and linewidth from their bare values $\gamma_{ac}$ and $\omega_{ac}$ (4)].

An OMIT/A spectrum (such as the one shown in Fig. 2B) can be characterized by its magnitude $A \equiv abs(X)$ and phase $\Psi \equiv arg(X)$, where $X \equiv 2a_{lor}/\gamma_{ac,eff}a_{bg}$. Figure 2, C and D, shows $A$ and $\Psi$ (derived from the best-fit values of $a_{bg}$, $a_{lor}$, and $\gamma_{ac,eff}$) as a function of the control beam's detuning $\Delta_{con}$ and power $P_{con}$. Both $A$ and $\Psi$ show a feature of width $\sim \kappa$ centered at $\Delta_{con} \sim \pm\omega_{ac}$, corresponding to the probe being resonant with the optical cavity. Figure 2, E and F, shows $\omega_{ac,eff}(\Delta_{con},P_{con})$ and $\gamma_{ac,eff}(\Delta_{con},P_{con})$, which exhibit the standard optical spring and damping (4).

The solid lines in Fig. 2, C to F, are the result of fitting all the OMIT/A data to the dynamical backaction model described in Ref. (28). The two fitting parameters are $g^{(0)}$, which represents the unitary electrostrictive interaction between the cavity mode and the acoustic mode; and $g_{pt}^{(0)}$, which represents an optomechanical interaction with a 90° phase lag between the optical

intensity and the resulting force on the acoustic mode. As described in Ref. (*28*), this arises from a photothermal effect in which optical absorption in the DBRs drives the acoustic mode, but with a bandwidth << $\omega_{ac}$. The best-fit values are $g^{(0)} = 2\pi \times (3.6 \pm 0.1)$ kHz and $g_{pt}^{(0)} = 2\pi \times (0.8 \pm 0.1)$ kHz (except where noted, errors correspond to the statistical uncertainty in the fits). This is consistent with the *a priori* calculation $g^{(0)} = 2\pi \times (3.9 \pm 0.2)$ kHz (see Supplementary Materials) and the value $g_{pt}^{(0)} = 2\pi \times (0.97 \pm 0.05)$ kHz found in a similar device (*28*).

When no drive is applied, the acoustic mode's thermal and quantum fluctuations still add motional sidebands to light interacting with the cavity. Standard optomechanics theory predicts that the acoustic mode's thermal fluctuations contribute equally to the red and blue motional sidebands, but that its quantum fluctuations contribute unequally (*4*). Specifically, when the blue sideband is converted to a photocurrent via heterodyne detection, its power spectral density $S_{ii}^{(bb)}$ is predicted to consist of a noise floor plus a peak that reproduces the acoustic mode's Lorentzian lineshape. When the photocurrent is appropriately calibrated (see below and the Supplementary Materials), the height of this peak $h_{bb}$ equals the mode's mean phonon number $n_{ac}$. The same holds for $S_{ii}^{(rr)}$ (the photocurrent spectrum resulting from the red sideband), except that its peak height $h_{rr} = n_{ac} + 1$. Furthermore, the spectrum of correlations between the two sidebands ($S_{ii}^{(rb)}$) is predicted to have a real part consisting of the same lineshape (with height $h_{rb,Re} = n_{ac} + \frac{1}{2}$) and an imaginary part with an antisymmetric lineshape of magnitude $h_{rb,Im} = \frac{1}{2}$. [Information equivalent to $S_{ii}^{(rb)}$ can also be extracted by measuring both quadratures of the reflected light (*6,11,40*)].

While various interpretations can be applied to these features (*20,21,22*), they are intrinsically quantum in nature as the perceived energy differences between $S_{ii}^{(bb)}$, $S_{ii}^{(rr)}$, and $S_{ii}^{(rb)}$ are set by the energy of a single phonon $\hbar\omega_{ac}$. It is convenient to parametrize the sidebands' quantum features via three parameters: $H_{AS} \equiv h_{rr} - h_{bb}$, $H_{Re} \equiv 2(h_{rb,Re} - h_{bb})$, and $H_{Im} \equiv 2h_{rb,Im}$. These are each predicted to be unity, independent of experimental parameters such as temperature and laser power.

The system described here operates well in the resolved sideband regime ($\omega_{ac} \approx 15\,\kappa$), so it is impractical to measure the two sidebands produced from a single beam (at least one will be strongly suppressed by the cavity's response). Instead, we apply two measurement beams to the

cavity: an "upper" beam with detuning (relative to the cavity resonance) $\Delta_u = \omega_{ac} + \delta$ and a "lower" beam with detuning $\Delta_\ell = -\omega_{ac} - \delta$ where $\delta$ is set to $2\pi \times 100$ kHz. As illustrated in the inset of Fig. 3 this ensures that two motional sidebands are approximately resonant with the cavity: the blue (red) sideband of the lower (upper) beam. The offset $\delta$ is chosen so that these sidebands do not overlap, but do lie within the measurement bandwidth. The sidebands are recorded simultaneously via a heterodyne measurement and $S_{ii}^{(bb)}$, $S_{ii}^{(rr)}$, and $S_{ii}^{(rb)}$ are computed from this record. Each of these records is calibrated (Supplementary Materials) so that the features in $S_{ii}^{(rr)}$, $S_{ii}^{(bb)}$, and $S_{ii}^{(rb)}$ should be related to $n_{ac}$ as described above.

Figure 3 shows a typical measurement of $S_{ii}^{(rr)}$ and $S_{ii}^{(bb)}$ (with their frequency-independent background subtracted) as well as $S_{ii}^{(rb)}$. The magnitude of the features in this data appear consistent with the quantum effects described above. To quantify this comparison we fit $S_{ii}^{(rr)}$, $S_{ii}^{(bb)}$, and $\text{Re}(S_{ii}^{(rb)})$ to $h_x / (1 + 4(\omega - \omega_{ac,eff})^2 / \gamma_{ac,eff}^2)$ with $x = \{rr; bb; rb,\text{Re}\}$, while $\text{Im}(S_{ii}^{(rb)})$ is fit to $h_{rb,\text{Im}} (\omega - \omega_{ac,eff})(2\gamma_{ac,eff})^{-1}(1 + 4(\omega - \omega_{ac,eff})^2 / \gamma_{ac,eff}^2)^{-1}$. These fits give $H_{AS} = 1.18 \pm 0.03$, $H_{Re} = 1.14 \pm 0.06$, $H_{Im} = 1.14 \pm 0.02$. The small but statistically significant deviation from unity is discussed below.

$H_{AS}$, $H_{Re}$, and $H_{Im}$ are defined so as to reflect only quantum effects; however the fit parameters from which they are determined ($h_{bb}$, $h_{rr}$, $h_{rb,Re}$, and $h_{rb,Im}$) reflect both thermal and quantum fluctuations. To compare the quantum and thermal signatures in the data, we measured heterodyne spectra similar to those in Fig. 3 over a range of $T_{MC}$ (the MC temperature) and $n_{circ}$ (the intracavity photon number). Figure 4, A and B, shows the inferred phonon number of the acoustic mode's bath, defined as $n_{th} = n_{ac}(\gamma_{ac,eff}/\gamma_{ac}) - n_O \gamma_O / \gamma_{ac}$. This expression was evaluated by fitting heterodyne spectra (as in Fig. 3) for $\gamma_{ac,eff}$ and $n_{ac}$ (for these measurements we use $n_{ac} = \frac{1}{2}(h_{bb} + h_{rr} - 1)$). Standard optomechanics theory (4) was used to calculate the phonon number associated with the quantum back-action $n_O$ and the optical damping rate $\gamma_O = \gamma_{ac,eff} - \gamma_{ac}$. For all the measurements described here $n_{th}$ nearly equals $n_{ac}$, as the "laser cooling" factor $\gamma_{ac,eff}/\gamma_{ac}$ ranges from 0.95 to 1.05 while the "quantum back-action" term $n_O \gamma_O / \gamma_{ac} < 1.1$ for these measurements. We plot $n_{th}$ (rather than $n_{ac}$) in Fig. 4, A and B to facilitate comparison with the thermal model discussed in the Supplementary Materials.

Figure 4A shows $n_{th}$ versus $T_{MC}$. For $T_{MC} \gtrsim 150$ mK $n_{th}$ tracks $T_{MC}$, while for $T_{MC} \lesssim 150$ mK $n_{th}$ does not track $T_{MC}$ and clearly depends on $n_{circ}$. Qualitatively similar behavior was found in Ref. (*28*), and was accounted for by a thermal model in which the He temperature was set by balancing optical absorption in the DBRs against the cooling provided by the slender superfluid region which linked that device to the MC. In contrast, the present device's more open geometry provides improved cooling, but the absence of a thermal bottleneck means that the temperature in the cavity is not uniform. We calculate the cavity's temperature distribution using standard models of thermal transport and convert this into an effective temperature for the mode $T_{eff}$ that depends upon $T_{MC}$ and $n_{circ}$ (Supplementary Materials). Fig. 4B shows the same values of $n_{th}$ as Fig. 4A, but plotted versus $T_{eff}$. In this case the data shows close agreement with the prediction $n_{th} = 1/(e^{\omega_{ac}/k_B T} - 1)$ over the full range of $T_{MC}$ and $n_{circ}$, indicating that this approach captures the main features of the device's thermal behavior. The small but systematic deviation from the prediction is roughly independent of $T_{MC}$ and $n_{circ}$, consistent with an imperfect calibration of the heterodyne signal (Supplementary Material, Section 1.3).

Figure 4C shows $H_{AS}$, $H_{Re}$, and $H_{Im}$ as a function of $T_{eff}$. The points in Fig. 4C are derived from data and fits similar to those in Fig. 3 (and from the same set of measurements used to produce Fig. 4, A and B). The uncertainty grows at higher $T_{eff}$ because of the rapid increase of $\gamma_{ac}$ with $T_{eff}$, which makes the motional sidebands harder to distinguish from the noise floor. The uncertainty also grows at the lowest values of $T_{eff}$ owing to the need to use low laser power. The data in Fig. 4C are consistent with the theoretical prediction (dashed line), indicating their origin in the coherent quantum dynamics of the cavity's acoustic and EM modes.

In conclusion, we have isolated a single normal mode of a liquid body and measured its quantum fluctuations. This result is distinct from the large body of work on the quantum aspects of superfluid He's bulk properties, which reflect the aggregate behavior of very many normal modes. It is also distinct from work on quantum effects directly related to the superfluid's wavefunction (such as persistent flow, quantized vortices, and Josephson effects). Although superfluidity greatly facilitates the experiments described here (by suppressing the viscous damping of the acoustic mode), the acoustic mode itself and its quantum dynamics are generic to any liquid.

# References

[1] M. H. Devoret, R. J. Schoelkopf, *Science* **339**, 1169-1174 (2013).

[2] D. M. Stamper-Kurn, in *Cavity Optomechanics,* M. Aspelmeyer, T. J. Kippenberg, F. Marquardt, Eds. (Springer, Heidelberg-New York-Dordrecht-London, 2014) pp. 283-325.

[3] Y. Tabuchi, S. Ishino, A. Noguchi, T. Ishikawa, R. Yamazaki, K. Usami, Y. Nakamura, *Science* **349**, 405-408 (2015).

[4] M. Aspelmeyer, T. Kippenberg, F. Marquardt, *Rev. Mod. Phys.* **86**, 1391 (2014).

[5] M. Underwood, D. Mason, D. Lee, H. Xu, L. Jiang, A. B. Shkarin, K. Børkje, S. M. Girvin, J. G. E. Harris, *Phys. Rev. A* **92**, 061801(R) (2015).

[6] T. P. Purdy, K. E. Grutter, K. Srinivasan, J. M. Taylor, *Science* **356**, 1265-1268 (2017).

[7] A. D. O'Connell, M. Hofheinz, M. Ansmann, R. C. Bialczak, M. Lenander, E. Lucero, M. Neeley, D. Sank, H. Wang, M. Weides, J. Wenner, J. M. Martinis, A. N. Cleland, *Nature* **464**, 697-703 (2010).

[8] T. A. Palomaki, J. D. Teufel, R. W. Simmonds, K. W. Lehnert, *Science* **342**, 710-713 (2013).

[9] C. F. Ockeloen-Korppi, E. Damskägg, J.-M. Pirkkalainen, A. A. Clerk, M. J. Woolley, M. A. Sillanpää, *Phys. Rev. Lett.* **117**, 140401 (2016).

[10] A. H. Safavi-Naeini, J. Chan, J. T. Hill, T. P. Mayer Alegre, A. Krause, O. Painter, *Phys. Rev. Lett.* **108**, 033602 (2012).

[11] T. P. Purdy, R. W. Peterson, C. A. Regal, *Science* **339**, 801-804 (2013).

[12] D. W. C. Brooks, T. Botter, T. P. Purdy, S. Schreppler, N. Brahms, D. M. Stamper-Kurn, *Nature* **488**, 476-480 (2012).

[13] V. Sudhir, D. J. Wilson, R. Schilling, H. Schuetz, S. A. Fedorov, A. H. Ghadimi, A. Nunnenkamp and T. J. Kippenberg, *Phys. Rev. X* **7**, 011001, (2017).

[14] J.-M. Pirkkalainen, S. U. Cho, Jian Li, G. S. Paraoanu, P. J. Hakonen, M. A. Sillanpää, *Nature* **494**, 211-215 (2013).

[15] Y. Chu, P. Kharel, W. H. Renninger, L. D. Burkhart, L. Frunzio, P. T. Rakich, R. J. Schoelkopf, ArXiv:1703.00342 (2017).

[16] C. B. Møller, R. A. Thomas, G. Vasilakis, E. Zeuthen, Y. Tsaturyan, K. Jensen, A. Schliesser, K. Hammerer, E. S. Polzik, *Nature* **547**, 191-195 (2017).

[17] D. A. Golter, T. Oo, M. Amezcua, K. A. Stewart, H. Wang, *Phys. Rev. Lett.* **116**, 143602 (2016).

[18] R. W. Andrews, R. W. Peterson, T. P. Purdy, K. Cicak, R. W. Simmonds, C. A. Regal, K. W. Lehnert, *Nat. Phys.* **10**, 321–326 (2014).

[19] J. Bochmann, A. Vainsencher, D. D. Awschalom, A. N. Cleland, *Nat. Phys.* **9**, 712-716 (2013).

[20] F. Ya. Khalili, H. Miao, H. Yang, A. H. Safavi-Naeini, O. Painter, Y. Chen, *Phys. Rev. A* **86**, 033840 (2012).

[21] A. J. Weinstein, C. U. Lei, E. E. Wollman, J. Suh, A. Metelmann, A. A. Clerk, K. C. Schwab, *Phys. Rev. X* **4**, 041003 (2014).

[22] K. Børkje, *Phys. Rev. A* **94**, 043816 (2016).

[23] N. Brahms, T. Botter, S. Schreppler, D. W. C. Brooks, D. M. Stamper-Kurn, *Phys. Rev. Lett.* **108**, 133601 (2012).

[24] T. P. Purdy, P. L. Yu, N. S. Kampel, R. W. Peterson, K. Cicak, R. W. Simmonds, C. A. Regal, *Phys. Rev. A* **92**, 031802(R) (2015).

[25] L. Childress, M. P. Schmidt, A. D. Kahkanova, C. D. Brown, G. I. Harris, A. Aiello, F. Marquardt, J. G. E. Harris, *ArXiv*:1708.01803 (2017).

[26] P. M. Platzman, M. I. Dykman, *Science* **284**, 1967-1969 (1999).

[27] L. A. De Lorenzo, K. C. Schwab, *New J. Phys.* **16**, 113020 (2014).

[28] A. D. Kashkanova, A. B. Shkarin, C. D. Brown, N. E. Flowers-Jacobs, L. Childress, S. W. Hoch, L. Hohmann, K. Ott, J. Reichel, J. G. E. Harris, *Nat. Phys.* **13**, 74-79 (2017).

[29] G. I. Harris, D. L. McAuslan, E. Sheridan, Y. Sachkou, C. Baker & W. P. Bowen, *Nat. Phys.* **12**, 788-793 (2016).

[30] F. Souris, X. Rojas, P. H. Kim, J. P. Davis, *Phys. Rev. Appl.* **7**, 044008 (2017).

[31] W. F. Vinen, M. Tsubota, A. Mitani, *Phys. Rev. Lett.* **91**, 135301 (2003).

[32] E. Kozik, B. Svistunov, *Phys. Rev. Lett.* **92**, 035301 (2004).

[33] G. Bahl, K. H. Kim, W. Lee, J. Liu, X. Fan, T. Carmon, *Nat. Comm.* **4**, 1994 (2013).

[34] S. Kaminski, L. L. Martin, S. Maayani, T. Carmon, *Nat. Phot.* **10**, 758-761 (2016).

[35] A. D. Kashkanova, A. B. Shkarin, C. D. Brown, N. E. Flowers-Jacobs, L. Childress, S. W. Hoch, L. Hohmann, K. Ott, J. Reichel, J. G. E. Harris, *J. Opt.* **19**, 034001 (2017).

[36] D. Hunger, C. Deutsch, R. J. Barbour, R. J. Warburton, J. Reichel, *AIP Adv.* **2**, 012119 (2012).

[37] N. E. Flowers-Jacobs, *et al.*, *Appl. Phys. Lett.* **101**, 221109 (2012).

[38] G. S. Agarwal, S. S. Jha, *Phys. Rev. A* **90**, 023812 (2014).

[39] S. Weis, *et al.*, *Science* **330**, 1520–1523 (2010).

[40] K. Børkje, A. Nunnenkamp, B. M. Zwickl, C. Yang, J. G. E. Harris, S.M. Girvin, *Phys. Rev. A* **82**, 013818 (2010).

**Supplementary Materials**

Measurement setup

Theoretical description of the measurement results

Displacement measurement calibration

Thermal model

Table S1

Figure S1 – S12

**Figure Captions**

**Fig. 1: Schematic of the experiment.** (**A**) Top: Illustration of the optomechanical device. The optical fibers (yellow) and ferrules (white) are attached to a glass plate (orange) inside a Cu cell (gray) which is mounted to the mixing chamber of a dilution refrigerator (DR, not shown). Liquid He (blue) is introduced to the cell via a capillary line. The fibers enter the cell via epoxy feedthroughs (black). Bottom: An enlarged view of the cavity. Red curve: the intensity profile of an optical mode; blue shading: the density profile of the acoustic mode to which it couples. The actual optical and acoustic modes used in this work have, respectively, 91 and 182 half-wavelengths along the cavity length. (**B**) Simplified layout of the measurement setup. Light from a tunable laser (TL) passes through a phase modulation system ($\phi$M) driven by a microwave source (MW). Light is delivered to (and collected from) the DR via a circulator (pink). The reflected light is collected on a photodiode (PD), and the resulting photocurrent is analyzed by a data acquisition system (DAQ). Details are given in the Supplementary Materials.

**Fig. 2: Optomechanical characterization.** (**A**) Illustration of the OMIT/A measurement scheme. Black curve: the cavity lineshape. Colored arrows: the "probe" and "control" laser tones described in the text (the far-detuned LO tone is not shown). (**B**) A typical OMIT/A measurement ($\Delta_{con} = 2\pi \times 320$ MHz, $P_{con} = 5.1$ µW.). Upper panel: abs($a_{OMIT}$) normalized so that the background $a_{bg} = 1$. Lower panel: arg($a_{OMIT}$). Both are shown as a function of the beat note frequency $\omega$. The solid line is the fit described in the text. The subsequent panels show the four parameters extracted from this type of fit, all as a function of $\Delta_{con}$ and $P_{con}$. (**C**) Amplitude of the OMIT/A signal $A$. (**D**) Phase of the OMIT/A signal $\Psi$. (**E**) Acoustic mode resonance frequency $\omega_{ac,eff}$. (**F**) Acoustic mode damping rate $\gamma_{ac,eff}$. The solid lines in (**C**) – (**F**) are a fit to all the data shown, using the model described in Ref (28). The only free parameter is the complex optomechanical coupling rate $g_0$. For all of these measurements, 25 mK < $T_{MC}$ < 60 mK.

**Fig. 3: Sidebands produced by the acoustic mode's fluctuations.** Inset: Illustration of the measurement scheme. Black curve: the cavity lineshape. Colored arrows: the laser tones described

in the text. Colored curves: the mechanical sidebands. Upper panel: the spectrum of the red and blue motional sidebands ($S_{ii}^{(rr)}$ and $S_{ii}^{(bb)}$) and the real part of their cross-correlation ($\text{Re}[S_{ii}^{(rb)}]$). A frequency-independent background has been subtracted from $S_{ii}^{(rr)}$ and $S_{ii}^{(bb)}$. Lower panel: the imaginary part of the cross-correlation ($\text{Im}[S_{ii}^{(rb)}]$). The data were normalized and fit as described in the text. For this measurement $T_{MC} = 50$ mK and $n_{circ} = 970$.


**Fig. 4: Thermal and quantum fluctuations of the acoustic mode.** (**A**) The mean phonon number $n_{th}$ associated with the acoustic mode's bath temperature, plotted versus $T_{MC}$. $n_{th}$ is extracted from measurements of $S_{ii}^{(rr)}$ and $S_{ii}^{(bb)}$ as described in the text. (**B**) The same measurements of $n_{th}$, but plotted as a function of $T_{eff}$, the effective device temperature calculated in the Supplementary Materials. The color of each marker encodes $n_{circ}$, the intracavity photon number. The black points are obtained by averaging data in 30 mK bins. In both (A) and (B) the dashed lines show the expected behavior $n_{th} = 1/(e^{\omega_{ac}/k_B T} - 1)$, while the grey area represents the systematic uncertainty resulting from the calibration of the heterodyne signal. (**C**) Three measures of the quantum features, plotted as a function of $T_{eff}$. The dashed line is the prediction of quantum optomechanics theory, and the grey area shows the systematic uncertainty resulting from the calibration of the heterodyne signal. Each data point is extracted from data and fits similar to Fig. 3. Error bars indicate the statistical uncertainty in these fits.

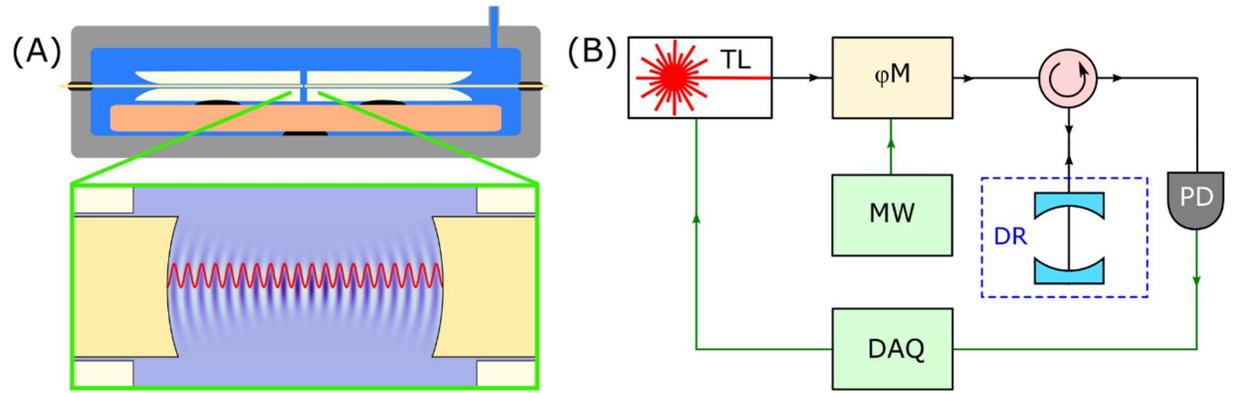

**Figure 1**

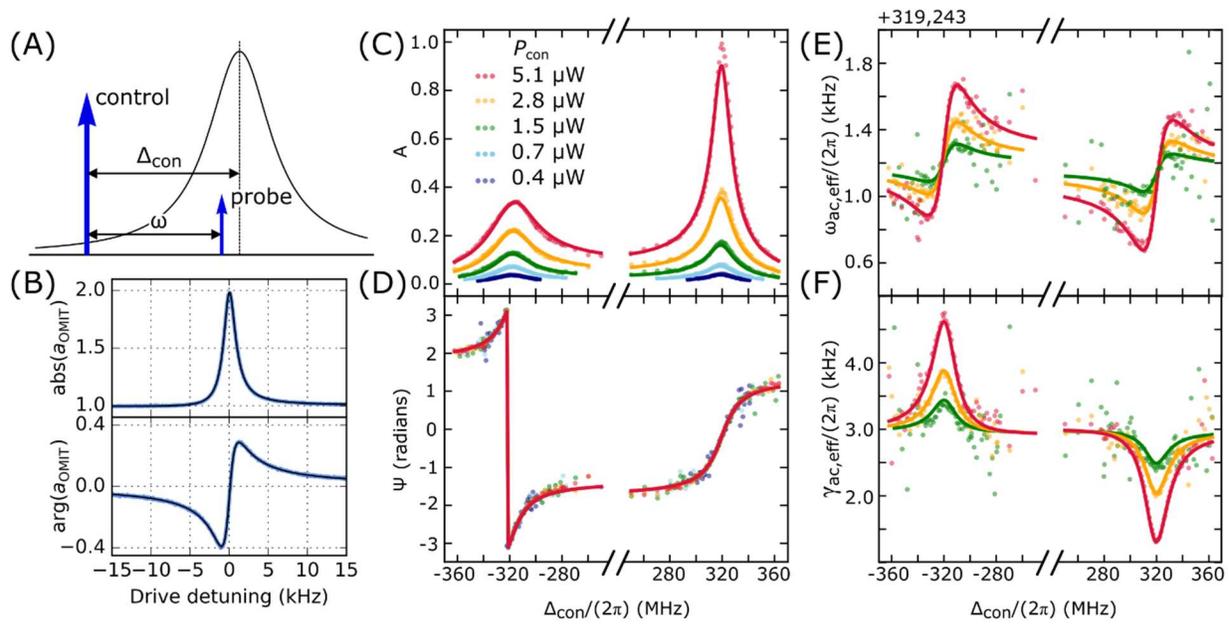

**Figure 2**

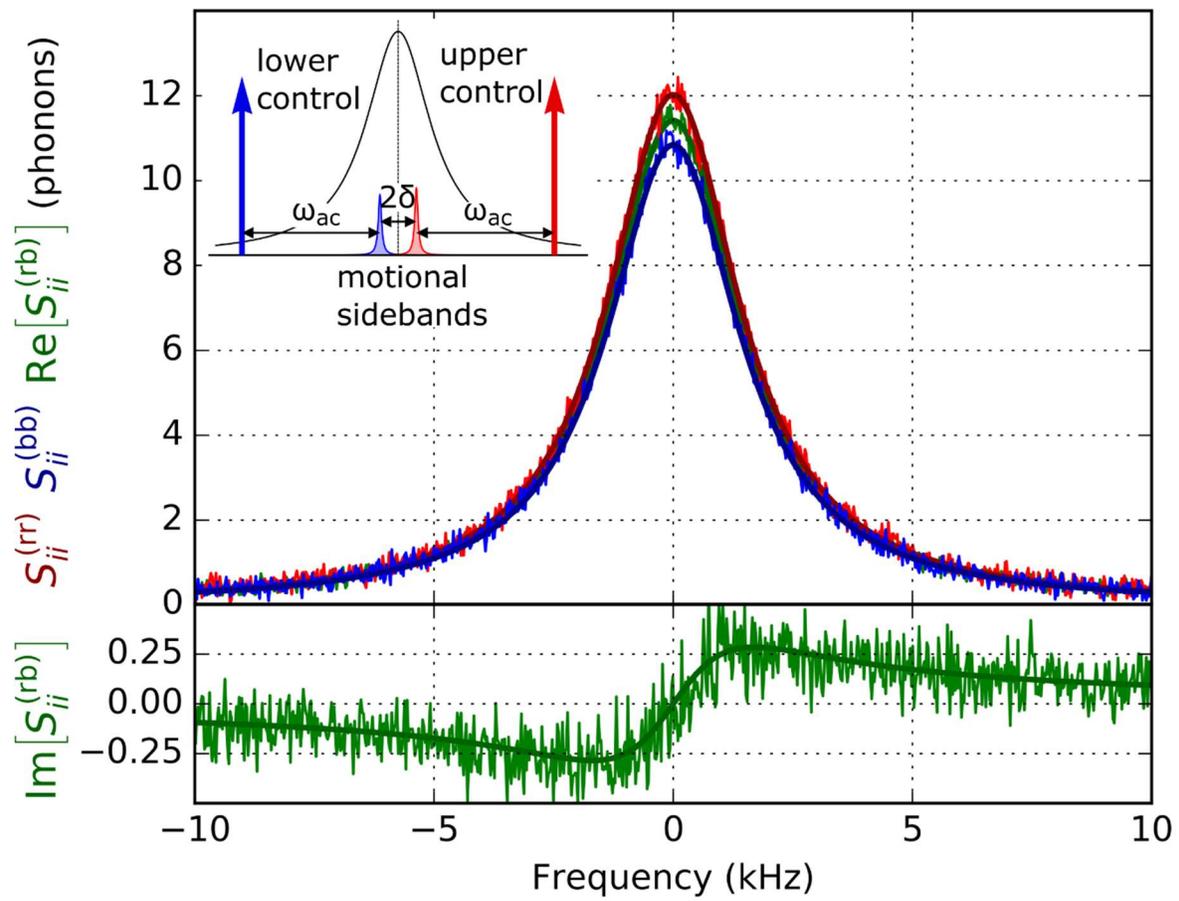

**Figure 3**

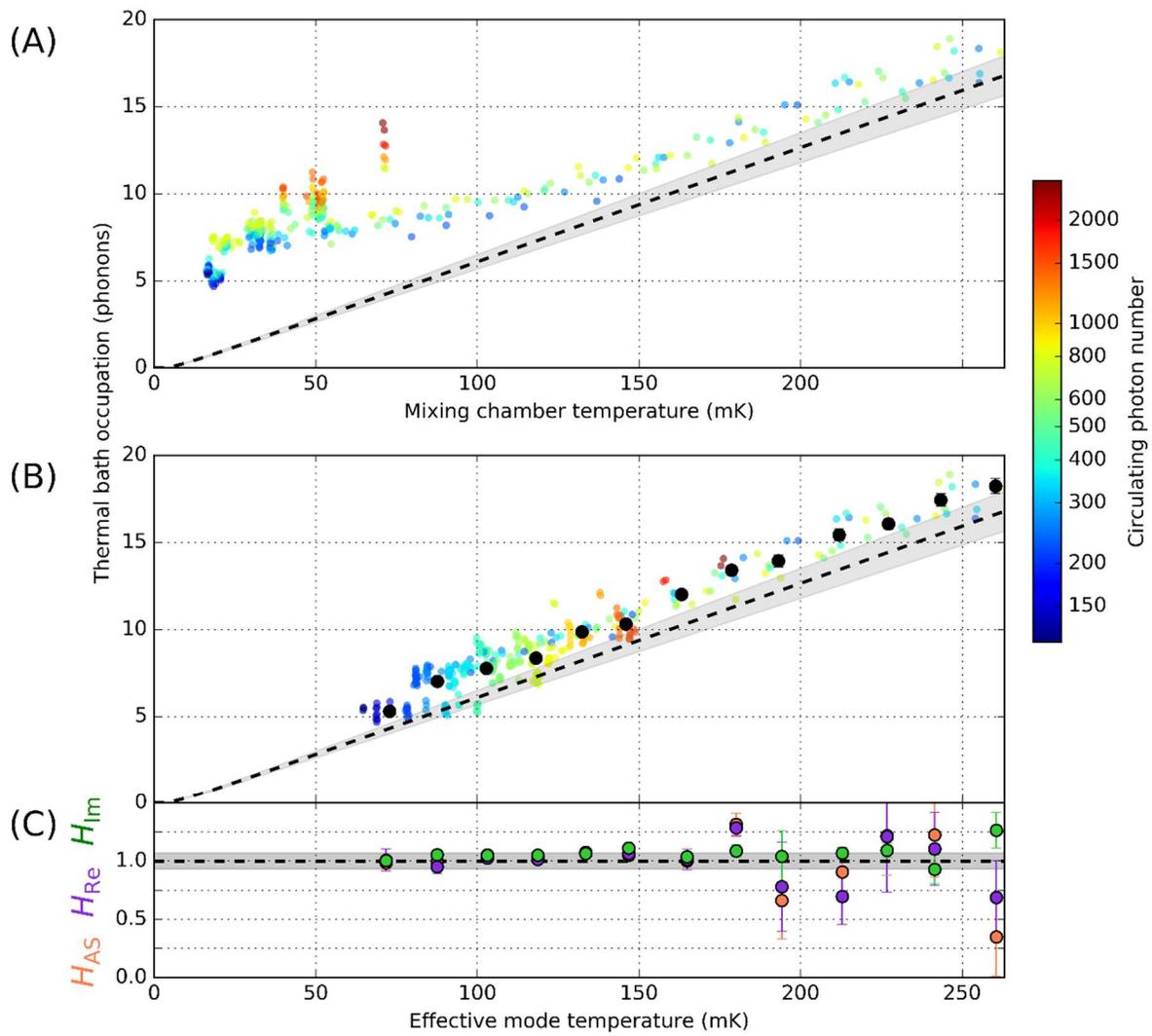

**Figure 4**



\maketitle 

\tableofcontents

\newpage

\section{Measurement setup}
\begin{sidewaysfigure}
\centering
\includegraphics[width=1\textwidth]{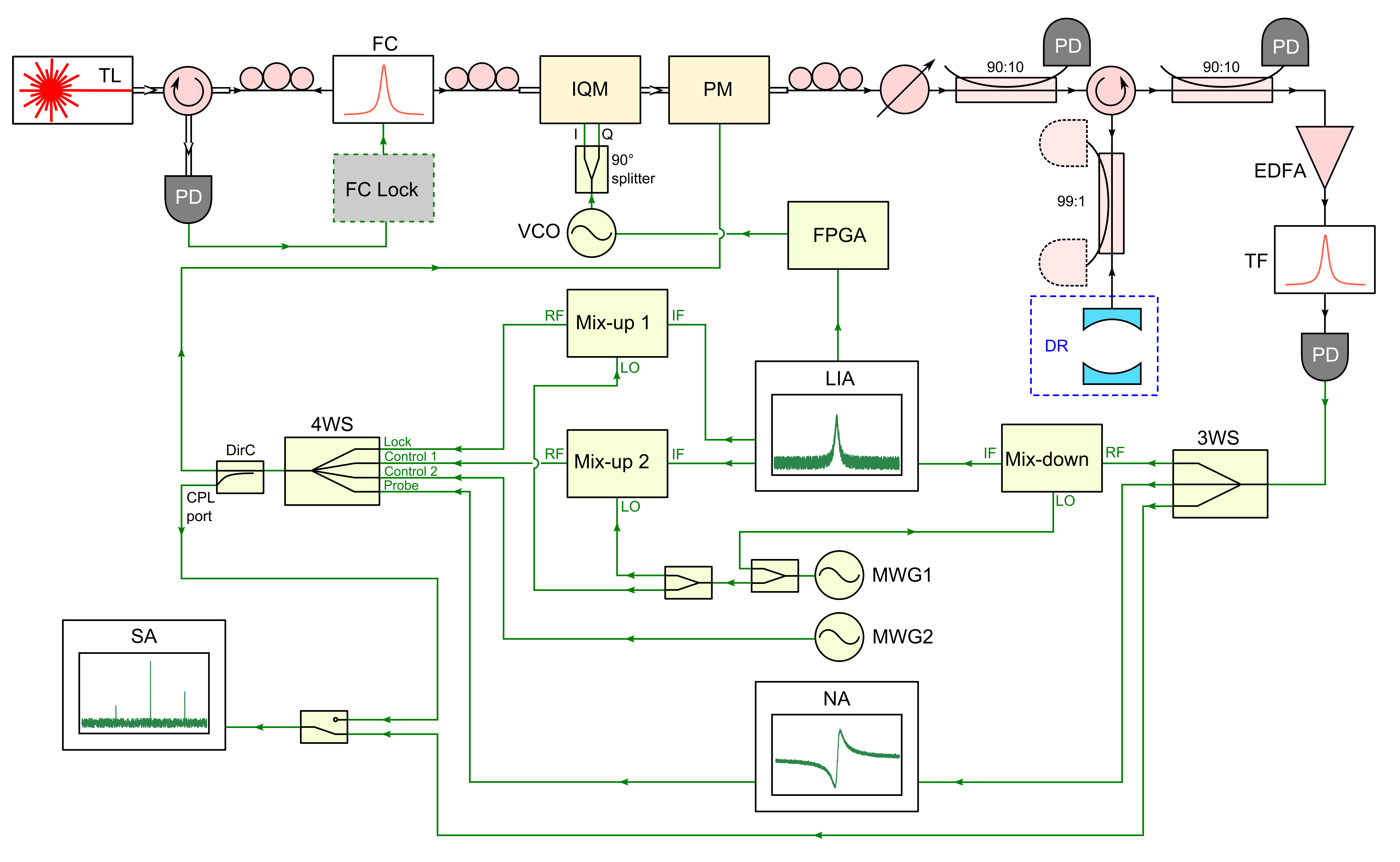} 
\caption{Measurement setup. SM fiber: black line.  PM fiber: double black lines. Electrical path: green lines. TL: tunable laser. FC: filter cavity. FC Lock: locking circuit for the filter cavity. IQM: IQ-modulator. PM: phase modulator. PD: photodiode. EDFA: erbium doped fiber amplifier. TF: tunable filter. DR: dilution refrigerator. 3WS: three-way splitter. NA: network analyzer. LIA: lock-in amplifier.  SA: spectrum analyzer. MWG: microwave generator. FPGA: field programmable gate array. VCO: voltage controlled oscillator. 4WS: four-way splitter. DirC: directional coupler. }
\label{fig:ms}
\end{sidewaysfigure}

The purpose of this section is to describe the experimental setup used in the experiment. A schematic of the setup is shown in figure \ref{fig:ms}.

\subsection{Optical setup}
Light is produced by a tunable laser (TL)\footnote{Pure Photonics PPCL200} and passes through a circulator and a filter cavity (FC)\footnote{MicronOptics FFP-TF, $\kappa/2\pi=30$ MHz, $\omega_\mathrm{FSR}/2\pi=15$ GHz}. The reflection from the FC is used to lock it to the frequency of the TL. Light transmitted through the FC passes through an IQ-modulator (IQM)\footnote {EOspace QPSK modulator IQ-0DKS-25-PFA-PFA-LV-UL} operating in the single sideband suppressed carrier mode. The IQM serves as a frequency shifter to lock the laser to the experimental cavity. The tone generated by the IQM is used as a local oscillator (LO) for the heterodyne detection.

\begin{figure}
\centering
\includegraphics[width=1\textwidth]{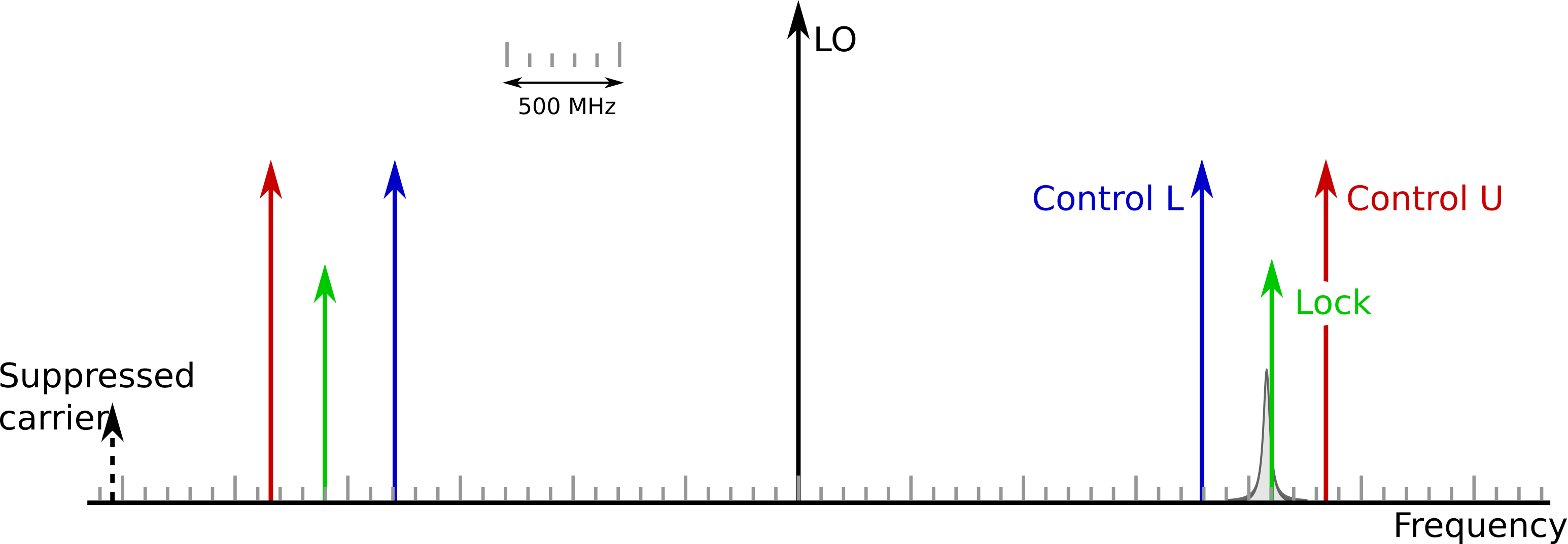} 
\caption{Beams incident on the cavity during the Brownian motion measurement. The horizontal axis is to scale. The phase modulator adds 2 control beams and a lock beam on either side of the LO beam. The laser is locked to the cavity using one of the lock beams.The cavity lineshape is shown in gray.}
\label{fig:beams}
\end{figure}

After the IQM, the frequency-shifted light passes through a phase modulator (PM)\footnote{EOSpace phase modulator PM-0KS-10-PFA-PFAP-UL}. The PM is driven by up to four different tones, originating from four microwave sources described in section \ref{subs:gen}. Each of these tones produces sidebands on the LO. The beams incident on the cavity during Brownian motion measurements are shown in figure \ref{fig:beams}. The relative power in all the sidebands as a function of the microwave signals driving the phase modulator was calibrated as described in section \ref{subs:pm_cal}. 

The light then goes through a variable attenuator. A 90:10 splitter sends 90\% of light to the experimental cavity via a circulator;  the remaining 10\% is monitored to control the incident power. The power incident on the cavity and reflected from the cavity is calibrated using a 99:1 splitter immediately before the dilution refrigerator (DR)\footnote{Janis DR500}. The light reflected from the cavity passes through the circulator and another 90:10 splitter, which sends 10\% of the power onto a photodiode and 90\% towards an Erbium Doped Fiber Amplifier (EDFA)\footnote{Nuphoton EDFA-CW-LNF-RS-10-40-FCA}, which amplifies the optical signal by a factor of 20-50 and adds $\approx 4$ dB noise. The noise figure of the EDFA was calibrated as described in section \ref{subs:EDFA_cal}.

The light leaving the EDFA goes through a broadband tunable filter (TF)\footnote{OzOptics TF100, 0.5 nm bandwidth}, which is used to suppress the amplified spontaneous emission (ASE) noise from the EDFA. The filtered light then lands on a photodiode (PD)\footnote{Thorlabs DET08CFC}. 
 
\subsection{Microwave setup}
It is convenient to separate the microwave setup into the generation part and the detection part.
\subsubsection{Generation}
\label{subs:gen}
Up to 4 microwave are tones used to drive the phase modulators:
\begin{itemize}
\item{The Lock beam is used to lock the laser to the experimental cavity. The beam is generated using a lock-in amplifier (LIA)\footnote{Zurich Instruments UHF}. A tone at 200 MHz from the LIA  is sent to a mix-up circuit. There it is mixed with a tone from a microwave generator (MWG1)\footnote{Vaunix Lab Brick LMS-232D} at 1{,}900 MHz. The mixed-up tone at $\omega_\mathrm{Lock}=2{,}100$ GHz is sent to the four-way splitter (4WS) where it is combined with other tones and then sent to the phase modulator. }

\item{The Control 1 signal is used to generate one of the control beams. It is generated by the LIA at $529.343$ MHz. This tone is mixed with the signal from MWG1 at $1,900$ MHz. The mixed up signal  at $\omega_\mathrm{Control 1}=2{,}429.343$ MHz is sent to the 4WS.}

\item{The Control 2 signal is used to generate the other control beam. It is generated directly by a different  microwave generator (MWG2)\footnote{Vaunix Lab Brick  LMS-232D} at $\omega_\mathrm{Control 2}=1{,}790.657$ MHz.}

\item{The Probe signal is only on for the OMIT/A measurements. It is generated by the network analyzer (NA)\footnote{Keysight HP 8722D} at a frequency $\omega_\mathrm{Control (1,2)}\pm\omega_\mathrm{ac} \pm \delta$}, where $\omega_\mathrm{ac} =319.243$ MHz is the frequency of the acoustic mode, and $\delta \approx 100$ kHz. 

\end{itemize}

The signal out of the 4WS is split and a small portion is sent to a spectrum analyzer (SA)\footnote{Rigol DSA1030A}, where the spectrum is recorded. This is done to measure the power in all the microwave tones incident on the PM. 

For the Brownian motion measurements, the control signals are filtered with high pass filter for Control 1 and low pass filter for Control 2. The filters are placed before the PM to block microwave noise that would produce laser noise near the cavity resonance frequency.

For the OMIT/A  measurements only one control beam is on and its frequency is swept, as described in the main text.

The typical optical powers in the measurements are: 
\begin{itemize}
\item{$P_\mathrm{Control 1}\approx P_\mathrm{Control 2}\approx 0.1 P_\mathrm{total}$}
\item{$P_\mathrm{Lock}\approx 10^{-6}P_\mathrm{total}$}
\item{$P_\mathrm{Probe}\approx 10^{-5}P_\mathrm{total}$}
\item{$P_\mathrm{LO}\approx 0.8 P_\mathrm{total}$}
\end{itemize}

The total incident power $P_\mathrm{total}$ is up to $100$ $\mu$W.

\subsubsection{Detection}
A heterodyne detection scheme is used. The signal from the PD consists primarily of beating between the LO and the sidebands. These beat notes occur at $2{,}100$ MHz (Lock), $2{,}429.343$ MHz (Control 1), $1{,}790.657$ MHz (Control 2), and $2{,}110 \pm 0.1$ MHz (motional sidebands of  the control beams). The signal is  sent to a three-way splitter (3WS).

The first part is mixed down with the signal from MWG1 (at $1{,}900$ MHz). During the Brownian motion measurements, the mixed-down signal is dominated by 5 frequencies: $200$ MHz (Lock), $529.343$ MHz (Control 1), $109.343$ MHz (Control 2), and $210\pm0.1$ MHz (motional sidebands of the control beams). It is sent to the LIA, where the spectra at $210+0.1$ MHz and $210-0.1$ MHz are recorded. The quadratures of the signal at $200$ MHz are sent to a field programmable gate array (FPGA)\footnote{National Instruments FPGA NI PXI-7854R}, which uses them to generate an error signal, which is then sent to the voltage controlled oscillator (VCO) to vary its output frequency between $3$ GHz and $3.5$ GHz in order to lock the laser to the experimental cavity. The Lock beam is typically detuned by $\approx 10$ MHz from the cavity resonance, as indicated in figure \ref{fig:beams}.

The second part is sent to the NA. It gives the response at the probe beam frequency when the probe beam is on (i.e. for OMIT/A measurements). 

The third part is sent to the SA to record the spectrum of the light coming from the experimental cavity.

\subsection{Calibrations}
\label{subs:pm_cal}
This section describes the calibration measurements. The power incident on the cavity is found as the geometric mean of the incident and reflected powers measured at the 99:1 splitter. After the filter cavity, the laser light is found to be shot noise limited at $300$ MHz for powers less than $\approx 1$ mW. Since the powers in the control beams are at most $10-20$ $\mu$W,  the classical laser noise is expected to be at most a few percent of the shot noise. 
\subsubsection{Calibration of the phase modulator}
\begin{figure}
\centering
\includegraphics[width=0.6\textwidth]{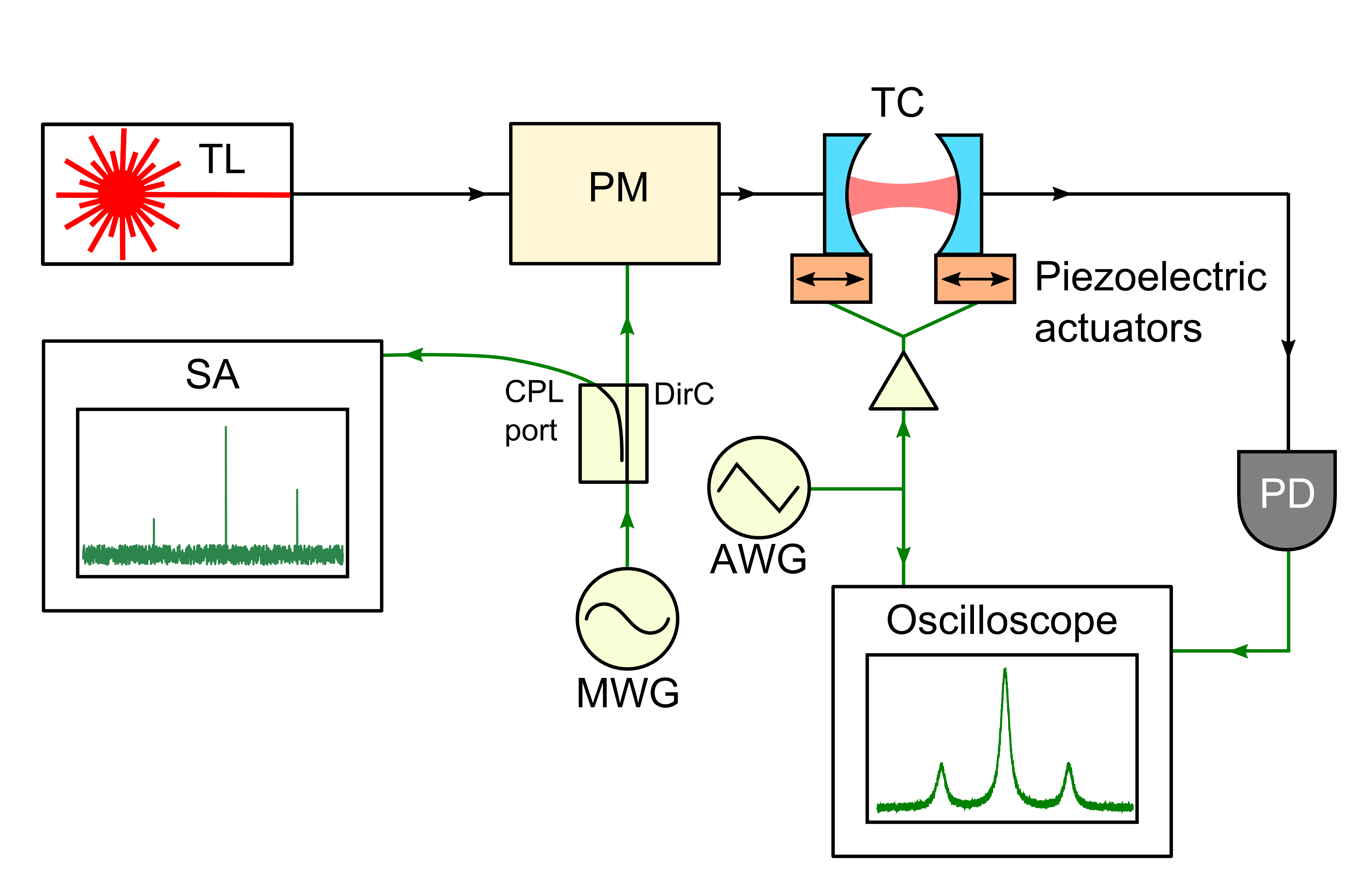} 
\caption{Calibration of the phase modulator. SM fiber: black line. Electrical path: green lines. TL: tunable laser. PM: phase modulator. TC: tunable cavity. PD: photodiode. AWG: arbitrary wave generator.  MWG: microwave generator. SA: spectrum analyzer.  DirC: directional coupler.}
\label{fig:cpm}
\end{figure}

The relative optical power in the beams after the phase modulator is calibrated using the setup shown in figure \ref{fig:cpm}.

Light from the TL passes through the PM and a tunable cavity (TC)\footnote{Homebuilt, $\kappa/2\pi=200$ MHz, $\omega_\mathrm{FSR}/2\pi=1.5$ THz}, which acts as an optical spectrum analyzer; the light transmitted through the TC lands on the PD.
The PM is driven by a microwave generator (MWG)\footnote{Agilent N9310A} with varying frequency and power. The SA records the power in the CPL port of the directional coupler (DirC), which is used in the actual experiment. The TC length is swept by applying a triangle wave from the arbitrary wave generator (AWG) to piezoelectric elements within the TC. The TC transmission is fit to a Lorentzian with two sidebands:
\begin{equation}
f(x)=\frac{E_0}{x^2+(\kappa/2)^2}+\frac{E_1}{(x-d_\mathrm{sb})^2+(\kappa/2)^2}+\frac{E_1}{(x+
d_\mathrm{sb})^2+(\kappa/2)^2}
\end{equation}
The ratio of the sidebands to the carrier is recorded (as a function of microwave drive power and frequency). This ratio is expected to be:
\begin{equation}
\frac{E_1}{E_0}=\frac{J_1(\pi V_\mathrm{rel})^2}{J_0(\pi V_\mathrm{rel})^2}
\end{equation}
Here $J_0$ and $J_1$ are Bessel functions of order 0 and 1. The drive voltage amplitude relative to the half-wave voltage $V_\pi$ is:
\begin{equation}
V_\mathrm{rel}=\frac{V}{V_\pi}=\left(\frac{P}{P_\pi}\right)^{1/2}=10^{(P_\mathrm{dBm}-P_{\pi\mathrm{dBm}})/20}
\end{equation}
Here $V$ is the voltage sent to the phase modulator; the  half-wave  voltage $V_\pi$ is the voltage necessary to induce a phase change of $\pi$. The values $P$ and $P_\pi$ are the corresponding powers in Watts and the value $P_\mathrm{dBm}$ and $P_{\pi\mathrm{dBm}}$ are the corresponding powers in dBm. The value of $P_{\pi\mathrm{dBm}}$ is given relative to the CPL port of the DirC, as that is what is measured during the experiment. This value is independent of microwave power, but varies with microwave frequency. We record its values for frequencies between 1,400 and 3,000 MHz, as that is the range of the microwave tones. 

During the experiment the optical power in the first order sideband, relative to the total power, is given by $J_1(\pi V_\mathrm{rel})^2$, where $P_{\pi\mathrm{dBm}}$ is known from the calibration and $P_\mathrm{dBm}$ is measured for each microwave tone using the SA.

\subsubsection{EDFA noise figure calibration}
\label{subs:EDFA_cal}
\begin{figure}
\centering
\includegraphics[width=0.6\textwidth]{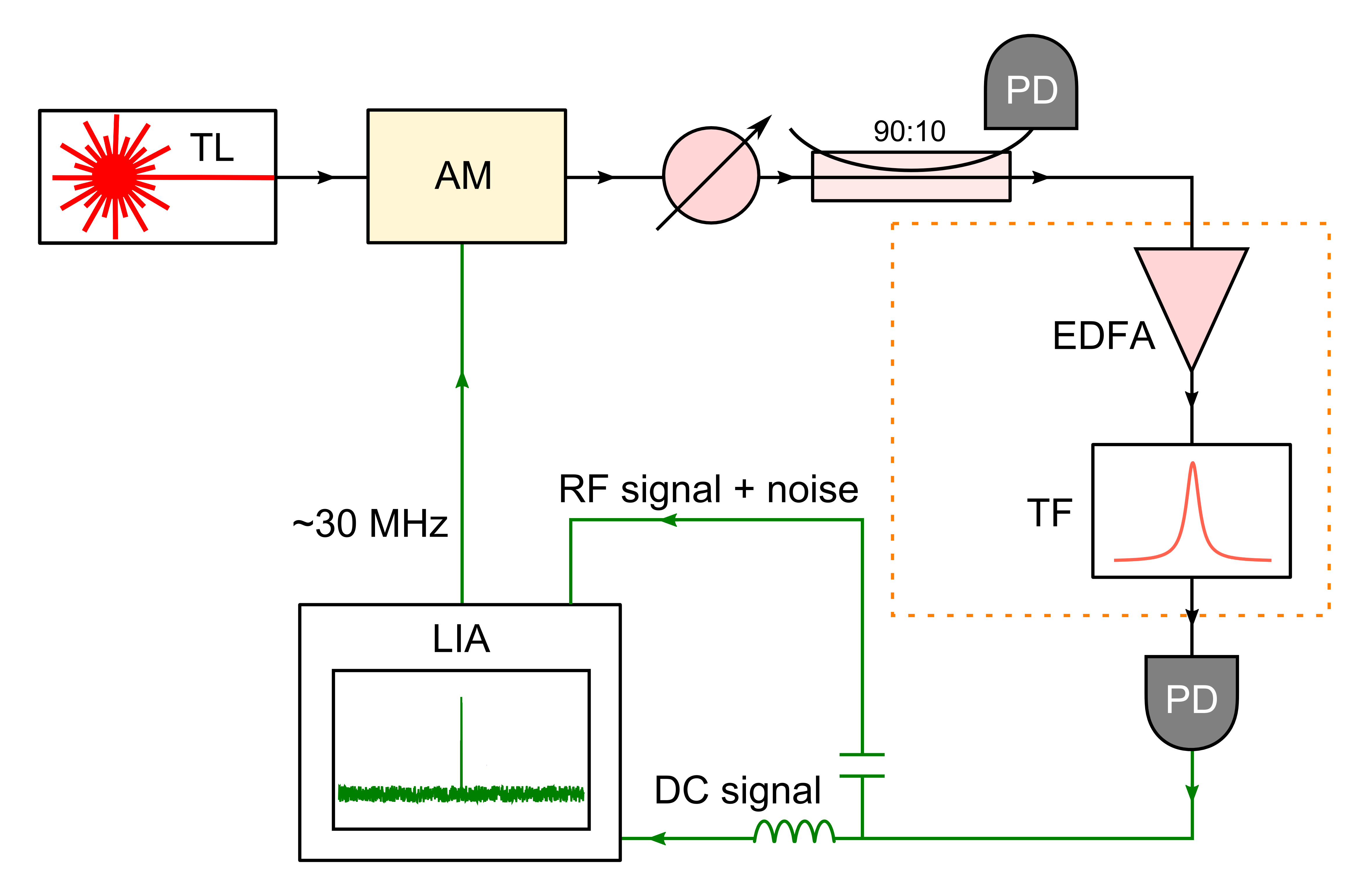} 
\caption{Calibration of the EDFA noise. SM fiber: black line. Electrical path: green lines. TL: tunable laser. AM: amplitude modulator.  PD: photodiode. EDFA: erbium doped fiber amplifier. TF: tunable filter. LIA: lock-in amplifier.}
\label{fig:edfa}
\end{figure}

The EDFA noise figure is calibrated as shown in figure \ref{fig:edfa}.

Light leaving the TL passes through an amplitude modulator (AM)\footnote{Thorlabs LN81S}, which puts small sidebands (``signal'') onto the beam. The amplitude modulator is driven at $\approx$ 30 MHz. At this frequency the laser amplitude noise is lower than shot noise, and the photodiode gain is within 5 \% of the DC photodiode gain. Light then passes through a variable attenuator and a 90:10 splitter, which is used to monitor the incident power. Then it either goes directly onto the PD, or passes through the EDFA and TF first.
During the measurement, the power of the AM sidebands and the background power spectral density are recorded as the incident laser power is changed using the attenuator. The ratio of the sideband power to the background gives the signal-to-noise ratio. The DC signal gives the record of laser power.

The gain of the photodiode and the laser noise are calibrated without the EDFA first (i.e., without the components inside the dashed orange square in figure \ref{fig:edfa}). The background grows linearly with increasing laser power, as expected for shot noise. The power of the AM sidebands grows quadratically. This measurement gives  SNR$_0$, the signal-to-noise ratio without the EDFA.

Then the EDFA and TF are put in, and the signal-to-noise ratio is measured again (SNR$_\mathrm{EDFA}$). 
The noise figure of the EFDA is calculated as:
\begin{equation}
\mathrm{NF}=10\log_{10}\left(\frac{\mathrm{SNR}_0}{\mathrm{SNR}_\mathrm{EDFA}}\right)
\end{equation}
NF was found to vary slightly with laser wavelength, so we measured it for a number of different wavelengths. For $1{,}529.7$ nm (the wavelength used for the Brownian motion measurements), the EDFA noise figure is 4 dB for laser powers below 30 $\mu$W. For laser powers below 100 $\mu$W (the maximum used in the experiment), the noise figure is smaller than 4.5 dB.

\FloatBarrier
\newpage

\section{Theoretical description of the measurement results}
\label{sec:theordesc}
\subsection{Two tone measurement scheme}

Here we consider a standard optomechanical system in which the two motional sidebands are measured using two beams detuned by $\pm\omegam$ from the cavity resonance.

The standard optomechanical Hamiltonian is
\begin{eqnarray}
\label{eq:SOMS_Hamiltonian}
\oHam
=\hbar\omegac\oad\oa+\hbar\omegam\ocd\oc+\hbar \grp(\ocd+\oc)\oad\oa+\oHamenv,
\end{eqnarray}
where $\oa$ and $\oc$ are annihilation operators for the optical and the acoustic modes respectively, $\omegac$ is the optical mode frequency (at zero acoustic mode amplitude), $\omegam$ is the acoustic mode frequency, and $\grp$ is the single-photon optomechanical coupling constant, i.e., the optical mode detuning per acoustic displacement equal to the acoustic zero-point fluctuations. Finally, $\oHamenv$ is the part of the Hamiltonian corresponding to the coupling to the environment (including both optical and acoustic mode noise and damping as well as the coherent optical drives). This Hamiltonian leads to the equations of motion
\begin{eqnarray}
\label{eq:SOMS_EOM_optical_orig}
\dot{\oa}
&=&-\left(\frac{\kappa}{2}+i\omegac\right)\oa
-i\grp(\ocd+\oc)\oa
+\sqrt{\kappaint}\oxiint
+\sqrt{\kappaext}(\aext+\oxiext)\\
\label{eq:SOMS_EOM_mechanical_orig}
\dot{\oc}
&=&-\left(\frac{\gammam}{2}+i\omegam\right)\oc
-i\grp\oad\oa
+\sqrt{\gammam}\oeta
\end{eqnarray}
Here $\kappaext$ and $\kappaint$ are the external and internal coupling rates for the optical mode with $\oxiext$ and $\oxiint$ being the corresponding optical noise operators, $\kappa=\kappaint+\kappaext$ is the total optical damping rate, $\aext$ denotes the optical drive, and $\gammam$ is the acoustic damping rate with the noise operator $\oeta$. The noise operators' correlations satisfy
\begin{eqnarray}
\label{eq:SOMS_xi_corr_1}
\ensavg{\oxi_i(t)\oxi_j(t')}
&=&0\\
\label{eq:SOMS_xi_corr_2}
\ensavg{\oxi_i^\dag(t)\oxi_j(t')}
&=&0\\
\label{eq:SOMS_xi_corr_3}
\ensavg{\oxi_i(t)\oxi_j^\dag(t')}
&=&\delta_{i,j}\delta(t-t')\\
\ensavg{\oeta(t)\oeta(t')}
&=&0\\
\ensavg{\oetad(t)\oeta(t')}
&=&\nth\delta(t-t')\\
\ensavg{\oeta(t)\oetad(t')}
&=&(\nth+1)\delta(t-t'),
\end{eqnarray}
The subscripts ``$i$'' and ``$j$'' stand for either ``int'' or ``ext'', and $\nth=(e^{\hbar\omegam/\kB T}-1)^{-1}$ is the thermal occupation of the acoustic mode bath (we assume that $\gammam\ll\omegam$, so that the frequency dependence of $\nth$ can be disregarded).

To simplify equation (\ref{eq:SOMS_EOM_optical_orig}), we can switch to a rotating frame for the optical mode to cancel the bare resonance frequency: $\oa \rightarrow \oa e^{-i\omegac t}$, with the corresponding transformations for $\oxiint$, $\oxiext$ and $\aext$. This doesn't affect correlation relations for the noise operators, since they are $\delta$-correlated. The equation of motion for the acoustic mode stays the same, while the optical one becomes
\begin{eqnarray}
\dot{\oa}
&=&-\frac{\kappa}{2}\oa
-i\grp(\ocd+\oc)\oa
+\sqrt{\kappaint}\oxiint
+\sqrt{\kappaext}(\aext+\oxiext)
\end{eqnarray}

Next, we specify the optical drive. We assume that it is comprised of two tones which we will call ``lower'' and ``upper'', with the corresponding detunings $\Deltal$ and $\Deltau$; the later discussion will assume that $\Deltal\approx -\omegam$ and $\Deltau\approx +\omegam$. Denoting the tones' amplitudes by $\aextl$ and $\aextu$, we can express the drive as $\aext=\aextl e^{-i\Deltal t}+\aextu e^{-i\Deltau t}$.

After that, we apply the usual expansion of $\oa$ in powers of $\grp$. The zeroth order only includes the coherent drive and not the vacuum noise, and results in the equations of motion
\begin{eqnarray}
\dot{a}_0
&=&-\frac{\kappa}{2}\ano
-i\grp(\cno+\cno^*)\ano
+\sqrt{\kappaext}\aext\\
\dot{c}_0
&=&-\left(\frac{\gammam}{2}+i\omegam\right)\cno
-i\grp \ano^*\ano
\end{eqnarray}
The radiation pressure force in the second equation $-i\grp \ano^*\ano$ has two components: one static and one at frequency $|\Deltau-\Deltal|\approx 2\omegam$. Since both of these are far away from the acoustic mode resonance, and the radiation pressure force is relatively small, we can ignore them in our case and simply assume $\cno=0$. To put it more quantitatively, these forces result in a dimensionless acoustic mode displacement on the order of $\zno\approx \frac{\grp}{\omegam}\nc$, where $\nc=\overline{|\ano|^2}$ is the average intracavity photon number. We can ignore this displacement when considering the optical mode if its contribution to the cavity detuning is less than a cavity linewidth: $\zno\grp\ll\kappa$, which results in $\nc\ll\frac{\kappa\omegam}{(\grp)^2}$. For our system this bound is about $4\cdot 10^8$, which is much higher than the maximum circulating photon number used in the experiment $\nc\lesssim 10^4$.
Thus, ignoring the static acoustic mode displacement is justified, and the zeroth order solution for the optical mode becomes
\begin{eqnarray}
\ano
&=&\anl e^{-i\Deltal t}+\anu e^{-i\Deltau t}\\
\anl
&=&\frac{\sqrt{\kappaext}\aextl}{\kappa/2-i\Deltal}\\
\anu
&=&\frac{\sqrt{\kappaext}\aextu}{\kappa/2-i\Deltau}
\end{eqnarray}

Next, the linearized equations of motion are
\begin{eqnarray}
\label{eq:SOMS_EOM_optical_lin}
\dot{\od}
&=&-\frac{\kappa}{2}\od
-i\grp(\ocd+\oc)\ano
+\sqrt{\kappaint}\oxiint
+\sqrt{\kappaext}\oxiext\\
\dot{\oc}
&=&-\left(\frac{\gammam}{2}+i\omegam\right)\oc
-i\grp(\ano^*\od+\odd \ano)
+\sqrt{\gammam}\oeta,
\end{eqnarray}
where $\od$ and $\oc$ are the first order expansion terms for the optical and acoustic modes respectively.

It is convenient to introduce a combined vacuum noise operator
\begin{eqnarray}
\label{eq:SOMS_xi}
\oxi=(\sqrt{\kappaext}\oxiext+\sqrt{\kappaint}\oxiint)/\sqrt{\kappa}
\end{eqnarray}
Because $\kappaint+\kappaext=\kappa$, this operator has the same correlation properties (\ref{eq:SOMS_xi_corr_1}-\ref{eq:SOMS_xi_corr_3}) as $\rmsub{\oxi}{int,ext}$. Equation (\ref{eq:SOMS_EOM_optical_lin}) for the optical mode can be rewritten as
\begin{eqnarray}
\dot{\od}
&=&-\frac{\kappa}{2}\od
-i\grp(\ocd+\oc)\ano
+\sqrt{\kappa}\oxi
\end{eqnarray}

The first order equations are linear in $\oc$ and $\od$, so we can solve them via Fourier transform, which is defined as
\begin{eqnarray}
\label{eq:SOMS_FT_definition}
\hat{x}[\omega]
=\lim_{T\rightarrow\infty}\frac{1}{\sqrt{T}}\int_{-T/2}^{T/2}\hat{x}(t)e^{i\omega t} dt,
\end{eqnarray}
so that the noise correlators become
\begin{eqnarray}
\ensavg{\oxi[\omega]\oxi[-\omega]}
=\ensavg{\oxid[\omega]\oxi[-\omega]}
&=&0\\
\ensavg{\oxi[\omega]\oxid[-\omega]}
&=&1\\
\ensavg{\oeta[\omega]\oeta[-\omega]}
&=&0\\
\ensavg{\oetad[\omega]\oeta[-\omega]}
&=&\nth\\
\ensavg{\oeta[\omega]\oetad[-\omega]}
&=&\nth+1
\end{eqnarray}
Note that in this notation the Hermitian conjugate is applied before the Fourier transform: $\hat{x}^\dag[\omega]=(\hat{x}[-\omega])^\dag$.

In the Fourier domain the equations of motion become
\begin{eqnarray}
\label{eq:SOMS_EOM_optical_FT}
\left(\frac{\kappa}{2}-i\omega\right)\od[\omega]
&=&-i\grp\left(\anl\left(\oc[\omega-\Deltal]+\ocd[\omega-\Deltal]\right)
+\anu\left(\oc[\omega-\Deltau]+\ocd[\omega-\Deltau]\right)\right)\flb
&&\qquad+\sqrt{\kappa}\oxi[\omega]\\
\label{eq:SOMS_EOM_mechanical_FT}
\left(\frac{\gammam}{2}-i(\omega-\omegam)\right)\oc[\omega]
&=&
-i\grp\left(\anl^*\od[\omega+\Deltal]+\anu^*\od[\omega+\Deltau]
+\anl\odd[\omega-\Deltal]+\anu\odd[\omega-\Deltau]\right)\flb
&&\qquad+\sqrt{\gammam}\oeta[\omega]
\end{eqnarray}

To solve equations (\ref{eq:SOMS_EOM_optical_FT}), (\ref{eq:SOMS_EOM_mechanical_FT}), we substitute the expression for $\od$ (i.e., equation (\ref{eq:SOMS_EOM_optical_FT})) into the equation for $\oc$ (i.e., equation (\ref{eq:SOMS_EOM_mechanical_FT})). This produces 16 terms containing acoustic motion ($\oc$ or $\ocd$), which we can divide into several categories. First, there are 8 terms involving $\ocd$. If the acoustic sidebands are far from each other ($|\Deltau-\Deltal-2\omegam|\gg \gammam$), these terms are off-resonant for the acoustic mode, and can be ignored. Of the remaining 8 terms, 4 include beating of the sideband of one control beam against the other beam, which would result in expressions like $\oc[\omega\pm(\Deltau-\Deltal)]$; since $\Deltau-\Deltal\approx 2\omegam\gg\gammam$, these terms are also very far off resonance and can be neglected. The last 4 terms produce a combination of the standard dynamic backaction effects of the beams (two terms per beam), and thus should be preserved. With the addition of the vacuum noise term, we obtain the following equation for the acoustic mode:
\begin{eqnarray}
\left(\frac{\gammam}{2}-i(\omega-\omegam)\right)\oc[\omega]
&=&
(\grp)^2\left(|\anl|^2(\chic[\omega-\Deltal]-\chic[\omega+\Deltal])\right.\flb
&&\left.\qquad\qquad+|\anu|^2(\chic[\omega-\Deltau]-\chic[\omega+\Deltau])\right)\oc[\omega]\flb
&&\qquad-i\grp\left(\anl^*\odoxi[\omega+\Deltal]+\anl\oddoxi[\omega-\Deltal]\right.\flb
&&\left.\qquad\qquad +\anu^*\odoxi[\omega+\Deltau]+\anu\oddoxi[\omega-\Deltau]\right)\flb
&&\qquad+\sqrt{\gammam}\oeta[\omega]
\end{eqnarray}
Here $\chic[\omega]=(\kappa/2-i\omega)^{-1}$ is the cavity susceptibility, and $\odoxi[\omega]=\chic[\omega]\sqrt{\kappa}\oxi[\omega]$ are the vacuum fluctuations of the intracavity field. 
Now we can rewrite the acoustic equation of motion as
\begin{eqnarray}
\label{eq:SOMS_solution_mech}
\oc[\omega]
&=&\chimeff[\omega]\left(-i\oFRPSN[\omega]+\sqrt{\gammam}\oeta[\omega]\right),
\end{eqnarray}
where the modified acoustic mode susceptibility is
\begin{eqnarray}
\chimeff[\omega]
=(\gammam/2-i(\omega-\omegam)+i\Sigma[\omega])^{-1}
\approx(\gammameff/2-i(\omega-\omegameff))^{-1},
\end{eqnarray}
with the acoustic linewidth and the acoustic frequency modified by the dynamic backaction:
\begin{eqnarray}
\gammameff
&=&\gammam-2\mathrm{Im}\Sigma[\omegameff]
=\gammam+\dgammamopt\\
\omegameff
&=&\omegam+\mathrm{Re}\Sigma[\omegameff]
=\omegam+\domegamopt
\end{eqnarray}
The self-energy $\Sigma[\omega]$ for the acoustic system is defined as
\begin{eqnarray}
\label{eq:SOMS_Sigma}
\Sigma[\omega]
=i(\grp)^2\left(|\anl|^2(\chic[\omega-\Deltal]-\chic[\omega+\Deltal])
+|\anu|^2(\chic[\omega-\Deltau]-\chic[\omega+\Deltau])\right),
\end{eqnarray}
and the radiation pressure force is defined as
\begin{eqnarray}
\label{eq:SOMS_FRPSN}
\oFRPSN[\omega]=\grp\left(\anl^*\odoxi[\omega+\Deltal]+\anl\oddoxi[\omega-\Deltal]+\anu^*\odoxi[\omega+\Deltau]+\anu\oddoxi[\omega-\Deltau]\right)
\end{eqnarray}
Note that this force is Hermitian: $\oFRPSN^\dag[\omega]=\oFRPSN[\omega]$.

Now we are ready to find an expression for the intracavity field. Because we focus on the part of the spectrum close to the optical resonance $\omega\approx 0$, we can neglect the other two sidebands: the red sideband of the lower control beam, which corresponds to $\ocd[\omega-\Deltal]\approx \ocd[+\omegam]\approx 0$, and the blue sideband of the upper control beam, corresponding to $\oc[\omega-\Deltau]\approx\oc[-\omegam]\approx 0$. The optical field thus becomes
\begin{eqnarray}
\od[\omega]
&\approx&\chic[\omega]\left( -i\grp\left(\anl\oc[\omega-\Deltal]+\anu\ocd[\omega-\Deltau]\right)+\sqrt{\kappa}\oxi \right)
\end{eqnarray}

Finally, the outgoing field can be calculated using the input-output relations:
\begin{eqnarray}
\label{eq:SOMS_dout}
\odout
&=&\oxiext-\sqrt{\kappaext}\od\flb
&=&\oxiext
-\sqrt{\kappaext}\chic[\omega]\left( -i\grp\left(\anl\oc[\omega-\Deltal]+\anu\ocd[\omega-\Deltau]\right)+\sqrt{\kappa}\oxi \right)
\end{eqnarray}
The acoustic mode annihilation operator spectrum has peaks at $+\omegameff$, while the creation operator (being its Hermitian conjugate) is peaked at $-\omegameff$. This means that the red sideband in the expression above (which comes from the $\ocd$ term) is located around $\omegar=\Deltau-\omegameff$, and the blue sideband (coming from the $\oc$ term) is located around $\omegab=\Deltal+\omegameff$. Because of the earlier choice $\Deltal\approx-\omegam$, $\Deltau\approx+\omegam$, both of these frequencies are close to zero.

\subsection{Detection modes}

It this section we describe how the acoustic sidebands are manifest in the photocurrent, and in the next section we use these results to calculate their power spectral densities and cross-correlations between them.

We consider heterodyne detection with a local oscillator (LO) at frequency $-\omegaLO$ with $\omegaLO>0$ (the case where the local oscillator's frequency is higher than the sidebands frequency is less convenient, since it leads to the photocurrent spectrum being flipped compared to the optical one). Ignoring the reflected control beams, the field incident on the photodiode after combining with the LO is $\oadet=\aOLO e^{+i\omegaLO t}+\odout$. Standard photodetection theory \cite{Carmichael1987} states that the (time-dependent) autocorrelation of the photocurrent $i(t)$ can be described as
\begin{eqnarray}
\Cii(t,\tau)
&\equiv& \ensavg{i(t+\tau/2)i(t-\tau/2)}\flb
&=&G^2\ensavg{:\oaddet(t+\tau/2)\oadet(t+\tau/2)\oaddet(t-\tau/2)\oadet(t-\tau/2):}\flb
&&\qquad+G^2\ensavg{\oaddet(t)\oadet(t)}\delta(\tau),
\end{eqnarray}
where $G$ is the photodetector gain and $::$ denotes normal and time ordering. Note that since $i(t)$ is a photocurrent, we take it to be classical and real, so $\Cii(\tau)$ is real and symmetric in $\tau$.

If we substitute the expression for $\oadet$ above and expand up to second order in $\od$ (keeping in mind that the first order terms average to zero), we get
\begin{eqnarray}
\label{eq:Det_Cii_general}
\Cii(t,\tau)
&\approx& G^2|\aOLO|^4
+G^2|\aOLO|^2\left(\ensavg{\oddout(t+\tau/2)\odout(t+\tau/2)}+\ensavg{\oddout(t-\tau/2)\odout(t-\tau/2)}\right)\flb
&&+G^2|\aOLO|^2\left(e^{i\omegaLO\tau}\ensavg{\oddout(t+\tau/2)\odout(t-\tau/2)}
+e^{-i\omegaLO\tau}\ensavg{\oddout(t-\tau/2)\odout(t+\tau/2)}\right)\flb
&&+G^2(\aOLO)^2e^{2i\omegaLO t}\ensavg{:\oddout(t+\tau/2)\oddout(t-\tau/2):}\flb
&&+G^2(\aOLO^*)^2e^{-2i\omegaLO t}\ensavg{:\odout(t+\tau/2)\odout(t-\tau/2):}\flb
&&+G^2|\aOLO|^2\delta(\tau)
\end{eqnarray}
The first line in equation (\ref{eq:Det_Cii_general}) is the DC component of the correlator, which is not relevant to the acoustic sideband spectrum and can be ignored. The next three lines reflect beating of the outgoing cavity field with the LO. Finally, the last line represents the unavoidable detector shot noise.

First, let us consider the power spectral density (PSD) of the photocurrent, which is the Fourier transform of the correlation function:
\begin{eqnarray}
\label{eq:Det_SfromC}
\Sii[\omega]
=\int_{-\infty}^{+\infty}\overline{\Cii(t,\tau)}e^{i\omega\tau}d\tau,
\end{eqnarray}
where $\overline{\Cii(t,\tau)}$ denotes that the correlator is averaged over the central time $t$. We assume that the correlators of the input field are stationary (or at least don't have components at $2\omegaLO$), and that the integration time is long enough that we can set $\overline{e^{2i\omegaLO t}}=0$. In this case, only the second and the last line in the correlator contribute to the PSD above, which can be re-expressed as
\begin{eqnarray}
\Sii[\omega]
=G^2|\aOLO|^2\left(\Soddod[\omegaLO+\omega]+\Soddod[\omegaLO-\omega]+1\right)
\end{eqnarray}
With the Fourier transform definition (\ref{eq:SOMS_FT_definition}), the spectrum of the outgoing field can be calculated in a straightforward way using the Wiener-Khinchin theorem:
\begin{eqnarray}
\label{eq:Det_WKT}
\Soddod[\omega]
=\int_{-\infty}^{+\infty}\overline{\ensavg{\oddout(t+\tau/2)\odout(t-\tau/2)}}e^{i\omega\tau}d\tau
=\ensavg{\oddout[\omega]\odout[-\omega]}
\end{eqnarray}

Now, let us consider what would be the photocurrent $i(t)$ and its corresponding Fourier transform (in the sense of equation (\ref{eq:SOMS_FT_definition})) $i[\omega]$. After mixing with the optical local oscillator, the two acoustic sidebands of interest will be located around $\omegaLO+{\rmsub{\omega}{r,b}}$. We can define the shifted ``local'' Fourier transforms
\begin{eqnarray}
\rmsub{i}{r,b}[\domega]
\equiv i[\omegaLO+\rmsub{\omega}{r,b}+\domega]
\end{eqnarray}
(note that unlike $i[\omega]$ these don't correspond to any real function of time, so in general $\ir[\omega]\neq (\ir[-\omega])^*$).
The PSDs of the sidebands are then described by
\begin{eqnarray}
\Siirr[\domega]
&\equiv&\ensavg{\ir[\domega](\ir[\domega])^*}
=\Sii[\omegaLO+\omegar+\domega]\flb
&=&G^2|\aOLO|^2(\Soddod[2\omegaLO+\omegar+\domega]+\Soddod[-\omegar-\domega]+1)\\
\Siibb[\domega]
&\equiv&\ensavg{\ib[\domega](\ib[\domega])^*}
=\Sii[\omegaLO+\omegab+\domega]\flb
&=&G^2|\aOLO|^2(\Soddod[2\omegaLO+\omegab+\domega]+\Soddod[-\omegab-\domega]+1)
\end{eqnarray}
Here $\Siirr[\domega]$ and $\Siibb[\domega]$ are the PSDs of the red and the blue sideband respectively, and $\domega$ is the frequency shift in the PSD from the sideband maximum.

While the second terms in the parentheses $\Soddod[-\rmsub{\omega}{r,b}-\domega]$ correspond to the optical spectrum close to the cavity resonance, the first terms probe the spectrum reoughly $2\omegaLO$ away from the cavity resonance, and therefore are insensitive to the cavity dynamics (more rigorously, the cavity susceptibility in the expression (\ref{eq:SOMS_dout}) is very small). Moreover, because of the normal ordering of the operators in $\Soddod$ the vacuum noise terms $\oxi$ don't contribute. Thus, it is clear that $\Soddod[2\omegaLO]\approx 0$, and the PSDs simplify to
\begin{eqnarray}
\label{eq:Det_Srr}
\Siirr[\domega]
&\approx&G^2|\aOLO|^2(\Soddod[-\omegar-\domega]+1)\\
\label{eq:Det_Sbb}
\Siibb[\domega]
&\approx&G^2|\aOLO|^2(\Soddod[-\omegab-\domega]+1)
\end{eqnarray}

Next, we turn to the correlations between the two sidebands.
It is natural to define them as
\begin{eqnarray}
\Siirb[\domega]
&\equiv&\ensavg{\ib[\domega]\ir[-\domega]}
=\ensavg{i[\omegaLO+\omegab+\domega]i[\omegaLO+\omegar-\domega]}
\end{eqnarray}
Note that $\ir$ isn't complex conjugated, because it comes from $\ocd$ rather than $\oc$. Similar to (\ref{eq:Det_SfromC}),
we can use the definition of the Fourier transform $i[\domega]$ to express the result above through the time correlator $\Cii(t,\tau)$:
\begin{eqnarray}
\Siirb[\domega]
&=&\int_{-\infty}^{+\infty}\overline{\Cii(t,\tau)e^{i(2\omegaLO+\omegar+\omegab)t}}
e^{i(\omegab/2-\omegar/2+\domega)\tau}d\tau\flb
&=&G^2(\aOLO^*)^2\int_{-\infty}^{+\infty}\overline{
\ensavg{:\left(e^{i\omegab(t+\tau/2)}\odout(t+\tau/2)\right)\left(e^{i\omegar (t-\tau/2)}\odout(t-\tau/2)\right):}}e^{i\domega\tau}d\tau
\end{eqnarray}
This expression can be greatly simplified if we recall from input-output theory that the commutation relations of the outgoing fields are the same as the incoming ones. This implies that $\odout$ (just like $\oxiext$) commutes at different times, so the time ordering inside the ensemble averaging is irrelevant. Therefore, we can apply the Wiener-Khinchin theorem again and arrive at
\begin{eqnarray}
\label{eq:Det_Srb}
\Siirb[\domega]
&=&G^2(\aOLO^*)^2\ensavg{\odout[\omegab+\domega]\odout[\omegar-\domega]}
\end{eqnarray}

\subsection{Correlators values and the interpretation}

In this section we calculate the sideband PSDs (\ref{eq:Det_Srr}), (\ref{eq:Det_Sbb}) and the cross-correlator (\ref{eq:Det_Srb}) for the optical field (\ref{eq:SOMS_dout}) obtained earlier.

We start with the sideband PSDs $\Siirr$ and $\Siibb$, which are proportional to $\Soddod[\omega]$. As noted before, due to the normal ordering the terms containing the vacuum noise $\oxi$ won't contribute. Thus, we're left with
\begin{eqnarray}
\label{eq:Corr_Sddd}
\Soddod[\omega]
&=&\ensavg{\oddout[\omega]\odout[-\omega]}\flb
&=&\kappaext|\chic[-\omega]|^2(\grp)^2
\left(|\anl|^2\Socdoc[\omega+\Deltal]+|\anu|^2\Sococd[\omega+\Deltau]\right)
\end{eqnarray}

We've also omitted two other terms involving the acoustic mode motion: $\anu^*\anl\ensavg{\oc[\omega+\Deltau]\oc[-\omega-\Deltal]}$ and its complex conjugate $\anl^*\anu\ensavg{\ocd[\omega+\Deltal]\ocd[-\omega-\Deltau]}$. While not strictly zero, these terms are nevertheless small because the acoustic susceptibilities of the two terms in the product don't overlap. For example, in the first expression the two acoustic terms are centered around $\omega=\omegam-\Deltau=-\omegar$ and $\omega=-\omegam-\Deltal=-\omegab$; as we're working in the assumption $|\omegar-\omegab|\gg\gammameff$ (non-overlapping sidebands), the product of these two terms is always small.

Now we need to calculate the acoustic motion correlators:
\begin{eqnarray}
\Socdoc[\omega]
&=&|\chimeff[-\omega]|^2(\SFFRPSN[\omega]+\SFdFth[\omega])\\
\Sococd[\omega]
&=&|\chimeff[+\omega]|^2(\SFFRPSN[\omega]+\SFFdth[\omega])
\end{eqnarray}
The PSD of the thermal force is:
\begin{eqnarray}
\SFdFth[\omega]
&\equiv& \ensavg{(\sqrt{\gammam}\oetad[\omega])(\sqrt{\gammam}\oeta[-\omega])}
=\gammam \nth\\
\SFFdth[\omega]
&\equiv& \ensavg{(\sqrt{\gammam}\oeta[\omega])(\sqrt{\gammam}\oetad[-\omega])}
=\gammam (\nth+1),
\end{eqnarray}
and the PSD of the radiation pressure is
\begin{eqnarray}
\SFFRPSN[\omega]
&\equiv&\ensavg{\oFRPSN[\omega]\oFRPSN[-\omega]}\flb
&=&(\grp)^2\kappa\left( |\anl|^2|\chic[\omega+\Deltal]|^2+|\anu|^2|\chic[\omega+\Deltau]|^2 \right)
\end{eqnarray}
(since $\oFRPSN$ is Hermitian, this is the only correlator that we need).

For the following discussion we note that
\begin{eqnarray}
\SFFdth[\omega]-\SFdFth[-\omega]
&=&\gammam\\
\SFFRPSN[\omega]-\SFFRPSN[-\omega]
&=&(\grp)^2\kappa\left(
|\anl|^2(|\chic[\omega+\Deltal]|^2-|\chic[-\omega+\Deltal]|^2)\right.\flb
&&\qquad\left.+|\anu|^2(|\chic[\omega+\Deltau]|^2-|\chic[-\omega+\Deltau]|^2)\right)\flb
&=&-2\mathrm{Im}\Sigma[\omega]
\equiv \dgammamopt
\end{eqnarray}
This shows that the antisymmetric part of the force noise spectrum (with appropriate ordering for a non-Hermitian noise operator) is equal to the dissipation rate associated with this force: either the intrinsic loss $\gammam$ for the environment acoustic mode noise $\oeta$, or the optomechanically induced damping $\dgammamopt$ for the radiation pressure shot noise. This is well-known from quantum noise theory \cite{Clerk2010}, where the positive and the negative parts of the force spectrum are associated with the tendency to (respectively) extract energy from or give energy to the system, so the difference between the two provides the net damping.

When substituting force spectra into the equations for $\Socdoc$ and $\Sococd$, we can simplify them by assuming that $\gammameff\ll\kappa$, so the radiation pressure noise spectrum is approximately flat over the acoustic resonance. This lets us write
\begin{eqnarray}
\Socdoc[\omega]
&\approx&|\chimeff[-\omega]|^2(\SFFRPSN[-\omegameff]+\SFdFth[-\omegameff])\flb
&=&|\chimeff[-\omega]|^2(\nRPSN\dgammamopt+\nth\gammam)
\\
\Sococd[\omega]
&\approx&|\chimeff[\omega]|^2(\SFFRPSN[\omegameff]+\SFFdth[\omegameff])\flb
&=&|\chimeff[\omega]|^2((\nRPSN+1)\dgammamopt+(\nth+1)\gammam),
\end{eqnarray}
where we've defined the effective phonon occupation of the RPSN bath $\nRPSN=\SFFRPSN[-\omegameff]/\dgammamopt$ analogously to the thermal bath occupation $\nth$. To determine the final mean energy of the acoustic mode, we can find the expectation value of the phonon number operator by integrating its PSD:
\begin{eqnarray}
\label{eq:Corr_nm}
\nm
&=&\ensavg{\ocd(t)\oc(t)}
=\frac{1}{2\pi}\int_{-\infty}^{+\infty} \Socdoc[\omega] d\omega\flb
&=&\frac{1}{2\pi}\int_{-\infty}^{+\infty} \frac{\nRPSN\dgammamopt+\nth\gammam}{\gammameff^2/4+(\omega+\omegameff)^2} d\omega
=\frac{\nRPSN\dgammamopt+\nth\gammam}{\gammameff}
\end{eqnarray}
This expression can be understood intuitively if we consider that the acoustic oscillator is coupled to two different baths (environment and radiation pressure force) with two different rates ($\gammam$ and $\dgammamopt$ respectively). This way, the final occupation of the oscillator is a weighted average of the occupations of the two baths, with the weights being proportional to the coupling rates.

With this expression for $\nm$ the PSDs simplify to
\begin{eqnarray}
\label{eq:Corr_Scdc}
\Socdoc[\omega]
&=&\frac{\nm\gammameff}{\gammameff^2/4+(\omega+\omegameff)^2}
\\
\label{eq:Corr_Sccd}
\Sococd[\omega]
&=&\frac{(\nm+1)\gammameff}{\gammameff^2/4+(\omega-\omegameff)^2}
\end{eqnarray}

Note that the difference in the magnitude between the two PSDs (which comes from the asymmetry of the force noise spectra) is directly related to the equal-time commutator of the acoustic mode creation and annihilation operators:
\begin{eqnarray}
[\oc(t),\ocd(t)]
&=&\ensavg{[\oc(t),\ocd(t)]}
=\ensavg{\oc(t)\ocd(t)}-\ensavg{\ocd(t)\oc(t)}
=(\nm+1)-\nm=1
\end{eqnarray}

With these spectra the PSD of the outgoing field is
\begin{eqnarray}
\Soddod[\omega]
&=&\ensavg{\oddout[\omega]\odout[-\omega]}\flb
&=&\kappaext|\chic[-\omega]|^2(\grp)^2
\left(|\anl|^2\frac{\nm\gammameff}{\gammameff^2/4+(\omega+\Deltal+\omegameff)^2}\right.\flb
&&\qquad\left.+|\anu|^2\frac{(\nm+1)\gammameff}{\gammameff^2/4+(\omega+\Deltau-\omegameff)^2}\right)\\
&=&\kappaext|\chic[-\omega]|^2(\grp)^2
\left(|\anl|^2\frac{\nm\gammameff}{\gammameff^2/4+(\omega+\omegab)^2}
+|\anu|^2\frac{(\nm+1)\gammameff}{\gammameff^2/4+(\omega+\omegar)^2}\right)
\end{eqnarray}
As expected, it is comprised of two Lorentzians centered at $\omega=-\omegar$ and $\omega=-\omegab$. From (\ref{eq:Det_Srr}), (\ref{eq:Det_Sbb}) the photocurrent PSDs of the individual sidebands are
\begin{eqnarray}
\label{eq:Corr_Srr}
\Siirr[\domega]
&\approx&G^2|\aOLO|^2\left(\kappaext|\chic[\omegar]|^2(\grp)^2|\anu|^2\frac{(\nm+1)\gammameff}{\gammameff^2/4+\domega^2} +1\right)\\
\label{eq:Corr_Sbb}
\Siibb[\domega]
&\approx&G^2|\aOLO|^2\left(\kappaext|\chic[\omegab]|^2(\grp)^2|\anl|^2\frac{\nm\gammameff}{\gammameff^2/4+\domega^2} +1\right)
\end{eqnarray}
Both are Lorentzians with shot noise background, and with area under the Lorentzian proportional to $\nm$ or $\nm+1$ for the blue and the red sideband respectively.

Now we switch to the cross-correlator (\ref{eq:Det_Srb}), which is proportional to $\ensavg{\odout[\omegab+\domega]\odout[\omegar-\domega]}$. Because the normal ordering is not enforced, there are terms involving the vacuum noise:
\begin{eqnarray}
\label{eq:Corr_Sdd}
\ensavg{\odout[\omegab+\domega]\odout[\omegar-\domega]}
&\approx&-\kappaext(\chic[\omegar]\chic[\omegab])(\grp)^2\anl\anu \Sococd[\omegameff+\domega]\flb
&&\qquad+i\left\langle\left(\oxiext[\omegab+\domega]-\sqrt{\kappaext\kappa}\chic[\omegab]\oxi[\omegab+\domega]\right)\right.\flb
&&\qquad\qquad\times\left.\left(\sqrt{\kappaext}\chic[\omegar]\grp\left(\anu\ocd[-\omegameff-\domega]\right)\right)\right\rangle
\end{eqnarray}

The first term is just the acoustic motion PSD, similar to the sidebands' PSDs (as before, we've assumed $\domega\sim\gammameff\ll|\omegar-\omegab|$ and neglected all off-resonant acoustic terms). The second term involves the correlations of the optical vacuum fluctuations with the acoustic mode motion, which are non-zero because the acoustic oscillator is driven by the radiation pressure shot noise arising from these vacuum fluctuations. Thus, this term directly represents the action of the radiation pressure shot noise on the acoustic oscillator.

Using expression (\ref{eq:SOMS_xi}) for $\oxi$ and (\ref{eq:SOMS_FRPSN}) for $\oFRPSN$, we get
\begin{eqnarray}
\ensavg{\oxi[\omegab+\domega]\oFRPSN[-\omegameff-\domega]}
&=&\grp \anl\ensavg{\oxi[\omegab+\domega]\oddoxi[-\omegab-\domega]}\flb
&=&\grp \anl\chic[-\omegab]\sqrt{\kappa}\\
\ensavg{\oxiext[\omegab+\domega]\oFRPSN[-\omegameff-\domega]}
&=&\sqrt{\kappaext/\kappa}\ensavg{\oxi[\omegab+\domega]\oFRPSN[-\omegameff-\domega]}\flb
&=&\grp \anl\chic[-\omegab]\sqrt{\kappaext},
\end{eqnarray}
so that
\begin{eqnarray}
&&\ensavg{\left(\oxiext[\omegab+\domega]-\sqrt{\kappaext\kappa}\chic[\omegab]\oxi[\omegab+\domega]\right)\ocd[-\omegameff-\domega]}\flb
&&\qquad=i(\chimeff[\omegameff+\domega])^*\ensavg{\left(\oxiext[\omegab+\domega]-\sqrt{\kappaext\kappa}\chic[\omegab]\oxi[\omegab+\domega]\right)\oFRPSN[-\omegameff-\domega]}\flb
&&\qquad=i(\chimeff[\omegameff+\domega])^*\sqrt{\kappaext}(1-\kappa\chic[\omegab])\grp \anl\chic[-\omegab]\flb
&&\qquad=-i(\chimeff[\omegameff+\domega])^*\sqrt{\kappaext}\grp \anl\chic[\omegab]
\end{eqnarray}
Note that this expression depends on the full complex acoustic susceptibility $\chimeff[\omega]$, unlike, for example, the acoustic PSD, where only $|\chimeff|^2$ is present. This implies that it is sensitive to the phase response of the acoustic oscillator, meaning that this term really is a correlator between the force and the displacement (simple force-force or displacement-displacement correlators wouldn't depend on the force-displacement phase shift).

The complete noise correlator becomes
\begin{eqnarray}
\label{eq:Corr_Sdd_final}
\ensavg{\odout[\omegab+\domega]\odout[\omegar-\domega]}
&\approx&G_{cc}
\left(\Sococd[\omegameff+\domega]
-(\chimeff[\omegameff+\domega])^*\right)\\
G_{cc}
&=&-\kappaext(\chic[\omegar]\chic[\omegab])(\grp)^2\anl\anu
\end{eqnarray}
where $G_{cc}$ simply is a conversion factor between the displacement and the outgoing field.

Finally, the photocurrent cross-correlator is
\begin{eqnarray}
\label{eq:Corr_Srb}
\Siirb[\domega]
&=&G^2(\aOLO^*)^2G_{cc}\left(\omegameff+\domega]
-(\chimeff[\omegameff+\domega])^*\right)\flb
&=&G^2(\aOLO^*)^2G_{cc}
\left(\frac{(\nm+1)\gammameff}{\gammameff^2/4+\domega^2}-\frac{\gammameff/2-i\domega}{\gammameff^2/4+\domega^2}\right)\flb
&=&G^2(\aOLO^*)^2G_{cc}\frac{(\nm+1/2)\gammameff-i\domega}{\gammameff^2/4+\domega^2}
\end{eqnarray}

This expression is different from (\ref{eq:Corr_Srr}) and (\ref{eq:Corr_Sbb}) in several important ways. First, there's no shot noise background present, as this noise is uncorrelated between the two sidebands (it's important to note that the measurement SNR is still affected by the shot noise; it just averages to zero instead of to some finite value). Second, the cross-correlator is complex, with an imaginary part that is antisymmetric in $\domega$. Finally, the real Lorentzian part of the cross-correlator is proportional not to $\nm$ (like in $\Siibb$) or $\nm+1$ (as in $\Siirr$), but to $\nm+1/2$. As was shown above in equation (\ref{eq:Corr_Sdd_final}), this additional half phonon in the real part together with the anti-Lorentzian imaginary part can be combined to produce a complex acoustic susceptibility. This susceptibility shows up because of the correlation between the random radiation pressure force noise and the acoustic mode displacement driven by this force, and thus is an unambiguous signature of the RPSN acting on the acoustic oscillator.

\subsection{Photothermal coupling}

In this section we consider the effect of the photothermal optomechanical coupling.

The quantum treatment of the photothermal coupling is similar to \cite{Restrepo2011}. We model it as an additional optical loss and an extra acoustic force whose magnitude is proportional to the optical power lost to that channel. To describe this quantitatively, we first introduce an optical loss channel with a rate $\kappapt$ and a corresponding vacuum noise $\oxipt$. This modifies the original equation of motion for the optical mode (\ref{eq:SOMS_EOM_optical_orig}) to
\begin{eqnarray}
\label{eq:PT_EOM_optical_orig}
\dot{\oa}
&=&-\frac{\kappa}{2}\oa
-i\grp(\ocd+\oc)\oa
+\sqrt{\kappaint}\oxiint
+\sqrt{\kappapt}\oxipt
+\sqrt{\kappaext}(\aext+\oxiext)
\end{eqnarray}
The total damping is now a combination of all three loss rates: $\kappa=\kappaint+\kappapt+\kappaext$. The vacuum noise $\oxipt$ is uncorrelated with any other noise and is described by the same correlation relations (\ref{eq:SOMS_xi_corr_1}-\ref{eq:SOMS_xi_corr_3}). The amplitude of the field lost to that channel can be found from the input-output relations, just like (\ref{eq:SOMS_dout}) for the external coupling:
\begin{eqnarray}
\oaoutpt
=\oxipt-\sqrt{\kappapt}\oa
\end{eqnarray}
The corresponding power is simply
\begin{eqnarray}
\oIoutpt
=\oadoutpt\oaoutpt
\end{eqnarray}
The photothermal force is proportional to this intensity. However, the thermal reaction rate may be slow compared to the characteristic frequencies of interest (i.e., $\omegameff$), so the force may be subject to a low-pass filtering. We can model this by writing
\begin{eqnarray}
\taupt\dot{\hat{F}}_\mathrm{pt}
=-\oFpt+\Apt\oIoutpt,
\end{eqnarray}
where $\Apt$ is the DC proportionality coefficient between the intensity and the photothermal force, and $\taupt$ is the time constant of the low-pass filter. The solution of this equation (in the Fourier domain) is
\begin{eqnarray}
\oFpt[\omega]
=\frac{\Apt\oIoutpt[\omega]}{1-i\omega\taupt}
\end{eqnarray}
Since we're only interested in the forces in a small frequency band around $\omegameff$, we can substitute $\omega\approx\omegameff$ in the denominator of the expression above and transform back to the time domain, getting
\begin{eqnarray}
\oFpt(t)
=\frac{\Apt}{1-i\omegameff\taupt}\oIoutpt(t)
=\Apteff\oIoutpt(t)
\end{eqnarray}
Note that this equation only holds for spectral components of the photothermal force near $\omegameff$.

With that result, we can modify the acoustic equation of motion (\ref{eq:SOMS_EOM_mechanical_orig}) and turn it into
\begin{eqnarray}
\label{eq:PT_EOM_mechanical_orig}
\dot{\oc}
&=&-\left(\frac{\gammam}{2}+i\omegam\right)\oc
-i\grp\oad\oa
-i\oFpt
+\sqrt{\gammam}\oeta
\end{eqnarray}

Next, we once again perform the first order expansion of the optical mode $\oa=\ano+\od$. This leads to the photothermal force
\begin{eqnarray}
\oFpt(t)
&=&\Apteff\oadoutpt\oaoutpt\flb
&=&\Apteff\left(\oxidpt-\sqrt{\kappapt}(\ano^*+\odd)\right)\left(\oxipt-\sqrt{\kappapt}(\ano+\od)\right)\flb
&\approx& \Apteff\kappapt|\ano|^2
+\Apteff\kappapt(\ano^*\od+\odd \ano)
-\Apteff\sqrt{\kappapt}(\ano^*\oxipt+\oxidpt \ano)\flb
&=&\gpt|\ano|^2+\gpt(\ano^*\od+\odd \ano)-\frac{\gpt}{\sqrt{\kappapt}}(\ano^*\oxipt+\oxidpt \ano),
\end{eqnarray}
where $\gpt=\Apteff\kappapt$ is a single-photon photothermal coupling rate. It is analogous to $\grp$, but it is in general complex (owing to the low-pass filtering) and appears only in the acoustic equation of motion, since its origin is non-unitary. In the following we ignore the static force term $\gpt|\ano|^2$ (this term is incorrect anyway, since we've used the low-passed proportionality coefficient $\Apteff$ instead of the static $\Apt$), just as for the radiation pressure.
The acoustic equation of motion then becomes
\begin{eqnarray}
\dot{\oc}
&=&-\left(\frac{\gammam}{2}+i\omegam\right)\oc
-i(\grp+\gpt)(\ano^*\od+\odd \ano)
+i\frac{\gpt}{\sqrt{\kappapt}}(\ano^*\oxipt+\oxidpt \ano)
+\sqrt{\gammam}\oeta
\end{eqnarray}

The rest follows fairly closely the derivation for case of pure radiation pressure. After transitioning to the Fourier domain and solving for $\oc[\omega]$, we find, similarly to (\ref{eq:SOMS_solution_mech})
\begin{eqnarray}
\oc[\omega]
&=&\chimeff[\omega]\left(-i\oFopt[\omega]+\sqrt{\gammam}\oeta[\omega]\right)
\end{eqnarray}
There are two modifications here. First, the expression for the acoustic susceptibility is still the same $\chimeff[\omega]
=(\gammam/2-i(\omega-\omegam)+i\Sigma[\omega])^{-1}$, but the self-energy is slightly different:
\begin{eqnarray}
\Sigma[\omega]
=i\grp(\grp+\gpt)\left(|\anl|^2(\chic[\omega-\Deltal]-\chic[\omega+\Deltal])
+|\anu|^2(\chic[\omega-\Deltau]-\chic[\omega+\Deltau])\right)
\end{eqnarray}
(this expression is proportional to $\grp(\grp+\gpt)$, in contrast with $\left(\grp\right)^2$ in the radiation pressure case (\ref{eq:SOMS_Sigma})).
Second, the radiation pressure force is replaced by a more general optical force:
\begin{eqnarray}
\oFopt[\omega]
&=&\grp\left(\anl^*\odoxi[\omega+\Deltal]+\anl\oddoxi[\omega-\Deltal]+\anu^*\odoxi[\omega+\Deltau]+\anu\oddoxi[\omega-\Deltau]\right)\flb
&&\quad+\gpt\left(\anl^*\odpt[\omega+\Deltal]+\anl\oddpt[\omega-\Deltal]+\anu^*\odpt[\omega+\Deltau]+\anu\oddpt[\omega-\Deltau]\right),
\end{eqnarray}
where the RPSN is associated with the same vacuum noise as before $\odoxi[\omega]=\chic[\omega]\sqrt{\kappa}\oxi[\omega]$, while for the photothermal noise it's modified:
\begin{eqnarray}
\odpt[\omega]
=\chic[\omega]\sqrt{\kappa}\oxi[\omega]
-\frac{\oxipt}{\sqrt{\kappapt}}
\end{eqnarray}
Since $\gpt$ is in general complex, the optical force is no longer Hermitian: $\oFdopt\neq\oFopt$.
Therefore, we need to calculate two different force noise spectra:
\begin{eqnarray}
\SFFdopt[\omega]
&\equiv&\ensavg{\oFopt[\omega]\oFdopt[-\omega]}\flb
&=&|\grp+\gpt|^2\kappa\left( |\anl|^2|\chic[\omega+\Deltal]|^2+|\anu|^2|\chic[\omega+\Deltau]|^2 \right)\flb
&&\qquad-2\mathrm{Re}\left[(\grp+\gpt)^*\gpt\left(|\anl|^2\chic[-\omega-\Deltal])+|\anu|^2\chic[-\omega-\Deltau]\right)\right]\flb
&&\qquad+\frac{|\gpt|^2}{\kappapt}\left(|\anl|^2+|\anu|^2\right)\flb
&=&(\grp)^2\kappa\left( |\anl|^2|\chic[\omega+\Deltal]|^2+|\anu|^2|\chic[\omega+\Deltau]|^2 \right)\flb
&&\qquad+2\mathrm{Re}\left[\grp\gpt\left(|\anl|^2\chic[\omega+\Deltal]+|\anu|^2\chic[\omega+\Deltau]\right)\right]\flb
&&\qquad+\frac{|\gpt|^2}{\kappapt}\left(|\anl|^2+|\anu|^2\right)\\
\SFdFopt[\omega]
&\equiv&\ensavg{\oFdopt[\omega]\oFopt[-\omega]}\flb
&=&(\grp)^2\kappa\left( |\anl|^2|\chic[\omega+\Deltal]|^2+|\anu|^2|\chic[\omega+\Deltau]|^2 \right)\flb
&&\qquad+2\mathrm{Re}\left[\grp\gpt\left(|\anl|^2\chic[-\omega-\Deltal]+|\anu|^2\chic[-\omega-\Deltau]\right)\right]\flb
&&\qquad+\frac{|\gpt|^2}{\kappapt}\left(|\anl|^2+|\anu|^2\right)
\end{eqnarray}
Nevertheless, the general property of the antisymmetric component of the noise spectrum still holds:
\begin{eqnarray}
\SFFdopt[\omega]
-\SFdFopt[-\omega]
&=&(\grp)^2\kappa\left(
|\anl|^2(|\chic[\omega+\Deltal]|^2-|\chic[-\omega+\Deltal]|^2)\right.\flb
&&\qquad\left.+|\anu|^2(|\chic[\omega+\Deltau]|^2-|\chic[-\omega+\Deltau]|^2)\right)\flb
&&\quad +2\mathrm{Re}\left[\grp\gpt\left(|\anl|^2(\chic[\omegam+\Deltal]-\chic[\omegam-\Deltal])\right.\right.\flb
&&\qquad\left.\left.+|\anu|^2(\chic[\omegam+\Deltau]-\chic[\omegam-\Deltau])\right)\right]\flb
&=&2\mathrm{Re}\left[\grp(\grp+\gpt)\left(|\anl|^2(\chic[\omegam+\Deltal]-\chic[\omegam-\Deltal])\right.\right.\flb
&&\quad\left.\left.+|\anu|^2(\chic[\omegam+\Deltau]-\chic[\omegam-\Deltau])\right)\right]\flb
&=&-2\mathrm{Im}\Sigma[\omegam]
=\dgammamopt
\end{eqnarray}
This result relies crucially on the presence of the $\oxipt/\sqrt{\kappapt}$ term in the photothermal force noise, and on the fact that this noise is partially correlated with the intracavity field. Ignoring it and simply replacing the $\grp$ by $\grp+\gpt$ in the equation of motion for $\oc$ (which is sufficient for a classical treatment) would ultimately result in $[\oc,\ocd]\neq 1$.

The relation above allows us to follow the same route as for a purely radiation pressure coupled system. We can still define the effective occupation of the bath associated with the optical force shot noise
\begin{eqnarray}
\nopt
=\frac{\SFdFopt[-\omegameff]}{\dgammamopt}
\end{eqnarray}
and obtain the equilibrium occupation of the acoustic mode in the same way as in the expression (\ref{eq:Corr_nm}) before:
\begin{eqnarray}
\nm
=\frac{\nopt\dgammamopt+\nth\gammam}{\gammameff}
\end{eqnarray}
With that, the expressions for the acoustic spectrum look the same as (\ref{eq:Corr_Scdc}) and (\ref{eq:Corr_Sccd}):
\begin{eqnarray}
\Socdoc[\omega]
&=&\frac{\nm\gammameff}{\gammameff^2/4+(\omega+\omegameff)^2}
\\
\Sococd[\omega]
&=&\frac{(\nm+1)\gammameff}{\gammameff^2/4+(\omega-\omegameff)^2}
\end{eqnarray}
The difference is concealed in the definitions of the optomechanical self-energy $\Sigma=\domegamopt-i\dgammamopt/2$ and the equilibrium acoustic occupation $\nm$.

One caveat about this difference is the additional optical force noise arising from the photothermal force:
\begin{eqnarray}
\delta(\nopt\dgammamopt)
&=&\SFdFopt[-\omegameff]-\SFFRPSN[-\omegameff]\flb
&=&2\mathrm{Re}\left[\grp\gpt\left(|\anl|^2\chic[-\omegameff+\Deltal]+|\anu|^2\chic[-\omegameff+\Deltau]\right)\right]\flb
&&\qquad+\frac{|\gpt|^2}{\kappapt}\left(|\anl|^2+|\anu|^2\right)
\end{eqnarray}
While the first term only depends on the measurable system parameters (and in our case it is much smaller than the RPSN owing to the fact that $\gpt$ is purely imaginary), the second term involves the photothermal channel loss rate $\kappapt$, which we can not access experimentally. To estimate the effect of this term, we can assume that $\kappapt$ is associated with the absorptive loss discussed in section \ref{subsubsec:absorp}. This means that its value can be calculated as $\kappapt=\alpha (\kappa-\kappaext)$, where $\alpha$ is defined in equation \eqref{eq:Q} and is estimated to be $\alpha=0.2$ (from room temperature measurements of mirrors' absorption, Section \ref{subsubsec:absorp}) or $\alpha=0.7\pm 0.1$ (from fitting the cryogenic data using the thermal model, Section \ref{sec:fitting}). Using these numbers and a total intracavity photon number $\ncirc=|\anl|^2+|\anu|^2=2500$ (which is the maximum photon number for the data shown in the main text) yields an extra phonon occupation of between 0.05 (for $\alpha=0.7$) and 0.2 (for $\alpha=0.2$) phonons; these values are between 3 and 10 times smaller than the RPSN effects. Because of this, and because this extra noise becomes even smaller for lower photon numbers (the majority of the data was taken at $\ncirc<1000$), we ignore it in the data analysis and assume $\nopt=\nRPSN$.

Since the equation of motion (equation \eqref{eq:SOMS_EOM_optical_lin}) for the optical mode doesn't change (except for an additional loss channel), the general expression (\ref{eq:Corr_Sddd}) for the PSD of the outgoing light still holds. Following that, the results for the PSDs of the red and blue sidebands are also the same as before (equations (\ref{eq:Corr_Srr}) and (\ref{eq:Corr_Sbb})).

To find $\Siirb$ we can still apply equation (\ref{eq:Corr_Sdd}). In order to do so, we once again need to calculate the correlations between the vacuum noise and the acoustic motion, which follow from the generalized optical force noise:
\begin{eqnarray}
\ensavg{\oxi[\omegab+\domega]\oFdopt[-\omegameff-\domega]}
&=&(\grp+\gpt)^* \anl\ensavg{\oxi[\omegab+\domega]\oddoxi[-\omegab-\domega]}\flb
&&\qquad-(\gpt)^* \frac{\anl}{\sqrt{\kappapt}}\ensavg{\oxi[\omegab+\domega]\oxipt[-\omegab-\domega]}\flb
&=&(\grp+\gpt) \anl\chic[-\omegab]\sqrt{\kappa}-(\gpt)^* \frac{\anl}{\sqrt{\kappa}}\\
\ensavg{\oxiext[\omegab+\domega]\oFdopt[-\omegameff-\domega]}
&=&(\grp+\gpt)^* \anl\chic[-\omegab]\sqrt{\kappaext}
\end{eqnarray}
These lead to
\begin{eqnarray}
&&\ensavg{\left(\oxiext[\omegab+\domega]-\sqrt{\kappaext\kappa}\chic[\omegab]\oxi[\omegab+\domega]\right)\oFdopt[-\omegameff-\domega]}\flb
&&\qquad=\sqrt{\kappaext}(1-\kappa\chic[\omegab])(\grp+\gpt)^* \anl\chic[-\omegab]-\sqrt{\kappaext}[\omegab](\gpt)^* \anl\flb
&&\qquad=-\sqrt{\kappaext}\grp \anl\chic[\omegab],
\end{eqnarray}
which is not dependent on the photothermal coupling. This means that the rest of the derivation follows the pure radiation pressure case, and we arrive at the same equation (\ref{eq:Corr_Srb}) as before.

\section{Displacement measurement calibration}
\label{sec:dispmeas}
In this section we describe the procedure for calibrating the acoustic motional sidebands in units of zero point fluctuations.

Consider the expression for the power spectral density of the blue acoustic sideband \eqref{eq:Corr_Sbb}
\begin{eqnarray}
\Sii^\mathrm{(bb)}[\domega]
&=&G^2|\aOLO|^2\left(
\kappaext|\chic[\omegab]|^2(\grp)^2|\anl|^2\frac{\nm\gammameff}{\gammameff^2/4+\domega^2}
+1\right)\flb
&=&G^2|\aOLO|^2\left(
4\frac{\kappaext}{\kappa}\frac{4(\grp)^2|\anl|^2}{\kappa\gammameff}\frac{1}{1+(2\domega/\gammameff)^2}\nm\frac{1}{1+(2\omegab/\kappa)^2}
+1\right)\flb
&=&G^2|\aOLO|^2\left(
4\etak\frac{\gammameasl}{\gammameff}\frac{1}{1+(2\omegab/\kappa)^2}\nm\frac{1}{1+(2\domega/\gammameff)^2}
+1\right)
\end{eqnarray}
Here $\gammameasl=\frac{4(\grp)^2|\anl|^2}{\kappa}$ is the measurement strength of the lower control beam, and $\etak=\frac{\kappaext}{\kappa}$ is the contribution to the quantum efficiency due to imperfect external coupling to the cavity. This power spectral density represents a Lorentzian with width $\gammameff$ on top of an approximately frequency-independent background (more precisely, the background is a Lorentzian with width $\kappa\gg\gammameff$). The height of the Lorentzian with respect to the background is
\begin{eqnarray}
\arelbb
=4\etak\frac{\gammameasl}{\gammameff}\frac{1}{1+(2\omegab/\kappa)^2}\nm
\end{eqnarray}

This expression is derived under the assumption of no additional loss or noise sources between the cavity output and the detector. Now, assume that there is a finite transmission from the cavity output to the photodetector $\etal$. It will affect the signal part of the PSD, but not the background, which will stay 1 in photon units. This is especially apparent in the normal-ordering description of the photodetection, where the background comes from the optical local oscillator, which is unaffected by the additional loss. Thus, the relative height is multiplied by $\etal$
\begin{eqnarray}
\arelbb
=4\etak\etal\frac{\gammameasl}{\gammameff}\frac{1}{1+(2\omegab/\kappa)^2}\nm
\end{eqnarray}

Next, let us consider an additional source of noise on the way from the cavity to the photodetector (in our case this comes from erbium-doped fiber amplifier, which has a noise figure of $\sim 4$ dB). We denote its strength relative to the vacuum noise as $\nadd=\frac{1}{\etan}-1$, were $\etan\leq 1$ represents a drop in quantum efficiency due to this additional noise. With that, the noise background becomes $1+\nadd=\frac{1}{\etan}$, and the relative height is further reduced to
\begin{eqnarray}
\arelbb
=4\etak\etal\etan\frac{\gammameasl}{\gammameff}\frac{1}{1+(2\omegab/\kappa)^2}\nm
\end{eqnarray}

Finally, there may be additional mechanisms reducing the signal-to-noise ratio which can't be readily attributed to loss or additional noise. We can denote the quantum efficiency reduction of these residual mechanisms as $\etar$ and get the final expression
\begin{eqnarray}
\arelbb
&=&4\etak\etal\etan\etar\frac{\gammameasl}{\gammameff}\frac{1}{1+(2\omegab/\kappa)^2}\nm\flb
&=&4\etat\frac{\gammameasl}{\gammameff}\frac{1}{1+(2\omegab/\kappa)^2}\nm,
\end{eqnarray}
where $\etat=\etak\etal\etan\etar$ is the combined quantum efficiency of the measurement process. For a sideband close to the optical resonance $|\omegab|\ll\kappa$ the expression above simplifies to
\begin{eqnarray}
\label{eq:cal_abb_final}
\arelbb
=4\etat\frac{\gammameasl}{\gammameff}\nm
\end{eqnarray}

To calibrate the measurement rate $\gammameasl$ we use the OMIT/A data. The expression for the normalized amplitude of the OMIT/A feature is derived in the Supplemental Information of Ref. \cite{Kashkanova2016}, which in the notation used in this paper can be written as
\begin{eqnarray}
\adrbb
&=&-\frac{\grp\gtot|\anl|^2\chic[\omegab]}{\gammameff/2}
=-\frac{4\grp\gtot|\anl|^2}{\kappa\gammameff}\frac{1}{1-2i\omegab/\kappa}\flb
&=&-\frac{\gtot}{\grp}\frac{\gammameasl}{\gammameff}\frac{1}{1-2i\omegab/\kappa}
\end{eqnarray}
The minus sign denotes that for a blue sideband (i.e., red-detuned control beam) the OMIT/A feature is a dip, so the Lorentzian is subtracted from the background. Similar to the motional sideband PSD, the expression can be further simplified for an on-resonance sideband:
\begin{eqnarray}
\adrbb
=-\frac{\gtot}{\grp}\frac{\gammameasl}{\gammameff}
\end{eqnarray}
If the photothermal coupling $\gpt=\gtot-\grp$ is known, then the measurement of $\adrbb$ can be used to extract the ratio $\gammameasl/\gammameff$. Knowing this ratio and $\etat$, one can then use formula (\ref{eq:cal_abb_final}) to relate the measured relative height of the Brownian motion peak $\arelbb$ to the acoustic mode occupation $\nm$, and consequently rescale the vertical axis in the motional PSD data in units of phonons. A similar calibration (using the same value of $\etat$, but a different individually determined measurement rate $\gammameasu$) is performed for the red acoustic sideband. Finally, to normalize the cross-correlator data $\Siirb$ we apply a scaling coefficient which is the geometric mean of the coefficients for the red and the blue sidebands.

The relevant contributions to the quantum efficiency in our setup are measured to be:
\begin{itemize}
\item Imperfect input cavity coupling $\etak=\kappaext/\kappa=0.44\pm0.03$.
\item Optical loss between the cavity output and the optical amplifier $\etal=0.44$.
\item Optical amplifier input noise $\etan=0.35\div0.40$ (depending on the total power incident on the amplifier).
\item Imperfection of the heterodyne detection. The idealized description of the heterodyne detection usually assumes that the power in the optical local oscillator (OLO) is much larger than in the rest of the optical field. If this assumption is relaxed, then the added noise background, which is proportional to the total laser power, becomes larger in comparison with the signal component, which comes only from the mixing with the OLO. As a result, the SNR degrades by an additional factor of $\etar=\rmsub{P}{OLO}/\rmsub{P}{tot}$, where $\rmsub{P}{OLO}$ is the power in the OLO and $\rmsub{P}{tot}$ is total power incident on the photodiode. In our measurements $\etar$ varies between $0.8$ and $0.95$, depending on the strength of the microwave drives used to create control beams.
\end{itemize}

Combining these contributions, the total quantum efficiency $\etat$ of the setup was between $0.05$ and $0.08$ depending on the measurement configuration. The relative error in its determination (which gives rise to the uncertainty in Figure 4 of the main text) is $7\%$, which almost entirely comes from the uncertainty in the relative input cavity coupling $\etak$.

\section{Thermal model}
\subsection{Introduction}

The temperature dependence of the speed of sound and acoustic damping
in liquid helium are well-studied. As a result, it is straightforward
to calculate the temperature dependence of the acoustic mode's frequency
$\omega_{\mathrm{ac}}$, damping $\gamma_{\mathrm{ac}}$, and mean
phonon number $n_{\mathrm{ac}}$ provided that the temperature is
uniform throughout the cavity. However in the present device the temperature
is not uniform. Here we calculate the expected temperature profile
within the cavity using well-known thermal properties of liquid helium
(Section \ref{sec:tempdist}). We then use this result to calculate
$\omega_{\mathrm{ac}}$, $\gamma_{\mathrm{ac}}$, and $n_{\mathrm{ac}}$
(Section \ref{sec:waveeq}). The results of these two sections are
then used to fit the data (Section \ref{sec:fitting})


\subsection{Temperature distribution in the cavity}

\label{sec:tempdist} In a superfluid-filled optical cavity, the helium's
temperature is set by the balance between heating (caused by optical
absorption in the cavity mirrors) and the transport of this heat through
the helium to the mixing chamber (MC). In prior work, \cite{Kashkanova2016}
this transport was limited by the thermal impedance of a narrow \char`\"{}sheath\char`\"{}
of helium that connected the cavity to the MC. The sheath's large
impedance ensured that the temperature drop between the cavity and
the MC occurred primarily in the sheath, leaving the cavity itself at an
approximately uniform temperature. As described in the main text,
the present device uses a more open geometry without a sheath. This
results in an improved thermal link to the MC and allows the cavity
to reach lower temperatures; however, the absence of a bottleneck
also means that the temperature within the cavity is less uniform
than in the device described in Ref. {\cite{Kashkanova2016}.

We model the temperature distribution in the present device by assuming
that the heating originates in sub-$\mu$m absorbers (located in the
DBR coatings) that overlap with the cavity's optical mode, and that
the resulting heat propagates outward through the helium. We find
that in the overwhelming majority of the cavity the temperature and
heat flux density are low enough that thermal transport is via ballistic
phonons. However within $\approx1$ $\mathrm{\mu}$m of each absorber
the heat flux density is high enough to produce turbulence, with the
result that thermal transport in this small region is described
by the Gorter-Mellinck model. Despite the smallness of the turbulent
region, we find that it plays an important role in the device's performance.

\subsubsection{Optical absorption}
\label{subsubsec:absorp}
A schematic illustration of the device is shown in figure \ref{fig:1}.
The cavity is formed between the end faces of two optical fibers,
each having a radius $r_{\mathrm{out}}=$ 100 $\mu$m. The separation
between the fibers (and hence the cavity length) is $d=69.1$ $\mu$m.
Laser light may be absorbed where the cavity's optical mode overlaps
with the mirrors. This corresponds to a disk-shaped region on the
fiber surfaces with radius $r_{\mathrm{opt}}\approx7$ $\mu$m. The
total heat flux from this absorption is: 
\begin{equation}
\dot{Q}=\hbar\omega_{\mathrm{l}}n_{\mathrm{circ}}\kappa_{\mathrm{int}}\alpha\label{eq:Q}
\end{equation}
Here $\hbar$ is the reduced Planck's constant, $\omega_{\mathrm{l}}/2\pi=196.0$
THz is the frequency of the optical mode, $n_{\mathrm{circ}}$ is
the circulating photon number, $\kappa_{\mathrm{int}}/2\pi=10$ MHz
is the internal cavity loss rate, and $\alpha$ is the fraction of
the internal loss that leads to heating of the mirrors (as distinguished from internal
loss due to photons that are elastically scattered out of the cavity
mode and absorbed in some distant part of the apparatus).

Some \textit{a priori} information about $\alpha$ is provided by
room-temperature calorimetry measurements, which give the mirror's
absorption coefficient $a=3$ ppm ($15\pm5$ ppm) for $\lambda=1{,}064$ nm ($532$ nm)\cite{Laseroptik}. To estimate $\alpha$ from this information,
we note that the probability for an intracavity photon to be absorbed
by a mirror is given by $P_{\mathrm{mir}}=a\frac{\mathcal{F}}{2\pi}\frac{\kappa}{\kappa_{\mathrm{int}}}$,
where the cavity finesse is $\mathcal{F}\simeq95,000$ and the cavity
linewidth $\kappa/2\pi=21$ MHz. Assuming $a=$ 3 ppm for $\lambda=
1{,}529.7$ nm (the wavelength used in the present experiment) gives $P_{\mathrm{mir}}=0.1$.
Since the cavity is defined by two mirrors, these assumptions would
give $\alpha=2P_{\mathrm{mir}}=0.2$. This estimate for $\alpha$
is necessarily rough, since the absorption coefficient was measured
at room temperature and for a somewhat different wavelength. In Section
\ref{sec:fitting}, $\alpha$ will be used as a fit parameter.

We assume that photons are absorbed in small (sub-$\mu$m) absorbers
distributed throughout the DBR layers (as illustrated by the small
red circles in figure \ref{fig:1}). Each absorber will produce an
average heat flux 
\begin{equation}
\dot{Q}_{1}=\dot{Q}/N\label{eq:Q1}
\end{equation}
where $N$ is the number of absorbers within the optical mode. The
heat from each absorber is assumed to spread isotropically into the
helium, since the optical fibers' thermal conductivity is extremely
low at the relevant temperatures.

\begin{figure}[h!]
\centering \includegraphics[width=0.8\textwidth]{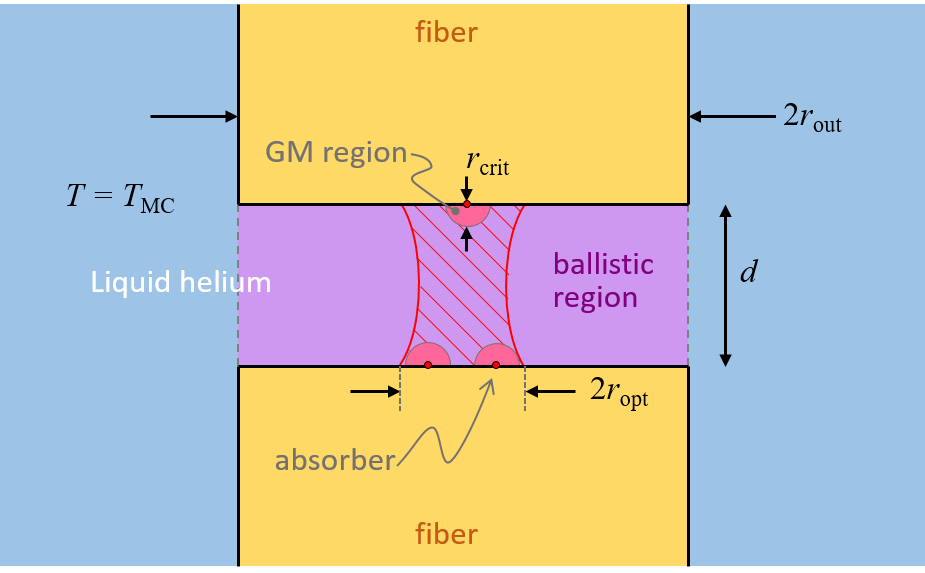}
\caption{Schematic illustration of the device highlighting the features relevant to the thermal model.}
\label{fig:1} 
\end{figure}


\subsubsection{Relevant regimes of thermal transport in liquid helium}

The character of thermal transport in helium II depends strongly on
the size of the helium channel, its temperature, and the heat flux
density \cite{Sciacca2014,Bertman1968}. As described below, we find
that thermal transport is in the ballistic regime throughout nearly
all of the cavity. The only exception is a small region around
each absorber, where thermal transport is in the Gorter-Mellinck regime.

To calculate the temperature profile, we assume that helium in the
region outside of the cavity (the blue region in figure \ref{fig:1})
is at the temperature of the mixing chamber $T_{\mathrm{MC}}$. For
all $T_{\mathrm{MC}}$ used in this work the phonon mean free path
$\lambda_{\mathrm{mfp}}$ \cite{Whitworth1958} is much larger than
any of the device's dimensions. Thus, at the boundary of the blue
region in Fig. \ref{fig:1} and for some distance inwards (i.e., towards
the absorbers) thermal transport is in the ballistic regime.

However the ballistic regime does not extend all the way to the absorbers.
This is because the heat flux density $\dot{q}=\dot{Q_{1}}/2\pi r^{2}$
increases as the distance $r$ from the absorber decreases. At some
distance $r_{\mathrm{crit}}$ from the absorber, $\dot{q}$ exceeds
the critical value $\dot{q}_{\mathrm{crit}}$ for generating turbulence.
Within the turbulent region (i.e., for $r<r_{\mathrm{crit}}$), thermal
transport is in the Gorter-Mellinck (GM) regime.

We calculate the temperature profile $T(r)$ throughout the cavity
by concatenating these two regimes. Specifically, we start with the
boundary condition $T(r_{\mathrm{out}})=T_{\mathrm{MC}}$, and integrate
the expressions describing ballistic transport towards decreasing
$r$, stopping when $\dot{q}=\dot{q}_{\mathrm{crit}}$ (or equivalently,
when $r=r_{\mathrm{crit}}$). We then use the calculated $T(r_{\mathrm{crit}})$
as a new boundary condition for integrating the GM expressions for
$r<r_\mathrm{crit}$. The following three subsections (\ref{subsec:critflux}
- \ref{subsec:gm}) provide a detailed description of these steps.

\subsubsection{The critical heat flux density}

\label{subsec:critflux} In the GM regime heat is carried by the turbulent
normal fluid. The onset of turbulence in helium II is typically estimated
in two different ways: (1) by considering the fluid velocity required
to produce vortices in the superfluid, or (2) by considering the Reynolds
number of the normal fluid. Below, we estimate $\dot{q}_{\mathrm{crit}}$
using both (1) and (2) (the corresponding estimates are labeled $\dot{q}_{\mathrm{crit,1}}$
and $\dot{q}_{\mathrm{crit,2}}$).

We assume throughout that the net mass flow is zero: 
\begin{equation}
\rho_{\mathrm{s}}{\bf v}_{\mathrm{s}}+\rho_{\mathrm{n}}{\bf v}_{\mathrm{n}}=0\label{eq:netmassflow}
\end{equation}
where $\rho_{\mathrm{s,n}}$ and ${\bf v}_{\mathrm{s,n}}$ are the
density and velocity of the superfluid and normal fluid. We also note
that regardless of the mechanism by which turbulence sets in, the
heat flux density $\dot{q}$ is given by

\begin{equation}
\dot{q}=s\rho Tv_{\mathrm{n}}\label{eq:qcritgen}
\end{equation}
where $s$ is the entropy and $\rho=145$ kg/m$^{3}$ is the mass
density of liquid He. The entropy can be evaluated using 
\begin{equation}
s(T)=\int_{0}^{T}\frac{C(T')}{T'}dT'\label{eq:entropy}
\end{equation}
and the following empirical expressions for the specific heat $C(T)$
\cite{VanSciver2012}: 
\begin{eqnarray}
C(T) & = & \zeta_{1}T^{3}\qquad\mathrm{for}\qquad T<0.6\ \mathrm{K}\nonumber \\
C(T) & = & \zeta_{2}T^{6.7}\qquad\mathrm{for}\qquad0.6<T<1.1\ \mathrm{K}\\
C(T) & = & \zeta_{3}T^{5.6}\qquad\mathrm{for}\qquad1.1<T<2.17\ \mathrm{K}\nonumber 
\end{eqnarray}
where $\zeta_{1}=20.4\ \mathrm{J/(kg\cdot K^{4})}$ , $\zeta_{2}=108\ \mathrm{J/(kg\cdot K^{7.7})}$
, and $\zeta_{3}=117\ \mathrm{J/(kg\cdot K^{6.6})}$ . 
\begin{enumerate}
\item { If the onset of turbulence is attributed to the production of vortices,
we use the result that vortex lines are produced for superfluid velocity
exceeding \cite{VanSciver2012} 
\begin{equation}
v_{\mathrm{s,crit}}\simeq\frac{\beta}{d^{1/4}}\label{eq:vcrit_sup}
\end{equation}
where $d$ is the channel diameter and the constant $\beta=0.03$
m$^{5/4}$/s. We assume $d=69.1$ $\mu$m (i.e., the spacing between
the fibers). Equation \ref{eq:vcrit_sup} can be combined with Eq.\ref{eq:netmassflow}
to give 
\begin{equation}
v_{\mathrm{n,crit}}\simeq\frac{\rho_{\mathrm{s}}\beta}{\rho_{\mathrm{n}}d^{1/4}}
\end{equation}
This may be rewritten (using Eq. \ref{eq:qcritgen}) as a critical
heat flux density: 
\begin{equation}
\dot{q}_{\mathrm{crit,1}}=s\rho T\frac{\rho_{\mathrm{s}}\beta}{\rho_{\mathrm{n}}d^{1/4}}\label{eq:qcrit1}
\end{equation}
In practice, we evaluate Eq. \ref{eq:qcrit1} using the following
expressions for $\rho_{\mathrm{s}}$ \& $\rho_{\mathrm{n}}$ (along
with Eq. \ref{eq:entropy}): 
\begin{equation}
\rho_{\mathrm{n}}(T)=\frac{s(T)}{s(T_{\lambda})}\rho
\end{equation}
\begin{equation}
\rho_{\mathrm{s}}(T)=\rho-\rho_{\mathrm{n}}(T)
\end{equation}
} 
\item { If the onset of turbulence is attributed to the dynamics of the
viscous normal fluid, this will occur when its Reynolds number 
\begin{equation}
\mathrm{Re}=\frac{\rho v_{\mathrm{n}}d}{\mu}\approx1200\label{eq:reynoldsnum}
\end{equation}
where $\mu$ is the viscosity of the normal fluid. Combining Eq. \ref{eq:qcritgen}
\& \ref{eq:reynoldsnum} gives 
\begin{equation}
\dot{q}_{\mathrm{crit,2}}=\frac{1200sT\mu}{d}\label{eq:qcrit2}
\end{equation}
In practice, we evaluate Eq. \ref{eq:qcrit2} using the empirical
expression for the viscosity (valid for $T<1.8$ K)\cite{VanSciver2012}
\begin{equation}
\mu\approx\nu_{5}T^{-5}+\nu_{0}
\end{equation}
} where $\nu_{5}=1.4\times10^{-6}\ \mathrm{Pa\cdot s/K^{5}}$ and
$\nu_{0}=1.4\times10^{-6}\mathrm{\ Pa\cdot s}$. 
\end{enumerate}
From equation \ref{eq:Q1}, we find: 
\begin{equation}
r_{\mathrm{crit,(1,2)}}=\sqrt{\frac{\dot{Q}_{1}}{2\pi\dot{q}_{\mathrm{crit,(1,2)}}}}=\eta_{(1,2)}\sqrt{n_{\mathrm{eff}}}
\end{equation}
Here we have defined $n_{\mathrm{eff}}=n_{\mathrm{circ}}\alpha/N$
(the number of photons absorbed by an individual absorber per cavity
lifetime) and $\eta_{(1,2)}$ is the value of $r_{\mathrm{crit},(1,2)}$
for $n_{\mathrm{eff}}=1$.

Figure \ref{fig:rc} shows $\eta_{(1,2)}(T)$. From this figure it
is apparent that both models (1) and (2) predict a modest temperature
dependence for $\eta$, and that the two models differ by less than
a factor of 4 for the relevant temperature range. Given these qualitative
features and the absence of a strong physical justification for choosing
one model over the other, we will assume in the following analysis
that $\eta$ is a constant. As described in Section \ref{sec:fitting},
$\eta$ will serve as a fit parameter. For the rough estimates presented
in section \ref{subsec:thermsummary} (i.e., prior to the fitting),
we will simply assume $\eta=1.0\times10^{-8}$ m for concreteness.
\begin{figure}[h!]
\centering \includegraphics[width=0.7\textwidth]{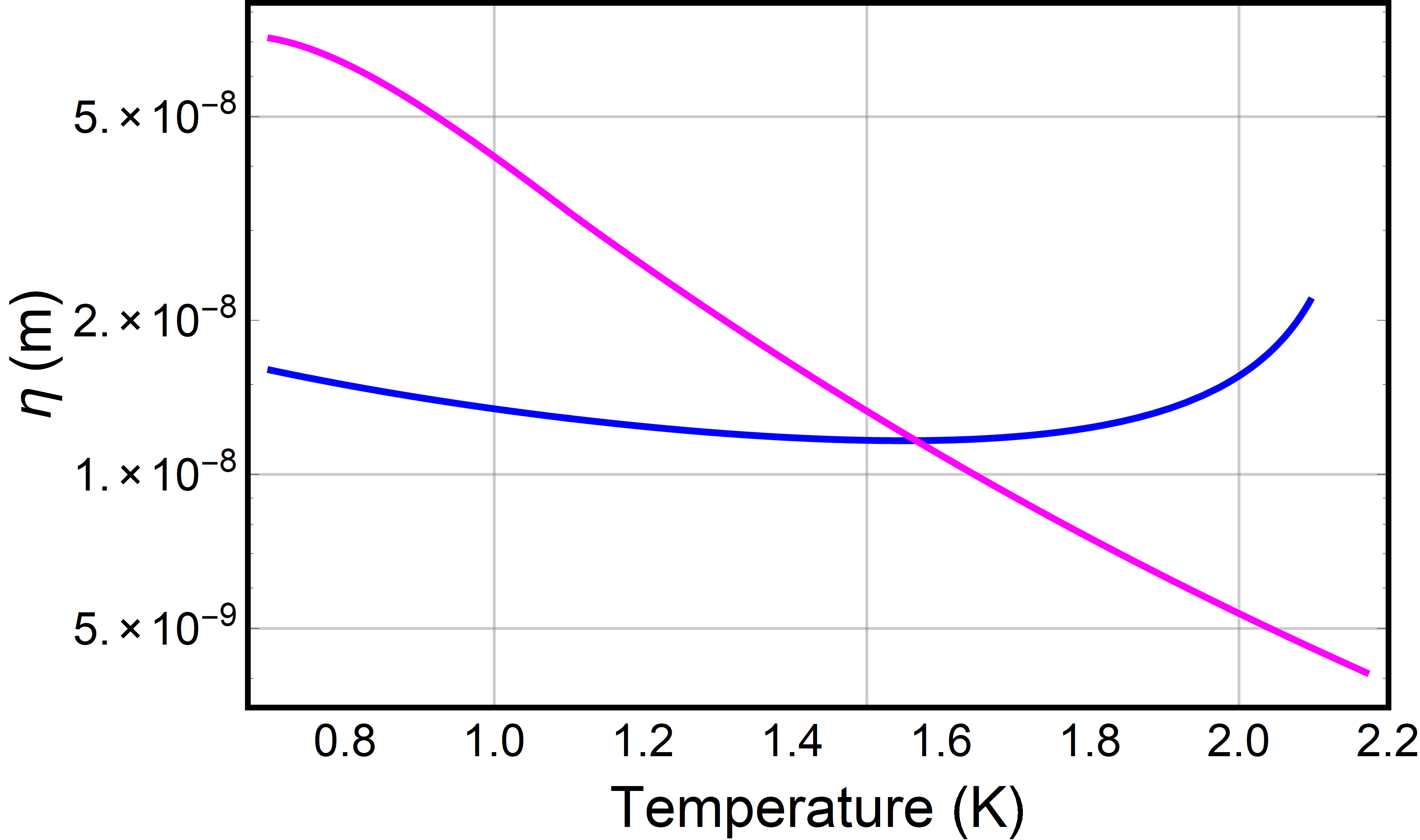}
\caption{The value of $\eta$ as a function of temperature. Blue line: $\eta_1$ (set by superfluid turbulence). Magenta line: $\eta_2$ (set by normal fluid turbulence). }
\label{fig:rc} 
\end{figure}

For the approach described in this section to be valid, one necessary
condition is that the spacing between the absorbers must be larger
than $r_{\mathrm{crit}}$: 
\begin{equation}
N<\frac{r_{\mathrm{opt}}^{2}}{r_{\mathrm{crit}}^{2}}
\end{equation}
In the present device, this is equivalent to the condition 
\begin{equation}
n_{\mathrm{circ}}\alpha<3\times10^{5}
\end{equation}
which is satisfied for all the measurements described here.

\subsubsection{Heat propagation in the ballistic regime}

\label{subsec:ballistic} As described above, heat propagation is
in the ballistic regime for $r>r_{\mathrm{crit}}$. In this regime,
the thermal conductivity of liquid helium in a channel is described
by the following equation \cite{Bertman1968}: 
\begin{equation}
k=\frac{1}{3}Cvd\frac{2-f}{f}\label{eq:heatcondbal}
\end{equation}
Here $C$ is the specific heat per unit volume, $v$ is the sound
velocity, $d$ is the channel diameter (which we take to be 70 $\mu$m
as above), and $f$ is the probability for a phonon to be diffusively
(as opposed to specularly) reflected from the fiber faces.

To estimate $f$, we note that the probability of diffusive scattering
from a rough interface is given by \cite{Bennett1961}: 
\begin{equation}
R_{\mathrm{d}}=1-R_{0}e^{-\frac{4\pi\sigma^{2}}{\lambda^{2}}}
\end{equation}
The constant $R_{0}$ is the interface's reflectivity in the absence
of roughness. We estimate $R_{0}=0.99$ based upon the acoustic impedances
of helium and the DBR materials \cite{Kashkanova2016}. The rms surface
roughness $\sigma$ of similarly prepared fibers was measured to be
0.24 nm \cite{Hunger2012}. For a thermal distribution of phonons
at $T=T_{\mathrm{MC}}$, the most likely wavelength $\lambda_{\mathrm{th}}$
in these experiments ranges from 20 nm (for $T_{\mathrm{MC}}=500$
mK) to 500 nm (for $T_{\mathrm{MC}}=20$ mK). As a result, $\lambda_{\mathrm{th}}\gg\sigma$
for all of the measurements described here, so $R_{\mathrm{d}}\approx1-R_{0}\approx0.01$.
Therefore we set $f=0.01$.

We can then rewrite expression \ref{eq:heatcondbal} as: 
\begin{equation}
k(T)=\xi T^{3}
\end{equation}
where the constant $\xi=3200\mathrm{\ W\cdot m^{-1}\cdot K^{-4}}$.
The expression relating heat flow to temperature gradient is \cite{Pobell1992}:
\begin{equation}
\frac{1}{2}\frac{\dot{Q}}{A}=-k(T)\frac{dT}{dr}\label{eq:2A}
\end{equation}
Here $A=2\pi r^{2}$ is the area over which the heat is distributed.
The factor of 1/2 accounts for the presence of two mirrors. Eq. \ref{eq:2A}
can be rewritten as 
\begin{equation}
\frac{\dot{Q}}{4\pi r^{2}}dr=-\xi T^{3}dT\label{eq:tempdiffeq}
\end{equation}
Assuming that $T(r_{\mathrm{out}})$ = $T_{\mathrm{MC}}$, the temperature
at $r$ can be found by integrating Eq. \ref{eq:tempdiffeq} : 
\begin{equation}
\frac{\dot{Q}}{4\pi}\int_{r_{\mathrm{out}}}^{r}\frac{1}{r^{2}}dr=-\xi\int_{T_{\mathrm{mc}}}^{T}T'^{3}dT'
\end{equation}
\begin{equation}
\frac{\dot{Q}}{4\pi}\left(\frac{1}{r_{\mathrm{out}}}-\frac{1}{r}\right)=-\frac{\xi}{4}(T^{4}-T_{\mathrm{mc}}^{4})
\end{equation}
\begin{equation}
T(r)=\left(T_{\mathrm{mc}}^{4}+\frac{\dot{Q}}{\pi\xi}\left(\frac{1}{r}-\frac{1}{r_{\mathrm{out}}}\right)\right)^{1/4}\label{eq:T}
\end{equation}

Figure \ref{fig:temp3} shows the temperature profile between $r_{\mathrm{crit}}$
and $r_{\mathrm{out}}$ for three different circulating photon numbers
and for $T_{\mathrm{MC}}=50$ mK. The red curve shows the most extreme
case used in this work ($T_{\mathrm{MC}}=50$ mK and $n_{\mathrm{circ}}=100,000$).
Higher $T_{\mathrm{MC}}$ or lower $n_{\mathrm{circ}}$ leads to more
uniform temperature throughout the cavity, as evidenced by figure
\ref{fig:TcrTmc}. The color scale in figure \ref{fig:TcrTmc} shows
the ratio $T(r_{\mathrm{crit}})/T_{\mathrm{MC}}$ for different values
of $T_{\mathrm{MC}}$ and the circulating photon number. The white
dots in the figure show the conditions under which the data in in
the main paper were taken.

\begin{figure}[h!]
\centering \includegraphics[width=0.5\textwidth]{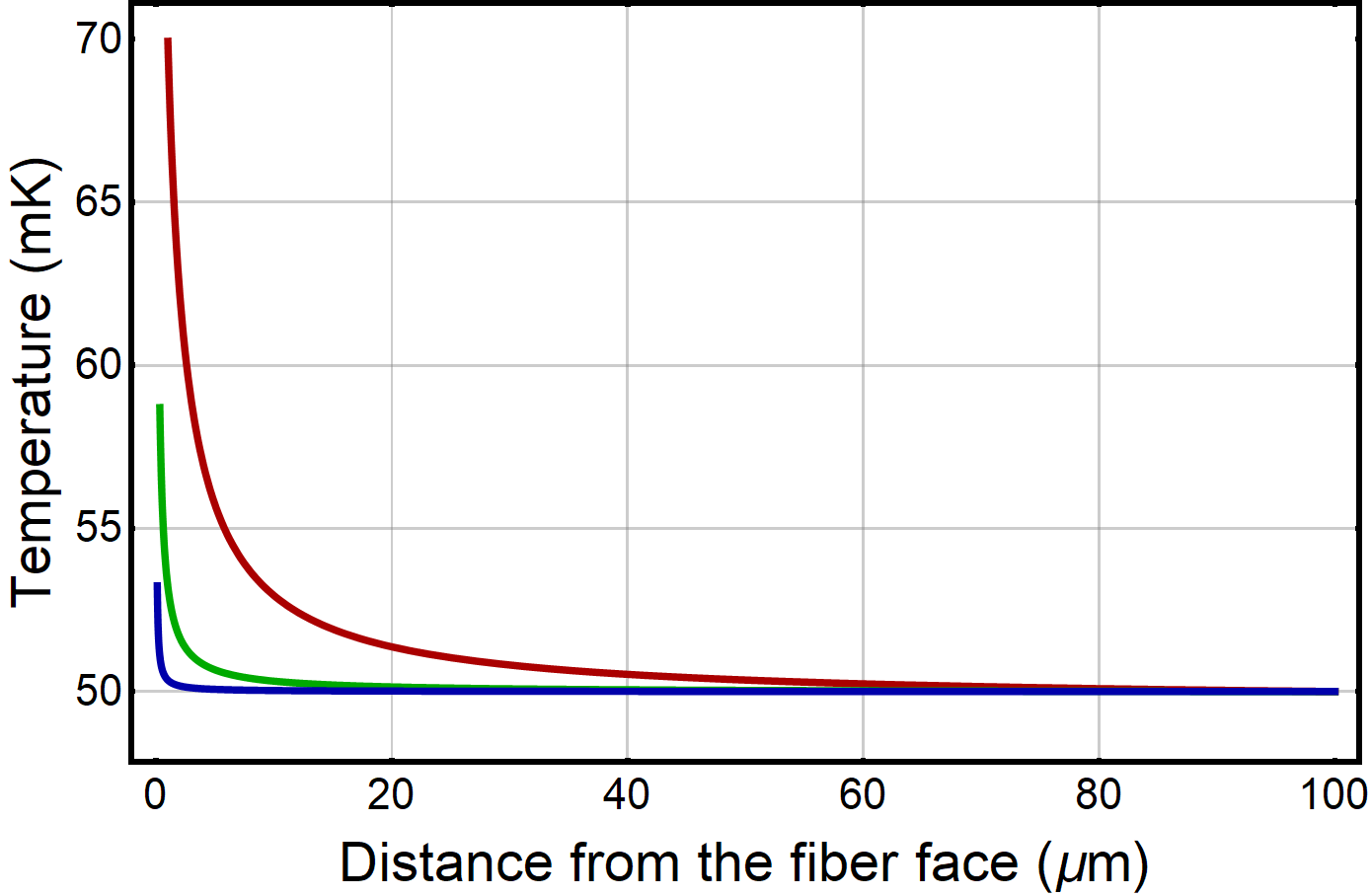}
\caption{The temperature profile between $r_{\mathrm{crit}}$ and $r_{\mathrm{out}}$
in three cases. Blue: $n_{\mathrm{circ}}=1,000$. Green: $n_{\mathrm{circ}}=10,000$.
Red: $n_{\mathrm{circ}}=100,000$. In all three cases $T_{\mathrm{MC}}=50$
mK, $\alpha=0.2$ and each fiber mirror absorbs the same amount of
light. }
\label{fig:temp3} 
\end{figure}

\begin{figure}[h!]
\centering \includegraphics[width=0.5\textwidth]{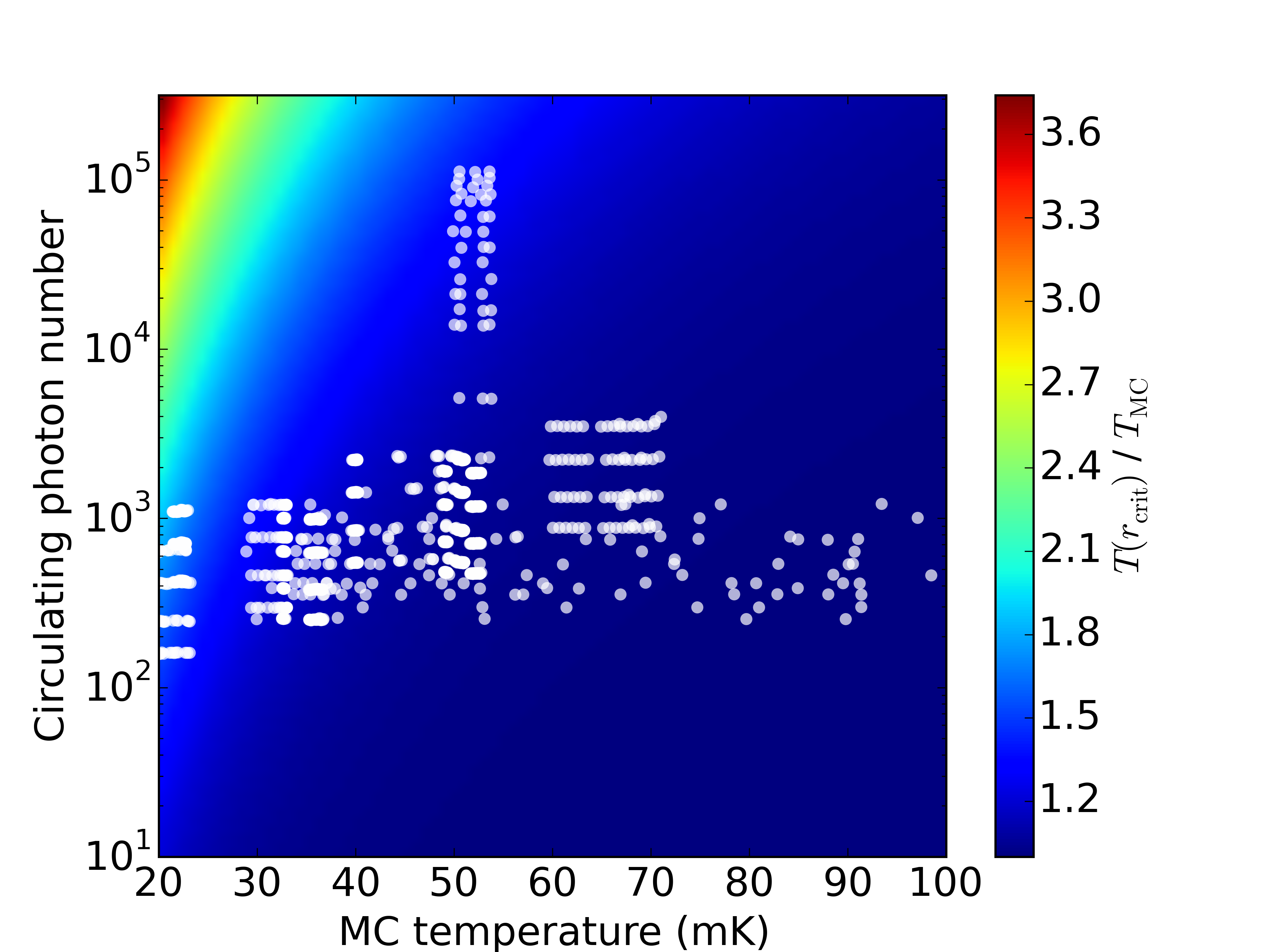}
\caption{Color scale: the ratio $T_{r_{\mathrm{crit}}}/T_{\mathrm{MC}}$ for different values
of mixing chamber temperature and circulating photon number, assuming
$\alpha=0.2$. White points: the conditions under which the data in the main text were taken. }
\label{fig:TcrTmc} 
\end{figure}


\subsubsection{Heat propagation in the Gorter-Mellink regime}

\label{subsec:gm} For heat flux above the critical value (i.e., for
$r<r_{\mathrm{crit}}$), the thermal conductance is described by the
Gorter-Mellink model. This regime is characterized by the following
equation \cite{Arp1970}: 
\begin{equation}
\left(\frac{\dot{Q}_{1}}{A}\right)^{3}=-g(T)\frac{dT}{dr}
\end{equation}
Note that the heat flux from a single absorber $\dot{Q}_{1}$ is used.
The function $g(T)$ is given by: 
\begin{equation}
g(T)=\frac{s^{4}\rho_{s}^{3}T^{3}}{A_{\mathrm{GM}}\rho_{n}}
\end{equation}
Experiments have given a range of values for $A_{\mathrm{GM}}$ \cite{Arp1970};
however there is general agreement that $A_{\mathrm{GM}}\propto T^{3}$.
Using the approximate average of the data in \cite{Arp1970} we take
$A_{\mathrm{GM}}=\alpha_{\mathrm{GM}}T^{3}$ with $\alpha_{\mathrm{GM}}\approx200$
m$\cdot$s/(kg$\cdot$K$^{3}$).

To find the temperature profile inside the critical radius, we integrate
the temperature from the critical radius inward:

\begin{equation}
\left(\frac{\dot{Q}_{1}}{2\pi}\right)^{3}\int_{r_{\mathrm{crit}}}^{r}\frac{1}{r'^{6}}dr'=-\int_{T_{\mathrm{crit}}}^{T}g(T')dT'
\end{equation}
\begin{equation}
\left(\frac{\dot{Q}_{1}}{2\pi}\right)^{3}\frac{1}{5}\left(\frac{1}{r^{5}}-\frac{1}{r_{\mathrm{crit}}^{5}}\right)=f(T)-f(T_{\mathrm{crit}})
\end{equation}
The function $f(T)$ is defined as the indefinite integral of $g(T)$.
It can be written analytically, but the expression is cumbersome so
instead we make use of the fact that it can be approximated (to within
a factor of 4) by:

\begin{equation}
f_{\mathrm{app}}(T)=\beta T^{18}
\end{equation}
Where $\beta=0.5\times10^{8}$ W$^{3}$/(m$^{5}\cdot$ K$^{18}$)}
for 0.7 K \textless{} $T$ \textless{} 2 K (Figure \ref{fig:3}).
\begin{figure}[h!]
\centering \includegraphics[width=0.6\textwidth]{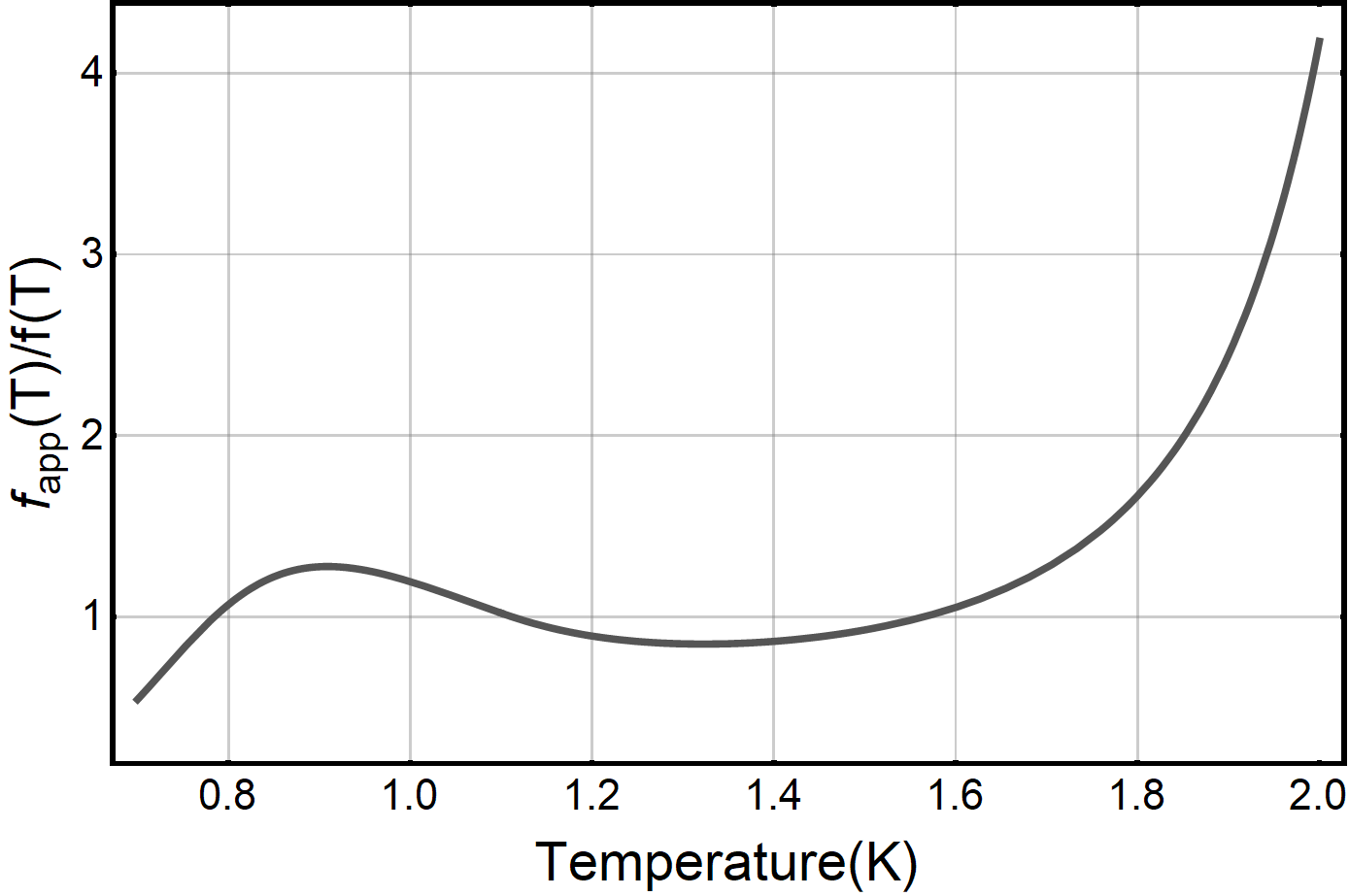}
\caption{The ratio $f_{\mathrm{app}}(T)/f(T)$.}
\label{fig:3} 
\end{figure}

Combining the preceding two equations gives: 
\begin{eqnarray}
T^{18} & = & T_{\mathrm{crit}}^{18}+\left(\frac{\dot{Q}_{1}}{2\pi}\right)^{3}\frac{1}{5\beta}\left(\frac{1}{r^{5}}-\frac{1}{r_{\mathrm{crit}}^{5}}\right)\label{eq:T18a}
\end{eqnarray}
Equation \ref{eq:T18a} can be further simplified by noting that $T_{\mathrm{crit}}$
is close to $T_{\mathrm{MC}}$ (as shown above), which is always smaller
than 300 mK. For $T_{\mathrm{crit}}=300$ mK and photon number $n=100$
photons (the lowest measurable), the second term on the right-hand-side
of equation \ref{eq:T18a} dominates in the the region $0<r<0.99r_{\mathrm{crit}}$),
so to a good approximation the temperature inside the critical radius
depends only on $n_{\mathrm{circ}}$ and not on $T_{\mathrm{MC}}$:

\begin{eqnarray}
T(r) & = & \left(\left(\frac{\dot{Q}_{1}}{2\pi}\right)^{3}\frac{1}{5\beta}\left(\frac{1}{r^{5}}-\frac{1}{r_{\mathrm{crit}}^{5}}\right)\right)^{1/18}\label{eq:t3t4}
\end{eqnarray}

\subsubsection{Temperature profile summary}

\label{subsec:thermsummary} 
\begin{figure}[h!]
\centering \includegraphics[width=0.6\textwidth]{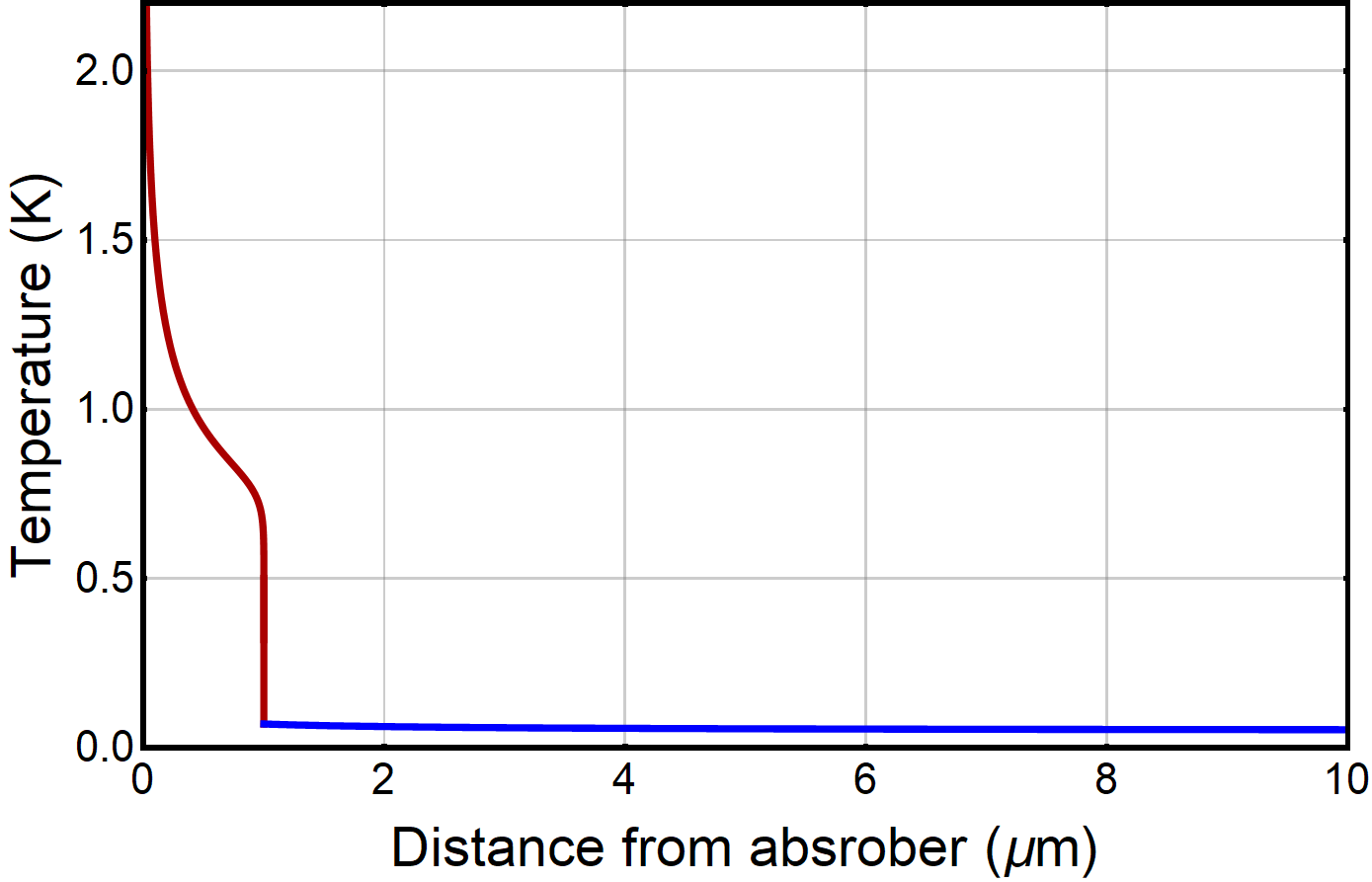}
\caption{The temperature profile inside the critical radius (red), and outside
the critical radius (blue) for $n_{\mathrm{circ}}=10^{5}$, $T_{\mathrm{MC}}=50$
mK, $\alpha=0.2$, $f=0.01$, and $N=2$.}
\label{fig:tempprof} 
\end{figure}

To summarize, the temperature profile
within the cavity is calculated via the following steps: 
\begin{itemize}
\item {Outside of the cylinder defined by the fibers, the temperature is
at the mixing chamber temperature $T(r\geq r_{\mathrm{out}})=T_{\mathrm{MC}}$.} 
\end{itemize}
\begin{itemize}
\item {The total heat radiating into the helium is $\dot{Q}=\hbar\omega_{\mathrm{l}}n_{\mathrm{circ}}\kappa_{\mathrm{int}}\alpha$.} 
\end{itemize}
\begin{itemize}
\item {The heat radiates isotropically into the helium from point-like
absorbers. The amount of heat radiated from each absorber is $\dot{Q}_{1}=\dot{Q}/N$.} 
\item {The heat flux density drops off with distance from the absorber
as $\dot{q}=\frac{\dot{Q}_{1}}{2\pi r^{2}}$. For $r<r_{\mathrm{crit}}$
thermal transport is in the Gorter-Mellink regime. For $r>r_{\mathrm{crit}}$
the propagation is in the ballistic regime.} 
\item { For $r>r_{\mathrm{crit}}$ the temperature is very close to the
mixing chamber temperature.} 
\item {For $r<r_{\mathrm{crit}}$ the temperature is roughly independent
of mixing chamber temperature} 
\end{itemize}
Using equations \ref{eq:T} and \ref{eq:t3t4}, the temperature profile
$T(r)$ can be calculated. Figure \ref{fig:tempprof} shows $T(r)$
for $n_{\mathrm{circ}}=10^{5}$, $T_{\mathrm{MC}}=50$ mK, $\alpha=0.2$,
$f=0.01$, and $N=2$. 


\subsection{Properties of the acoustic mode}

\label{sec:waveeq} The speed of sound $c$, density $\rho$, and
acoustic damping rate $\gamma$ in liquid helium are all functions
of temperature. As a result, the spatial variation of the temperature
(calculated in section \ref{sec:tempdist}) leads to spatial variation
of $c$, $\rho$, and $\gamma$. In this section we calculate how
this influences the frequency $\omega_{\mathrm{ac}}$, linewidth $\gamma_{\mathrm{ac}}$,
and phonon number $n_{\mathrm{ac}}$ of the paraxial acoustic mode
that is the focus of the main text. In this section we start with
the wave equation describing the propagation of sound in an inhomogeneous
fluid (subsection \ref{subsec:waveeqderiv}), find approximate solutions
relevant to the experiments described in the main text (subsection
\ref{subsec:waveeqpert}), and finally use these solutions to provide
expressions for $\omega_{\mathrm{ac}}$, $\gamma_{\mathrm{ac}}$,
and $n_{\mathrm{ac}}$ in terms of the experimentally controlled parameters
$n_{\mathrm{circ}}$ and $T_{\mathrm{MC}}$ (subsections \ref{subsec:waveeqfreq}
- \ref{subsec:waveeqphononnumber}).

\subsubsection{Wave equation for a non-uniform medium}

\label{subsec:waveeqderiv} As described in ref.~\cite{Pierce1981},
linearizing the hydrodynamic equations gives the following expression
for a small-amplitude pressure fluctuation $p({\bf x},t)$ propogating
through a fluid with spatially varying (but static) density $\rho({\bf x})$
and local speed of sound $c({\bf x})$: 
\begin{equation}
\rho({\bf x})\nabla\cdot\left(\frac{1}{\rho({\bf x})}\nabla p({\bf x},t)\right)-\frac{1}{c({\bf x})^{2}}\frac{\partial^{2}p({\bf x},t)}{\partial t^{2}}+\frac{2\delta_{\mathrm{cl}}({\bf x})}{c({\bf x})^{4}}\frac{\partial^{3}p({\bf x},t)}{\partial t^{3}}=0
\end{equation}
where $\delta_{\mathrm{cl}}({\bf x})$ is proportional to the fluid's
viscosity $\mu({\bf x})$~\cite{Pierce1981}. Assuming a solution
of the form $p({\bf x},t)=\pi({\bf x})e^{-i\tilde{\omega}t}$ gives
the following equation for the eigenmode $\pi({\bf x})$ and the (complex)
eigenvalue $\tilde{\omega}$: 
\begin{equation}
\rho({\bf x})\nabla\cdot\left(\frac{1}{\rho({\bf x})}\nabla\pi({\bf x})\right)+\frac{1}{c({\bf x})^{2}}\tilde{\omega}^{2}\pi({\bf x})+i\frac{\gamma({\bf x})}{c({\bf x})^{2}}\tilde{\omega}\pi({\bf x})=0\label{eq:timeind}
\end{equation}
where $\mu({\bf x})$ has been rewritten in terms of the local acoustic
damping rate $\gamma({\bf x}).$

In the following subsections, we assume that the eigenmode is normalized:
\begin{equation}
\int\pi^{2}({\bf x})d^{3}{\bf x}=1
\end{equation}
We also make use of the fact that for our system the mode is confined
by the optical fibers. This imposes the boundary condition $\frac{\partial\pi({\bf x})}{\partial z}=0$
at the fiber surface, where $z$ is the direction along the cavity
axis (and normal to the fiber surfaces).

\subsubsection{Perturbative solutions of the wave equation}

\label{subsec:waveeqpert} Exact solutions to equation \ref{eq:timeind}
are not available unless $c({\bf x})$, $\rho({\bf x})$, and $\gamma({\bf x})$
have very simple forms. To find approximate solutions for arbitrary
$c({\bf x})$, $\rho({\bf x})$, and $\gamma({\bf x})$, we write:
\begin{equation}
c({\bf x})=c_{0}+c_{1}({\bf x})\label{eq:c(x)}
\end{equation}
\begin{equation}
\rho({\bf x})=\rho_{0}+\rho_{1}({\bf x})
\end{equation}
\begin{equation}
\gamma({\bf x})=\gamma_{1}({\bf x})
\end{equation}
\begin{equation}
\pi({\bf x})=\pi_{0}({\bf x})+\pi_{1}({\bf x})
\end{equation}
\begin{equation}
\tilde{\omega}=\omega_{0}+\tilde{\omega}_{1}\label{eq:eigenvalue}
\end{equation}
The spatial variations $c_{1}({\bf x})$, $\rho_{1}({\bf x})$, and
$\gamma_{1}({\bf x})$ are assumed to be small perturbations. Specifcally,
we assume $c_{1}({\bf x})\ll c_{0}$, $\rho_{1}({\bf x})\ll\rho_{0}$,
and $\gamma_{1}({\bf x})\ll\omega_{0}$, and we assume that these
perturbations lead to a small change to the eigenmode ($\pi_{1}({\bf x})$)
and eigenvalue ($\tilde{\omega}_{1}$). The unperturbed eigenmode
$\pi_{0}({\bf x})$ and the unperturbed eigenvalue $\omega_{0}$ are
assumed to solve the wave equation for the uniform lossless fluid (i.e., equation
\ref{eq:timeind} with $c_{1}({\bf x})=\rho_{1}({\bf x})=\gamma_{1}({\bf x})=0$).

By combining equation \ref{eq:timeind} with equations \ref{eq:c(x)}
- \ref{eq:eigenvalue}, and keeping only terms that are first-order
in the perturbations, it is straightforward to find the shifts in
the mode's frequency of oscillation and damping rate that are due
to $c_{1}({\bf x})$, $\rho_{1}({\bf x})$, and $\gamma_{1}({\bf x})$.
These expressions are discussed in the following two subsections.

\subsubsection{Mode frequency}

\label{subsec:waveeqfreq} The perturbation theory described in subsection
\ref{subsec:waveeqpert} gives the first-order change in $\omega_{\mathrm{ac}}$
as: 
\begin{equation}
\delta\omega_{\mathrm{ac}}=\mathrm{Re}[\tilde{\omega}_{1}]=\frac{\omega_{0}}{c_{0}}\int c_{1}({\bf x})\pi_{0}^{2}({\bf x})d^{3}{\bf x}+\frac{c_{0}^{2}}{2\rho_{0}\omega_{0}}\int\pi_{0}({\bf x})(\nabla\rho_{1}({\bf x}))\cdot(\nabla\pi_{0}({\bf x}))d^{3}{\bf x}\label{eq:freqshift}
\end{equation}
The spatial variation in the speed of sound and density arise from
the spatial variation of the temperature: i.e., $c({\bf x})=c(T({\bf x}))$
and $\rho({\bf x})=\rho(T({\bf x}))$. As a result, we can use two
of the main results from section \ref{sec:tempdist} (which are summarized
in subsection \ref{subsec:thermsummary}) to write equation \ref{eq:freqshift}
in a more intuitive form.

First, we assume that in the ballistic region (i.e., whenever the
distance from an absorber is greater than $r_{\mathrm{crit}}$) the
temperature is simply equal to $T_{\mathrm{MC}}$ (and hence independent
of ${\bf x}$ and $n_{\mathrm{circ}}$). Second, we assume that inside
any GM region the temperature is given by equation \ref{eq:t3t4}
(and so depends upon ${\bf x}$ and $n_{\mathrm{circ}}$ but not $T_{\mathrm{MC}}$).
The justification for these assumptions is given in section \ref{sec:tempdist}.

With these assumptions, the mode's frequency of oscillation is conveniently
written as 
\begin{equation}
\omega_{\mathrm{ac}}(n_{\mathrm{circ}},T_{\mathrm{MC}})=\omega_{\mathrm{ac},0}+\delta\omega_{\mathrm{ac}}=\omega_{\mathrm{ac},0}+\delta\omega_{\mathrm{ac,ball}}(T_{\mathrm{MC}})+\delta\omega_{\mathrm{ac,GM}}(n_{\mathrm{circ}})
\end{equation}
In the final expression the first term ($\omega_{\mathrm{ac,0}}$)
is the mode frequency for a uniform lossless fluid with the constants $c$
and $\rho$ set to their $T=0$ values. It is used as a fit parameter.

The second term ($\delta\omega_{\mathrm{ac,ball}}$) is given by equation
\ref{eq:freqshift} but with the integration carried out only over
the ballistic region. In this region $c({\bf x})=c(T_{\mathrm{MC}})$
and $\rho({\bf x})=\rho(T_{\mathrm{MC}})$ are both constants. Combined
with the fact that the ballistic region's volume is much greater than
the GM regions' means that $\delta\omega_{\mathrm{ac,ball}}/\omega_{\mathrm{ac,0}}=c(T_{\mathrm{MC}})/c(T=0)$.
For the range of $T_{\mathrm{MC}}$ used here ($T_{\mathrm{MC}}<300$
mK) theory predicts $c(T)-c(T=0)\propto T^{4},$ or equivalently:
\begin{equation}
\delta\omega_{\mathrm{ac,ball}}=b_{\omega}T_{\mathrm{MC}}^{4}
\end{equation}
The constant $b_{\omega}$ is used as a fit parameter.

The third term ($\delta\omega_{\mathrm{ac,GM}}$) is given by equation
\ref{eq:freqshift} but with the integration carried out only over
the GM regions: 
\begin{align}
\delta\omega_{\mathrm{ac,GM}} & =N\frac{\omega_{0}}{c_{0}}\int_{V_{\mathrm{GM}}}c_{1}(T(r({\bf x})))\pi_{0}^{2}({\bf x})d^{3}{\bf x}\label{eq:freqGM}\\
 & +N\frac{c_{0}^{2}}{2\rho_{0}\omega_{0}}\int_{V_{\mathrm{GM}}}\pi_{0}({\bf x})(\nabla\rho_{1}(T(r({\bf x}))))\cdot(\nabla\pi_{0}({\bf x}))d^{3}{\bf x}
\end{align}
where $r$ is the distance from the absorber, and the factor of $N$
accounts for the total number of absorbers. In practice, we evaluate
equation \ref{eq:freqGM} by: (1) combining equation \ref{eq:t3t4}
(which gives $T(r)$) with interpolations of the data for $c(T)$
and $\rho(T)$ given in \cite{Abraham1969} and \cite{Donnelly1998};
(2) using the approximate one-dimensional form for the unperturbed
eigenmode $\pi_{0}({\bf x})=\sqrt{\frac{2}{d}}\mathrm{cos}(z\omega_{0}/c_{0})$;
and (3) performing the integration numerically over a hemisphere of
radius $r_{\mathrm{crit}}$.

This approach introduces the following fit parameters: $\omega_{\mathrm{ac,0}}$
(the mode's ``bare'' frequency) and $b_{\omega}$ (which appears
in $\delta\omega_{\mathrm{ac,ball}}$). The parameters $N$, $\alpha$,
and $\eta$ (introduced in section \ref{sec:tempdist}) appear in
$\delta\omega_{\mathrm{ac,GM}}$.

\subsubsection{Mode linewidth}

\label{subsec:waveeqdamp} The perturbation theory described in subsection
\ref{subsec:waveeqpert} gives the first-order change in $\gamma_{\mathrm{ac}}$
as: 
\begin{equation}
\delta\gamma_{\mathrm{ac}}=\mathrm{Im}[\tilde{\omega}_{1}]=\int\gamma_{1}({\bf x})\pi_{0}^{2}({\bf x})d^{3}{\bf x}\label{eq:gammachange}
\end{equation}
As in the previous subsection, we write the mode's damping rate as
the sum of three contributions: 
\begin{equation}
\gamma_{\mathrm{ac}}(n_{\mathrm{circ}},T_{\mathrm{MC}})=\gamma_{\mathrm{ac},0}+\delta\gamma_{\mathrm{ac}}=\gamma_{\mathrm{ac},0}+\delta\gamma_{\mathrm{ac,ball}}(T_{\mathrm{MC}})+\delta\gamma_{\mathrm{ac,GM}}(n_{\mathrm{circ}})
\end{equation}
The first term, $\gamma_{\mathrm{ac,0}}$, is the mode's $T=0$ damping
rate, which is due to acoustic radiation from the liquid helium into
the optical fibers (as discussed in Ref. \cite{Kashkanova2016}).

The second term, $\delta\gamma_{\mathrm{ac,ball}}$, is given by equation
\ref{eq:gammachange} but with the integration carried out only over
the ballistic region. For liquid helium at a uniform temperature the
acoustic damping rate is $\propto T^{4}$, so we have 
\begin{equation}
\delta\gamma_{\mathrm{ac,ball}}=b_{\gamma}T_{\mathrm{MC}}^{4}
\end{equation}
The constant $b_{\gamma}$ is used as a fit parameter.

The third term, $\delta\gamma_{\mathrm{ac,GM}}$, is given by equation
\ref{eq:gammachange} but with the integration carried out only over
the GM region. The procedure for evaluating this term is the same
as for evaluating $\delta\omega_{\mathrm{ac,GM}}$: the temperature
profile $T(r)$ is given by equation \ref{eq:t3t4}, while $\gamma(T)$
is given by the theoretical expressions in Ref. \cite{Abraham1969}
for $T<1.7$ K (where theory and experiment show close agreement)
and by interpolating the measurements in Ref. \cite{StPeters1970}
for $T>1.7$ K.

This approach introduces two fit parameters: $\gamma_{\mathrm{ac,0}}$
(the mode's ``bare'' damping) and $b_{\gamma}$ (which appears in
$\delta\gamma_{\mathrm{ac,ball}}$). The parameters $N$, $\alpha$,
and $\eta$ (which were introduced in section \ref{sec:tempdist}) appear in
$\delta\gamma_{\mathrm{ac,GM}}$.  

\subsubsection{The mode phonon number}

\label{subsec:waveeqphononnumber} 
In the main paper the acoustic mode's mean phonon number $n_{\mathrm{ac}}$ is determined from the optical heterodyne signal as $n_{\mathrm{ac}} = (h_{\mathrm{rr}} + h_{\mathrm{bb}} - 1)/2$. In order to facilitate comparison with the thermal model described in this section, we removed the optical damping ("laser cooling") and RPSN contributions from $n_{\mathrm{ac}}$ by plotting $n_{\mathrm{th}} = n_{\mathrm{ac}} (\gamma_{\mathrm{ac,eff}}/\gamma_{\mathrm{ac}}) - n_{\mathrm{O}}\gamma_{\mathrm{ac}}/\gamma_{\mathrm{ac}}$ on the vertical axes of Fig. 4, A and B. The quantity $n_{\mathrm{th}}$ represents the mean number of phonons in the acoustic mode's mechanical bath, as inferred from the optical heterodyne signal.

In this subsection, we extract an estimate of $n_{\mathrm{th}}(T_{\mathrm{MC}},n_{\mathrm{circ}})$ from the thermal model described above. In Fig. 4, B and C of the main paper, this estimate is converted to an effective temperature of the acoustic mode $T_{\mathrm{eff}}=\frac{\hbar \omega_{\mathrm{ac}}}{k_{\mathrm{B}} \ln(1+n_{\mathrm{th}}^{-1})}$ and used as the horizontal axis.

To begin, we note that if the temperature throughout the helium in the cavity were uniform, the acoustic mode's mean phonon number would be 
\begin{equation}
n_{\mathrm{th}}=\frac{n_{\mathrm{fib}}\gamma_{\mathrm{ac,0}}+n_{\mathrm{0}}\gamma_{\mathrm{0}}}{\gamma_{\mathrm{ac,0}}+\gamma_{\mathrm{0}}}\label{eq:uniformphonon}
\end{equation}
where  $n_{0}=1/(e^{\hbar\omega_{\mathrm{ac}}/(k_{\mathrm{B}}T_{0})}-1)$, $n_{\mathrm{fib}}=1/(e^{\hbar\omega_{\mathrm{ac}}/(k_{\mathrm{B}}T_\mathrm{fib})}-1)$, $T_{0}$ is the uniform temperature of the helium in this hypothetical case, $T_{\mathrm{fib}}$ is the temperature of the optical fiber, and $\gamma_{0}=\gamma(T_{0})$.
Since the helium's temperature and damping rate are both non-uniform, we
rewrite equation \ref{eq:uniformphonon} as 
\begin{equation}
n_{\mathrm{th}}=\frac{n_{\mathrm{fib}}\gamma_{\mathrm{ac,0}}+\int n_{\bf x}(T({\bf x}))\gamma(T({\bf x}))\pi_{0}^{2}({\bf x})d^{3}{\bf x}}{\gamma_{\mathrm{ac,0}}+\int\gamma(T({\bf x}))\pi_{0}^{2}({\bf x})d^{3}{\bf x}}\label{eq:distrubtedphonons}
\end{equation}
where 
\begin{equation}
n_{\bf x}(T({\bf x}))=1/(e^{\hbar\omega_{\mathrm{ac}}/(k_{\mathrm{B}}T({\bf x}))}-1) 
\end{equation}
As in subsections \ref{subsec:waveeqfreq} and \ref{subsec:waveeqdamp},
we separate the integrals in equation \ref{eq:distrubtedphonons}
into one integral over the ballistic region and another over the GM
region. This gives 
\begin{equation}
n_{\mathrm{th}}=\frac{n_{\mathrm{fib}}(T_{\mathrm{MC}},n_{\mathrm{circ}})\gamma_{\mathrm{ac,0}}+n_{\mathrm{ball}}(T_{\mathrm{MC}})\delta\gamma_{\mathrm{ac,ball}}(T_{\mathrm{MC}})+f_{\mathrm{GM}}(n_{\mathrm{circ}})}{\gamma_{\mathrm{ac,0}}+\delta\gamma_{\mathrm{ac,ball}}(T_{\mathrm{MC}})+\delta\gamma_{\mathrm{ac,GM}}(n_{\mathrm{circ}})}
\end{equation}
where the dependences upon $T_{\mathrm{MC}}$ and $n_{\mathrm{circ}}$
are noted explicitly, and $n_{\mathrm{ball}}=1/(e^{\hbar\omega_{\mathrm{ac}}/(k_{\mathrm{B}}T_\mathrm{MC})}-1)$.
The function 
\begin{equation}
f_{\mathrm{GM}}=N\int_{V_{\mathrm{GM}}}n_{\bf x}(T(r({\bf x})))\gamma(T(r({\bf x})))\pi_{0}^{2}({\bf x})d^{3}{\bf x}
\end{equation}
We estimate $T_{\mathrm{fib}}$ by assuming that the fiber's thermal
conductivity $\propto T^{k}$, which gives 
\begin{equation}
T_{\mathrm{fib}}=(T_{\mathrm{mc}}^{k+1}+\sigma^{k+1}n_{\mathrm{circ}})^{1/(k+1)}
\end{equation}
The constant $\sigma$ parameterizes how much each circulating photon
contributes to heating of the fiber. Measurements of the thermal conductivity
of amorphous silica at low temperatures \cite{Pobell1992} give $k=1.91$.
Both $\sigma$ and $k$ are used as fitting parameters.




\subsection{Fitting data}

\label{sec:fitting} In the work described here, three properties
of the acoustic mode were measured (i.e. by fitting heterodyne noise spectra
and OMIT/A spectra): $\omega_{\mathrm{ac}}$, $\gamma_{\mathrm{ac}}$,
and $n_{\mathrm{th}}$. They were measured as a function of two externally
controlled parameters: $n_{\mathrm{circ}}$ and $T_{\mathrm{MC}}$.
Expressions for $\omega_{\mathrm{ac}}(n_{\mathrm{circ}},T_{\mathrm{MC}})$,
$\gamma_{\mathrm{ac}}(n_{\mathrm{circ}},T_{\mathrm{MC}})$, and $n_{\mathrm{th}}(n_{\mathrm{circ}},T_{\mathrm{MC}})$
were derived in Sections \ref{sec:tempdist} and \ref{sec:waveeq}, and in Eq. (1) of the main text.
The complete set of measurements of $\omega_{\mathrm{ac}}(n_{\mathrm{circ}},T_{\mathrm{MC}})$,
$\gamma_{\mathrm{ac}}(n_{\mathrm{circ}},T_{\mathrm{MC}})$, and $n_{\mathrm{th}}(n_{\mathrm{circ}},T_{\mathrm{MC}})$
was fit to these expressions in two steps.

\begin{figure}[h!]
\centering \includegraphics[width=0.8\textwidth]{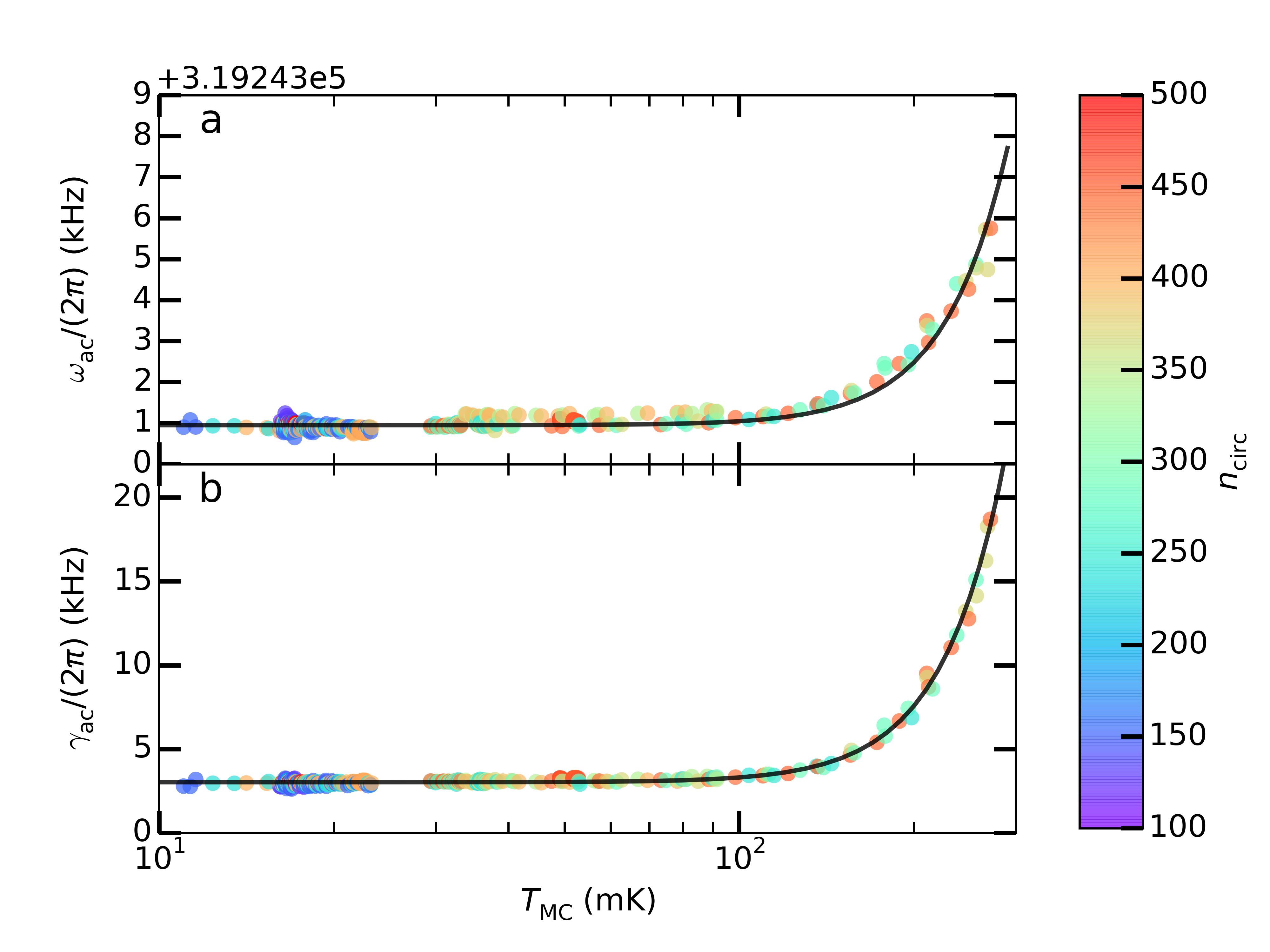}
\caption{(a) Frequency and (b) linewidth vs. $T_{\mathrm{MC}}$ for $n_{\mathrm{circ}}<500$. The dots represent data, with the color corresponding to the circulating photon number $n_{\mathrm{circ}}$.
The solid lines are the fits described in the text.}
\label{fig:firstfits} 
\end{figure}

For the first step, we considered measurements of $\omega_{\mathrm{ac}}(n_{\mathrm{circ}},T_{\mathrm{MC}})$
and $\gamma_{\mathrm{ac}}(n_{\mathrm{circ}},T_{\mathrm{MC}})$ for
which $n_{\mathrm{circ}}<500$. As shown in Fig.~\ref{fig:firstfits},
these measurements are approximately independent of $n_{\mathrm{circ}}$.
From this observation we conclude that for $n_{\mathrm{circ}}<500$
no appreciable heating occurs from optical absorption, so we fit this
data to the expressions derived above with $n_{\mathrm{circ}}$ set
to zero: 
\begin{equation}
\omega_{\mathrm{ac}}(0,T_{\mathrm{MC}})=\omega_{\mathrm{ac},0}+b_{\mathrm{\omega}}T_{\mathrm{MC}}^{4}
\end{equation}
\begin{equation}
\gamma_{\mathrm{ac}}(0,T_{\mathrm{MC}})=\gamma_{\mathrm{ac,0}}+b_{\gamma}T_{\mathrm{MC}}^{4}
\end{equation}
The advantage of this approach is that it employs only four fitting
parameters: $\omega_{\mathrm{ac},0}$, $\gamma_{\mathrm{ac,0}}$, $b_{\mathrm{\omega}}$,
and $b_{\gamma}$. The resulting fits are shown in Fig.~\ref{fig:firstfits}.
The best-fit values of $\omega_{\mathrm{ac},0}$,$\gamma_{\mathrm{ac,0}}$,
$b_{\mathrm{\omega}}$, and $b_{\gamma}$ are listed in Table \ref{table:only},
along with their \emph{a priori} expected values.

For the second step, the complete set of measurements of $\omega_{\mathrm{ac}}(n_{\mathrm{circ}},T_{\mathrm{MC}})$,
$\gamma_{\mathrm{ac}}(n_{\mathrm{circ}},T_{\mathrm{MC}})$, and $n_{\mathrm{th}}(n_{\mathrm{circ}},T_{\mathrm{MC}})$
was fit to the expressions derived in Sections \ref{sec:tempdist}
and \ref{sec:waveeq}: 
\begin{equation}
\omega_{\mathrm{ac}}(n_{\mathrm{circ}},T_{\mathrm{MC}})=\omega_{\mathrm{ac,0}}+\delta\omega_{\mathrm{ac,GM}}(n_{\mathrm{circ}})+b_{\mathrm{\omega}}T_{\mathrm{MC}}^{4}
\end{equation}
\begin{equation}
\gamma_{\mathrm{ac}}(n_{\mathrm{circ}},T_{\mathrm{MC}})=\gamma_{\mathrm{ac,0}}+\delta\gamma_{\mathrm{ac,GM}}(n_{\mathrm{circ}})+b_{\gamma}T_{\mathrm{MC}}^{4}
\end{equation}
\begin{equation}
n_{\mathrm{th}}(n_{\mathrm{circ}},T_{\mathrm{MC}})=\frac{n_{\mathrm{fib}}\gamma_\mathrm{ac,0}+n_\mathrm{ball}b_{\gamma}T_{\mathrm{MC}}^{4}+f_{\mathrm{GM}}(n_{\mathrm{circ}})}{\gamma_{\mathrm{ac,0}}+b_{\gamma}T_{\mathrm{MC}}^{4}+\gamma_{\mathrm{ac,GM}}(n_{\mathrm{circ}})}
\end{equation}
For these fits, the parameters $\omega_{\mathrm{ac},0}$, $\gamma_{\mathrm{ac,0}}$,
$b_{\mathrm{\omega}}$, and $b_{\gamma}$ are fixed to the best-fit
values determined in the first step. This leaves five fitting parameters:
$N$, $\alpha$, $\eta$, $k$, and $\sigma$. The data were fit using these five parameters, with $N$ constrained to be a positive integer and $\alpha$ constrained $\leq 1$. The best fit values are listed in Table~\ref{table:only}, along with the expected values. Although the best fit was achieved with $N=1$, qualitatively similar fits were achieved with $N=2$ and $N=3$. For $N \geq 4$ the fits do not reproduce the qualitative trends in the data.

\begin{table}
\begin{center}
{\renewcommand{\arraystretch}{1.2}
\begin{tabular}{| l V{5} l | l| }
 \hline			
  Parameter [units] 						& Best fit value 					& Expected value \\ \hlineB{5}
 $\omega_\mathrm{bare}/2\pi$ [MHz] 				& $319.24\pm 5\times 10^{-6}$			&$319.24$ \\ \hline
 $\gamma_\mathrm{bare}/2\pi$ [Hz]			& $3026 \pm 6$ 						& $4000\pm2400$ \\\hline
 $b_\mathrm{\omega}/2\pi$ [Hz/K$^{3}$]			& $(0.93 \pm 0.01)\times 10^6$ 		& $1 \times 10^6$ \\\hline
 $b_\mathrm{\gamma}/2\pi$ [Hz/K$^{4}$]		& $(2.79 \pm 0.01)\times 10^6$ 	 	& $2.70\times 10^6$ \\\hline
 $\sigma$ [K]						& $(2.2\pm0.2)\times 10^{-2}$ 	 	& - \\\hline
 $k$ 									& $3.0\pm 0.1$ 					& 1.91 \\\hline
 $\alpha$ 								& $0.69\pm 0.02$ 					& $0.2$ \\\hline
  $N \in \mathbb{Z}^{+}$									& $1$ 							& $\geq 1$ \\\hline
 $\eta$ [m] 						& $(1.01 \pm 0.01)\times 10^{-8}$ 			& $ 5\times 10^{-9}<\eta<7\times 10^{-8}$\\ 
 \hline  
\end{tabular}}
\caption{Fit parameters and the expected values.}
 \label{table:only}
\end{center}
\end{table}

\begin{figure}[h!]
\centering \includegraphics[width=0.8\textwidth]{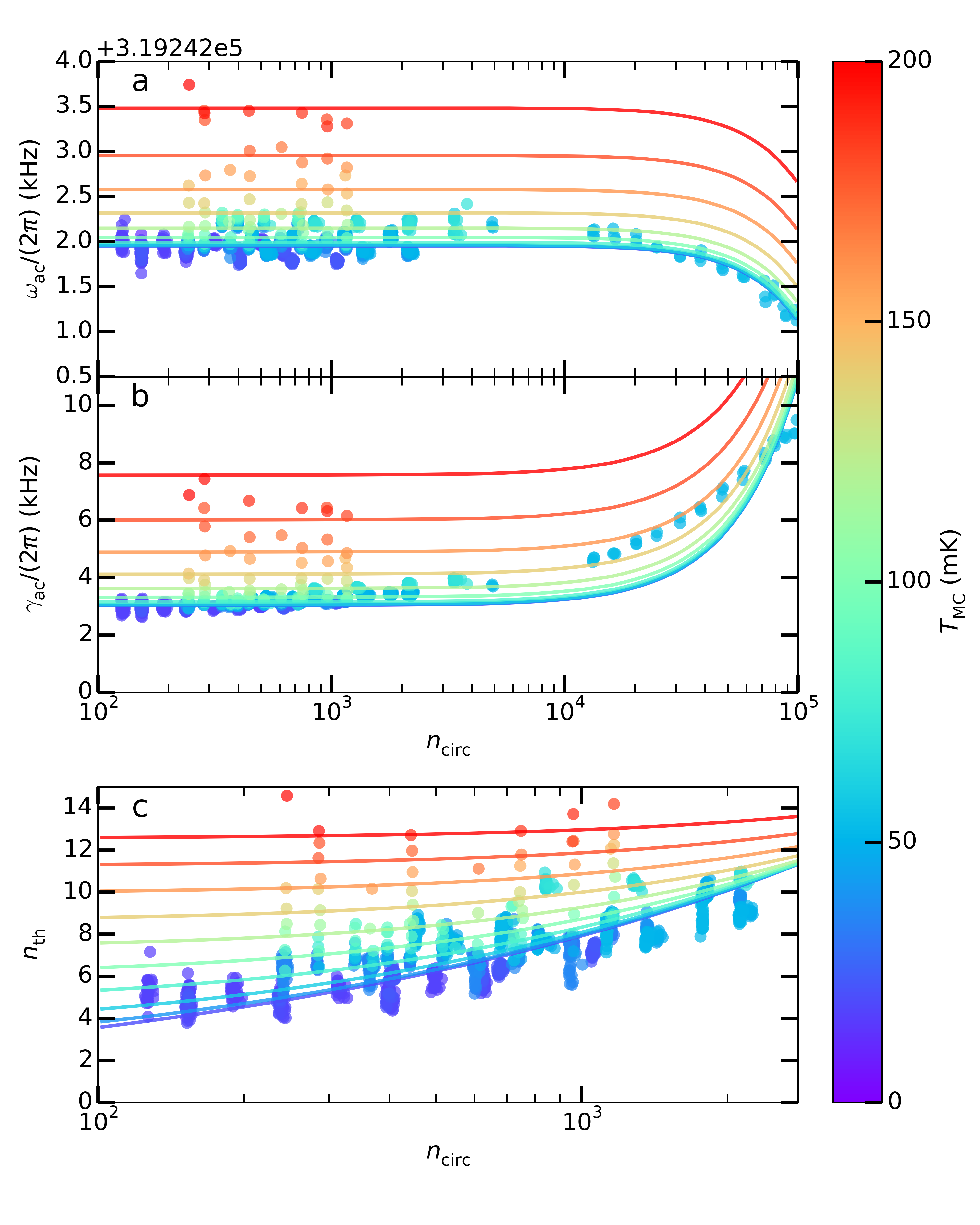}
\caption{(a) Frequency, (b) linewidth and (c) mode phonon number $n_\mathrm{th}$ vs. $n_{\mathrm{circ}}$
for $T_{\mathrm{MC}}<200$ mK. The dots represent data, with the color
corresponding to $T_{\mathrm{MC}}$. The solid lines are the fits
described in the text, with the color corresponding to $T_{\mathrm{MC}}$
as well.}
\label{fig:allfits} 
\end{figure}

\FloatBarrier


\begin{thebibliography}{10}

\bibitem{Carmichael1987}
H.~J. Carmichael.
\newblock {Spectrum of squeezing and photocurrent shot noise: a normally
  ordered treatment}.
\newblock {\em Journal of the Optical Society of America B}, 4(10):1588, oct
  1987.

\bibitem{Clerk2010}
A.~A. Clerk, S.~M. Girvin, F.~Marquardt, R.~J. Schoelkopf, and M.~H. Devoret.
\newblock {Introduction to quantum noise, measurement, and amplification}.
\newblock {\em Reviews of Modern Physics}, 82(2):1155--1208, apr 2010.

\bibitem{Restrepo2011}
J.~Restrepo, J.~Gabelli, C.~Ciuti, and I.~Favero.
\newblock {Classical and quantum theory of photothermal cavity cooling of a
  mechanical oscillator}.
\newblock {\em Comptes Rendus Physique}, 12(9-10):860--870, 2011.

\bibitem{Kashkanova2016}
A.~D. Kashkanova, A.~B. Shkarin, C.~D. Brown, N.~E. Flowers-Jacobs,
  L.~Childress, S.~W. Hoch, L.~Hohmann, K.~Ott, J.~Reichel, and J.~G.~E.
  Harris.
\newblock {Superfluid Brillouin optomechanics}.
\newblock {\em Nat. Phys.}, 1(10):449--450, 2016.

\bibitem{Laseroptik}
{Correspondence with LaserOptik mirror coating company}, 2014.

\bibitem{Sciacca2014}
M.~Sciacca, A.~Sellitto, and D.~Jou.
\newblock {Transition to ballistic regime for heat transport in helium II}.
\newblock {\em Phys. Lett. A}, 378(34):2471--2477, 2014.

\bibitem{Bertman1968}
B.~Bertman and T.~A. Kitchens.
\newblock {Heat transport in superfluid filled capillaries}.
\newblock {\em Cryogenics (Guildf).}, 8(1):36--41, 1968.

\bibitem{Whitworth1958}
R.~W. Whitworth.
\newblock {Experiments on the Flow of Heat in Liquid Helium below 0.7 degrees
  K}.
\newblock {\em Proc. R. Soc. A Math. Phys. Eng. Sci.}, 246(1246):390--405,
  1958.

\bibitem{VanSciver2012}
S.~W. {Van Sciver}.
\newblock {\em {Helium Cryogenics}}.
\newblock Springer New York, New York, NY, 2012.

\bibitem{Bennett1961}
H.~E. Bennett and J.~O. Porteus.
\newblock {Relation between surface roughness and spectral reflectance at
  normal incidence}.
\newblock {\em J. Opt. Soc. Am.}, 51(2):123--129, 1961.

\bibitem{Hunger2012}
D.~Hunger, C.~Deutsch, R.~J. Barbour, R.~J. Warburton, and J.~Reichel.
\newblock {Laser micro-fabrication of concave, low-roughness features in
  silica}.
\newblock {\em AIP Adv.}, 2(1):012119, 2012.

\bibitem{Pobell1992}
F.~Pobell.
\newblock {\em {Matter and methods at low temperatures}}.
\newblock Springer-Verlag Berlin Heidelberg, 2007.

\bibitem{Arp1970}
{V. Arp}.
\newblock {Heat Transport Through Helium II}.
\newblock {\em Cryogenics (Guildf).}, 10:96--105, 1970.

\bibitem{Pierce1981}
A.~D. Pierce.
\newblock {\em {Acoustics}}.
\newblock McGraw-Hill Book Company, New York, NY, USA, 1982.

\bibitem{Abraham1969}
B.~M. Abraham, Y.~Eckstein, J.~B. Ketterson, M.~Kuchnir, and J.~Vignos.
\newblock {Sound propagation in liquid He4}.
\newblock {\em Phys. Rev.}, 181(1):347--373, 1969.

\bibitem{Donnelly1998}
R.~J. Donnelly and C.~F. Barenghi.
\newblock {The observed properties of liquid helium at the saturated vapour
  pressure.}
\newblock {\em J. Phys. Chem. Ref. Data}, 27(6), 1998.

\bibitem{StPeters1970}
R.~L. {St. Peters}, T.~J. Greytak, and G.~B. Benedek.
\newblock {Brillouin scattering measurements of the velocity and attenuation of
  high frequency sound waves in superfluid helium}.
\newblock {\em Opt. Commun.}, 1:412--416, 1970.

\end{thebibliography}
\end{document}